\documentclass[onecollarge]{svjour2}       
\usepackage{graphicx,color,epsfig,rotating}
\usepackage{epsf}
\usepackage{latexsym}
\usepackage{subfigure}
\usepackage{amssymb}

\usepackage{amssymb,latexsym}
\usepackage{graphicx}
\usepackage{epstopdf}
\usepackage{amsmath}
\usepackage{amscd}
\usepackage{bbm}
\usepackage{ifpdf}
\usepackage{booktabs}
\usepackage[vcentermath,stdtext]{youngtab}

\long\def\comment#1{}

\newfont{\bbb}{msbm10 scaled 700}

\newfont{\bb}{msbm10 scaled 1100}

\newcommand{\pv}{{\bf p}}

\newcommand{\sv}{{\bf s}}

\newcommand{\xv}{{\bf x}}
\newcommand{\yv}{{\bf y}}
\newcommand{\zv}{{\bf z}}

\newcommand{\Xc}{{\cal X}}
\newcommand{\Yc}{{\cal Y}}

\definecolor{AliceBlue}{rgb}{0.94,0.97,1.00}
\definecolor{AntiqueWhite1}{rgb}{1.00,0.94,0.86}
\definecolor{AntiqueWhite2}{rgb}{0.93,0.87,0.80}
\definecolor{AntiqueWhite3}{rgb}{0.80,0.75,0.69}
\definecolor{AntiqueWhite4}{rgb}{0.55,0.51,0.47}
\definecolor{AntiqueWhite}{rgb}{0.98,0.92,0.84}
\definecolor{BlanchedAlmond}{rgb}{1.00,0.92,0.80}
\definecolor{BlueViolet}{rgb}{0.54,0.17,0.89}
\definecolor{CadetBlue1}{rgb}{0.60,0.96,1.00}
\definecolor{CadetBlue2}{rgb}{0.56,0.90,0.93}
\definecolor{CadetBlue3}{rgb}{0.48,0.77,0.80}
\definecolor{CadetBlue4}{rgb}{0.33,0.53,0.55}
\definecolor{CadetBlue}{rgb}{0.37,0.62,0.63}
\definecolor{CornflowerBlue}{rgb}{0.39,0.58,0.93}
\definecolor{DarkBlue}{rgb}{0.00,0.00,0.55}
\definecolor{DarkCyan}{rgb}{0.00,0.55,0.55}
\definecolor{DarkGoldenrod1}{rgb}{1.00,0.73,0.06}
\definecolor{DarkGoldenrod2}{rgb}{0.93,0.68,0.05}
\definecolor{DarkGoldenrod3}{rgb}{0.80,0.58,0.05}
\definecolor{DarkGoldenrod4}{rgb}{0.55,0.40,0.03}
\definecolor{DarkGoldenrod}{rgb}{0.72,0.53,0.04}
\definecolor{DarkGray}{rgb}{0.66,0.66,0.66}
\definecolor{DarkGreen}{rgb}{0.00,0.39,0.00}
\definecolor{DarkGrey}{rgb}{0.66,0.66,0.66}
\definecolor{DarkKhaki}{rgb}{0.74,0.72,0.42}
\definecolor{DarkMagenta}{rgb}{0.55,0.00,0.55}
\definecolor{DarkOliveGreen1}{rgb}{0.79,1.00,0.44}
\definecolor{DarkOliveGreen2}{rgb}{0.74,0.93,0.41}
\definecolor{DarkOliveGreen3}{rgb}{0.64,0.80,0.35}
\definecolor{DarkOliveGreen4}{rgb}{0.43,0.55,0.24}
\definecolor{DarkOliveGreen}{rgb}{0.33,0.42,0.18}
\definecolor{DarkOrange1}{rgb}{1.00,0.50,0.00}
\definecolor{DarkOrange2}{rgb}{0.93,0.46,0.00}
\definecolor{DarkOrange3}{rgb}{0.80,0.40,0.00}
\definecolor{DarkOrange4}{rgb}{0.55,0.27,0.00}
\definecolor{DarkOrange}{rgb}{1.00,0.55,0.00}
\definecolor{DarkOrchid1}{rgb}{0.75,0.24,1.00}
\definecolor{DarkOrchid2}{rgb}{0.70,0.23,0.93}
\definecolor{DarkOrchid3}{rgb}{0.60,0.20,0.80}
\definecolor{DarkOrchid4}{rgb}{0.41,0.13,0.55}
\definecolor{DarkOrchid}{rgb}{0.60,0.20,0.80}
\definecolor{DarkRed}{rgb}{0.55,0.00,0.00}
\definecolor{DarkSalmon}{rgb}{0.91,0.59,0.48}
\definecolor{DarkSeaGreen1}{rgb}{0.76,1.00,0.76}
\definecolor{DarkSeaGreen2}{rgb}{0.71,0.93,0.71}
\definecolor{DarkSeaGreen3}{rgb}{0.61,0.80,0.61}
\definecolor{DarkSeaGreen4}{rgb}{0.41,0.55,0.41}
\definecolor{DarkSeaGreen}{rgb}{0.56,0.74,0.56}
\definecolor{DarkSlateBlue}{rgb}{0.28,0.24,0.55}
\definecolor{DarkSlateGray1}{rgb}{0.59,1.00,1.00}
\definecolor{DarkSlateGray2}{rgb}{0.55,0.93,0.93}
\definecolor{DarkSlateGray3}{rgb}{0.47,0.80,0.80}
\definecolor{DarkSlateGray4}{rgb}{0.32,0.55,0.55}
\definecolor{DarkSlateGray}{rgb}{0.18,0.31,0.31}
\definecolor{DarkSlateGrey}{rgb}{0.18,0.31,0.31}
\definecolor{DarkTurquoise}{rgb}{0.00,0.81,0.82}
\definecolor{DarkViolet}{rgb}{0.58,0.00,0.83}
\definecolor{DeepPink1}{rgb}{1.00,0.08,0.58}
\definecolor{DeepPink2}{rgb}{0.93,0.07,0.54}
\definecolor{DeepPink3}{rgb}{0.80,0.06,0.46}
\definecolor{DeepPink4}{rgb}{0.55,0.04,0.31}
\definecolor{DeepPink}{rgb}{1.00,0.08,0.58}
\definecolor{DeepSkyBlue1}{rgb}{0.00,0.75,1.00}
\definecolor{DeepSkyBlue2}{rgb}{0.00,0.70,0.93}
\definecolor{DeepSkyBlue3}{rgb}{0.00,0.60,0.80}
\definecolor{DeepSkyBlue4}{rgb}{0.00,0.41,0.55}
\definecolor{DeepSkyBlue}{rgb}{0.00,0.75,1.00}
\definecolor{DimGray}{rgb}{0.41,0.41,0.41}
\definecolor{DimGrey}{rgb}{0.41,0.41,0.41}
\definecolor{DodgerBlue1}{rgb}{0.12,0.56,1.00}
\definecolor{DodgerBlue2}{rgb}{0.11,0.53,0.93}
\definecolor{DodgerBlue3}{rgb}{0.09,0.45,0.80}
\definecolor{DodgerBlue4}{rgb}{0.06,0.31,0.55}
\definecolor{DodgerBlue}{rgb}{0.12,0.56,1.00}
\definecolor{FloralWhite}{rgb}{1.00,0.98,0.94}
\definecolor{ForestGreen}{rgb}{0.13,0.55,0.13}
\definecolor{GhostWhite}{rgb}{0.97,0.97,1.00}
\definecolor{GreenYellow}{rgb}{0.68,1.00,0.18}
\definecolor{HotPink1}{rgb}{1.00,0.43,0.71}
\definecolor{HotPink2}{rgb}{0.93,0.42,0.65}
\definecolor{HotPink3}{rgb}{0.80,0.38,0.56}
\definecolor{HotPink4}{rgb}{0.55,0.23,0.38}
\definecolor{HotPink}{rgb}{1.00,0.41,0.71}
\definecolor{IndianRed1}{rgb}{1.00,0.42,0.42}
\definecolor{IndianRed2}{rgb}{0.93,0.39,0.39}
\definecolor{IndianRed3}{rgb}{0.80,0.33,0.33}
\definecolor{IndianRed4}{rgb}{0.55,0.23,0.23}
\definecolor{IndianRed}{rgb}{0.80,0.36,0.36}
\definecolor{LavenderBlush1}{rgb}{1.00,0.94,0.96}
\definecolor{LavenderBlush2}{rgb}{0.93,0.88,0.90}
\definecolor{LavenderBlush3}{rgb}{0.80,0.76,0.77}
\definecolor{LavenderBlush4}{rgb}{0.55,0.51,0.53}
\definecolor{LavenderBlush}{rgb}{1.00,0.94,0.96}
\definecolor{LawnGreen}{rgb}{0.49,0.99,0.00}
\definecolor{LemonChiffon1}{rgb}{1.00,0.98,0.80}
\definecolor{LemonChiffon2}{rgb}{0.93,0.91,0.75}
\definecolor{LemonChiffon3}{rgb}{0.80,0.79,0.65}
\definecolor{LemonChiffon4}{rgb}{0.55,0.54,0.44}
\definecolor{LemonChiffon}{rgb}{1.00,0.98,0.80}
\definecolor{LightBlue1}{rgb}{0.75,0.94,1.00}
\definecolor{LightBlue2}{rgb}{0.70,0.87,0.93}
\definecolor{LightBlue3}{rgb}{0.60,0.75,0.80}
\definecolor{LightBlue4}{rgb}{0.41,0.51,0.55}
\definecolor{LightBlue}{rgb}{0.68,0.85,0.90}
\definecolor{LightCoral}{rgb}{0.94,0.50,0.50}
\definecolor{LightCyan1}{rgb}{0.88,1.00,1.00}
\definecolor{LightCyan2}{rgb}{0.82,0.93,0.93}
\definecolor{LightCyan3}{rgb}{0.71,0.80,0.80}
\definecolor{LightCyan4}{rgb}{0.48,0.55,0.55}
\definecolor{LightCyan}{rgb}{0.88,1.00,1.00}
\definecolor{LightGoldenrod1}{rgb}{1.00,0.93,0.55}
\definecolor{LightGoldenrod2}{rgb}{0.93,0.86,0.51}
\definecolor{LightGoldenrod3}{rgb}{0.80,0.75,0.44}
\definecolor{LightGoldenrod4}{rgb}{0.55,0.51,0.30}
\definecolor{LightGoldenrodYellow}{rgb}{0.98,0.98,0.82}
\definecolor{LightGoldenrod}{rgb}{0.93,0.87,0.51}
\definecolor{LightGray}{rgb}{0.83,0.83,0.83}
\definecolor{LightGreen}{rgb}{0.56,0.93,0.56}
\definecolor{LightGrey}{rgb}{0.83,0.83,0.83}
\definecolor{LightPink1}{rgb}{1.00,0.68,0.73}
\definecolor{LightPink2}{rgb}{0.93,0.64,0.68}
\definecolor{LightPink3}{rgb}{0.80,0.55,0.58}
\definecolor{LightPink4}{rgb}{0.55,0.37,0.40}
\definecolor{LightPink}{rgb}{1.00,0.71,0.76}
\definecolor{LightSalmon1}{rgb}{1.00,0.63,0.48}
\definecolor{LightSalmon2}{rgb}{0.93,0.58,0.45}
\definecolor{LightSalmon3}{rgb}{0.80,0.51,0.38}
\definecolor{LightSalmon4}{rgb}{0.55,0.34,0.26}
\definecolor{LightSalmon}{rgb}{1.00,0.63,0.48}
\definecolor{LightSeaGreen}{rgb}{0.13,0.70,0.67}
\definecolor{LightSkyBlue1}{rgb}{0.69,0.89,1.00}
\definecolor{LightSkyBlue2}{rgb}{0.64,0.83,0.93}
\definecolor{LightSkyBlue3}{rgb}{0.55,0.71,0.80}
\definecolor{LightSkyBlue4}{rgb}{0.38,0.48,0.55}
\definecolor{LightSkyBlue}{rgb}{0.53,0.81,0.98}
\definecolor{LightSlateBlue}{rgb}{0.52,0.44,1.00}
\definecolor{LightSlateGray}{rgb}{0.47,0.53,0.60}
\definecolor{LightSlateGrey}{rgb}{0.47,0.53,0.60}
\definecolor{LightSteelBlue1}{rgb}{0.79,0.88,1.00}
\definecolor{LightSteelBlue2}{rgb}{0.74,0.82,0.93}
\definecolor{LightSteelBlue3}{rgb}{0.64,0.71,0.80}
\definecolor{LightSteelBlue4}{rgb}{0.43,0.48,0.55}
\definecolor{LightSteelBlue}{rgb}{0.69,0.77,0.87}
\definecolor{LightYellow1}{rgb}{1.00,1.00,0.88}
\definecolor{LightYellow2}{rgb}{0.93,0.93,0.82}
\definecolor{LightYellow3}{rgb}{0.80,0.80,0.71}
\definecolor{LightYellow4}{rgb}{0.55,0.55,0.48}
\definecolor{LightYellow}{rgb}{1.00,1.00,0.88}
\definecolor{LimeGreen}{rgb}{0.20,0.80,0.20}
\definecolor{MediumAquamarine}{rgb}{0.40,0.80,0.67}
\definecolor{MediumBlue}{rgb}{0.00,0.00,0.80}
\definecolor{MediumOrchid1}{rgb}{0.88,0.40,1.00}
\definecolor{MediumOrchid2}{rgb}{0.82,0.37,0.93}
\definecolor{MediumOrchid3}{rgb}{0.71,0.32,0.80}
\definecolor{MediumOrchid4}{rgb}{0.48,0.22,0.55}
\definecolor{MediumOrchid}{rgb}{0.73,0.33,0.83}
\definecolor{MediumPurple1}{rgb}{0.67,0.51,1.00}
\definecolor{MediumPurple2}{rgb}{0.62,0.47,0.93}
\definecolor{MediumPurple3}{rgb}{0.54,0.41,0.80}
\definecolor{MediumPurple4}{rgb}{0.36,0.28,0.55}
\definecolor{MediumPurple}{rgb}{0.58,0.44,0.86}
\definecolor{MediumSeaGreen}{rgb}{0.24,0.70,0.44}
\definecolor{MediumSlateBlue}{rgb}{0.48,0.41,0.93}
\definecolor{MediumSpringGreen}{rgb}{0.00,0.98,0.60}
\definecolor{MediumTurquoise}{rgb}{0.28,0.82,0.80}
\definecolor{MediumVioletRed}{rgb}{0.78,0.08,0.52}
\definecolor{MidnightBlue}{rgb}{0.10,0.10,0.44}
\definecolor{MintCream}{rgb}{0.96,1.00,0.98}
\definecolor{MistyRose1}{rgb}{1.00,0.89,0.88}
\definecolor{MistyRose2}{rgb}{0.93,0.84,0.82}
\definecolor{MistyRose3}{rgb}{0.80,0.72,0.71}
\definecolor{MistyRose4}{rgb}{0.55,0.49,0.48}
\definecolor{MistyRose}{rgb}{1.00,0.89,0.88}
\definecolor{NavajoWhite1}{rgb}{1.00,0.87,0.68}
\definecolor{NavajoWhite2}{rgb}{0.93,0.81,0.63}
\definecolor{NavajoWhite3}{rgb}{0.80,0.70,0.55}
\definecolor{NavajoWhite4}{rgb}{0.55,0.47,0.37}
\definecolor{NavajoWhite}{rgb}{1.00,0.87,0.68}
\definecolor{NavyBlue}{rgb}{0.00,0.00,0.50}
\definecolor{OldLace}{rgb}{0.99,0.96,0.90}
\definecolor{OliveDrab1}{rgb}{0.75,1.00,0.24}
\definecolor{OliveDrab2}{rgb}{0.70,0.93,0.23}
\definecolor{OliveDrab3}{rgb}{0.60,0.80,0.20}
\definecolor{OliveDrab4}{rgb}{0.41,0.55,0.13}
\definecolor{OliveDrab}{rgb}{0.42,0.56,0.14}
\definecolor{OrangeRed1}{rgb}{1.00,0.27,0.00}
\definecolor{OrangeRed2}{rgb}{0.93,0.25,0.00}
\definecolor{OrangeRed3}{rgb}{0.80,0.22,0.00}
\definecolor{OrangeRed4}{rgb}{0.55,0.15,0.00}
\definecolor{OrangeRed}{rgb}{1.00,0.27,0.00}
\definecolor{PaleGoldenrod}{rgb}{0.93,0.91,0.67}
\definecolor{PaleGreen1}{rgb}{0.60,1.00,0.60}
\definecolor{PaleGreen2}{rgb}{0.56,0.93,0.56}
\definecolor{PaleGreen3}{rgb}{0.49,0.80,0.49}
\definecolor{PaleGreen4}{rgb}{0.33,0.55,0.33}
\definecolor{PaleGreen}{rgb}{0.60,0.98,0.60}
\definecolor{PaleTurquoise1}{rgb}{0.73,1.00,1.00}
\definecolor{PaleTurquoise2}{rgb}{0.68,0.93,0.93}
\definecolor{PaleTurquoise3}{rgb}{0.59,0.80,0.80}
\definecolor{PaleTurquoise4}{rgb}{0.40,0.55,0.55}
\definecolor{PaleTurquoise}{rgb}{0.69,0.93,0.93}
\definecolor{PaleVioletRed1}{rgb}{1.00,0.51,0.67}
\definecolor{PaleVioletRed2}{rgb}{0.93,0.47,0.62}
\definecolor{PaleVioletRed3}{rgb}{0.80,0.41,0.54}
\definecolor{PaleVioletRed4}{rgb}{0.55,0.28,0.36}
\definecolor{PaleVioletRed}{rgb}{0.86,0.44,0.58}
\definecolor{PapayaWhip}{rgb}{1.00,0.94,0.84}
\definecolor{PeachPuff1}{rgb}{1.00,0.85,0.73}
\definecolor{PeachPuff2}{rgb}{0.93,0.80,0.68}
\definecolor{PeachPuff3}{rgb}{0.80,0.69,0.58}
\definecolor{PeachPuff4}{rgb}{0.55,0.47,0.40}
\definecolor{PeachPuff}{rgb}{1.00,0.85,0.73}
\definecolor{PowderBlue}{rgb}{0.69,0.88,0.90}
\definecolor{RosyBrown1}{rgb}{1.00,0.76,0.76}
\definecolor{RosyBrown2}{rgb}{0.93,0.71,0.71}
\definecolor{RosyBrown3}{rgb}{0.80,0.61,0.61}
\definecolor{RosyBrown4}{rgb}{0.55,0.41,0.41}
\definecolor{RosyBrown}{rgb}{0.74,0.56,0.56}
\definecolor{RoyalBlue1}{rgb}{0.28,0.46,1.00}
\definecolor{RoyalBlue2}{rgb}{0.26,0.43,0.93}
\definecolor{RoyalBlue3}{rgb}{0.23,0.37,0.80}
\definecolor{RoyalBlue4}{rgb}{0.15,0.25,0.55}
\definecolor{RoyalBlue}{rgb}{0.25,0.41,0.88}
\definecolor{SaddleBrown}{rgb}{0.55,0.27,0.07}
\definecolor{SandyBrown}{rgb}{0.96,0.64,0.38}
\definecolor{SeaGreen1}{rgb}{0.33,1.00,0.62}
\definecolor{SeaGreen2}{rgb}{0.31,0.93,0.58}
\definecolor{SeaGreen3}{rgb}{0.26,0.80,0.50}
\definecolor{SeaGreen4}{rgb}{0.18,0.55,0.34}
\definecolor{SeaGreen}{rgb}{0.18,0.55,0.34}
\definecolor{SkyBlue1}{rgb}{0.53,0.81,1.00}
\definecolor{SkyBlue2}{rgb}{0.49,0.75,0.93}
\definecolor{SkyBlue3}{rgb}{0.42,0.65,0.80}
\definecolor{SkyBlue4}{rgb}{0.29,0.44,0.55}
\definecolor{SkyBlue}{rgb}{0.53,0.81,0.92}
\definecolor{SlateBlue1}{rgb}{0.51,0.44,1.00}
\definecolor{SlateBlue2}{rgb}{0.48,0.40,0.93}
\definecolor{SlateBlue3}{rgb}{0.41,0.35,0.80}
\definecolor{SlateBlue4}{rgb}{0.28,0.24,0.55}
\definecolor{SlateBlue}{rgb}{0.42,0.35,0.80}
\definecolor{SlateGray1}{rgb}{0.78,0.89,1.00}
\definecolor{SlateGray2}{rgb}{0.73,0.83,0.93}
\definecolor{SlateGray3}{rgb}{0.62,0.71,0.80}
\definecolor{SlateGray4}{rgb}{0.42,0.48,0.55}
\definecolor{SlateGray}{rgb}{0.44,0.50,0.56}
\definecolor{SlateGrey}{rgb}{0.44,0.50,0.56}
\definecolor{SpringGreen1}{rgb}{0.00,1.00,0.50}
\definecolor{SpringGreen2}{rgb}{0.00,0.93,0.46}
\definecolor{SpringGreen3}{rgb}{0.00,0.80,0.40}
\definecolor{SpringGreen4}{rgb}{0.00,0.55,0.27}
\definecolor{SpringGreen}{rgb}{0.00,1.00,0.50}
\definecolor{SteelBlue1}{rgb}{0.39,0.72,1.00}
\definecolor{SteelBlue2}{rgb}{0.36,0.67,0.93}
\definecolor{SteelBlue3}{rgb}{0.31,0.58,0.80}
\definecolor{SteelBlue4}{rgb}{0.21,0.39,0.55}
\definecolor{SteelBlue}{rgb}{0.27,0.51,0.71}
\definecolor{VioletRed1}{rgb}{1.00,0.24,0.59}
\definecolor{VioletRed2}{rgb}{0.93,0.23,0.55}
\definecolor{VioletRed3}{rgb}{0.80,0.20,0.47}
\definecolor{VioletRed4}{rgb}{0.55,0.13,0.32}
\definecolor{VioletRed}{rgb}{0.82,0.13,0.56}
\definecolor{WhiteSmoke}{rgb}{0.96,0.96,0.96}
\definecolor{YellowGreen}{rgb}{0.60,0.80,0.20}
\definecolor{aliceblue}{rgb}{0.94,0.97,1.00}
\definecolor{antiquewhite}{rgb}{0.98,0.92,0.84}
\definecolor{aquamarine1}{rgb}{0.50,1.00,0.83}
\definecolor{aquamarine2}{rgb}{0.46,0.93,0.78}
\definecolor{aquamarine3}{rgb}{0.40,0.80,0.67}
\definecolor{aquamarine4}{rgb}{0.27,0.55,0.45}
\definecolor{aquamarine}{rgb}{0.50,1.00,0.83}
\definecolor{azure1}{rgb}{0.94,1.00,1.00}
\definecolor{azure2}{rgb}{0.88,0.93,0.93}
\definecolor{azure3}{rgb}{0.76,0.80,0.80}
\definecolor{azure4}{rgb}{0.51,0.55,0.55}
\definecolor{azure}{rgb}{0.94,1.00,1.00}
\definecolor{beige}{rgb}{0.96,0.96,0.86}
\definecolor{bisque1}{rgb}{1.00,0.89,0.77}
\definecolor{bisque2}{rgb}{0.93,0.84,0.72}
\definecolor{bisque3}{rgb}{0.80,0.72,0.62}
\definecolor{bisque4}{rgb}{0.55,0.49,0.42}
\definecolor{bisque}{rgb}{1.00,0.89,0.77}
\definecolor{black}{rgb}{0.00,0.00,0.00}
\definecolor{blanchedalmond}{rgb}{1.00,0.92,0.80}
\definecolor{blue1}{rgb}{0.00,0.00,1.00}
\definecolor{blue2}{rgb}{0.00,0.00,0.93}
\definecolor{blue3}{rgb}{0.00,0.00,0.80}
\definecolor{blue4}{rgb}{0.00,0.00,0.55}
\definecolor{blueviolet}{rgb}{0.54,0.17,0.89}
\definecolor{blue}{rgb}{0.00,0.00,1.00}
\definecolor{brown1}{rgb}{1.00,0.25,0.25}
\definecolor{brown2}{rgb}{0.93,0.23,0.23}
\definecolor{brown3}{rgb}{0.80,0.20,0.20}
\definecolor{brown4}{rgb}{0.55,0.14,0.14}
\definecolor{brown}{rgb}{0.65,0.16,0.16}
\definecolor{burlywood1}{rgb}{1.00,0.83,0.61}
\definecolor{burlywood2}{rgb}{0.93,0.77,0.57}
\definecolor{burlywood3}{rgb}{0.80,0.67,0.49}
\definecolor{burlywood4}{rgb}{0.55,0.45,0.33}
\definecolor{burlywood}{rgb}{0.87,0.72,0.53}
\definecolor{cadetblue}{rgb}{0.37,0.62,0.63}
\definecolor{chartreuse1}{rgb}{0.50,1.00,0.00}
\definecolor{chartreuse2}{rgb}{0.46,0.93,0.00}
\definecolor{chartreuse3}{rgb}{0.40,0.80,0.00}
\definecolor{chartreuse4}{rgb}{0.27,0.55,0.00}
\definecolor{chartreuse}{rgb}{0.50,1.00,0.00}
\definecolor{chocolate1}{rgb}{1.00,0.50,0.14}
\definecolor{chocolate2}{rgb}{0.93,0.46,0.13}
\definecolor{chocolate3}{rgb}{0.80,0.40,0.11}
\definecolor{chocolate4}{rgb}{0.55,0.27,0.07}
\definecolor{chocolate}{rgb}{0.82,0.41,0.12}
\definecolor{coral1}{rgb}{1.00,0.45,0.34}
\definecolor{coral2}{rgb}{0.93,0.42,0.31}
\definecolor{coral3}{rgb}{0.80,0.36,0.27}
\definecolor{coral4}{rgb}{0.55,0.24,0.18}
\definecolor{coral}{rgb}{1.00,0.50,0.31}
\definecolor{cornflowerblue}{rgb}{0.39,0.58,0.93}
\definecolor{cornsilk1}{rgb}{1.00,0.97,0.86}
\definecolor{cornsilk2}{rgb}{0.93,0.91,0.80}
\definecolor{cornsilk3}{rgb}{0.80,0.78,0.69}
\definecolor{cornsilk4}{rgb}{0.55,0.53,0.47}
\definecolor{cornsilk}{rgb}{1.00,0.97,0.86}
\definecolor{cyan1}{rgb}{0.00,1.00,1.00}
\definecolor{cyan2}{rgb}{0.00,0.93,0.93}
\definecolor{cyan3}{rgb}{0.00,0.80,0.80}
\definecolor{cyan4}{rgb}{0.00,0.55,0.55}
\definecolor{cyan}{rgb}{0.00,1.00,1.00}
\definecolor{darkblue}{rgb}{0.00,0.00,0.55}
\definecolor{darkcyan}{rgb}{0.00,0.55,0.55}
\definecolor{darkgoldenrod}{rgb}{0.72,0.53,0.04}
\definecolor{darkgray}{rgb}{0.66,0.66,0.66}
\definecolor{darkgreen}{rgb}{0.00,0.39,0.00}
\definecolor{darkgrey}{rgb}{0.66,0.66,0.66}
\definecolor{darkkhaki}{rgb}{0.74,0.72,0.42}
\definecolor{darkmagenta}{rgb}{0.55,0.00,0.55}
\definecolor{darkolive}{rgb}{0.33,0.42,0.18}
\definecolor{darkorange}{rgb}{1.00,0.55,0.00}
\definecolor{darkorchid}{rgb}{0.60,0.20,0.80}
\definecolor{darkred}{rgb}{0.55,0.00,0.00}
\definecolor{darksalmon}{rgb}{0.91,0.59,0.48}
\definecolor{darksea}{rgb}{0.56,0.74,0.56}
\definecolor{darkslate}{rgb}{0.18,0.31,0.31}
\definecolor{darkslate}{rgb}{0.18,0.31,0.31}
\definecolor{darkslate}{rgb}{0.28,0.24,0.55}
\definecolor{darkturquoise}{rgb}{0.00,0.81,0.82}
\definecolor{darkviolet}{rgb}{0.58,0.00,0.83}
\definecolor{deeppink}{rgb}{1.00,0.08,0.58}
\definecolor{deepsky}{rgb}{0.00,0.75,1.00}
\definecolor{dimgray}{rgb}{0.41,0.41,0.41}
\definecolor{dimgrey}{rgb}{0.41,0.41,0.41}
\definecolor{dodgerblue}{rgb}{0.12,0.56,1.00}
\definecolor{firebrick1}{rgb}{1.00,0.19,0.19}
\definecolor{firebrick2}{rgb}{0.93,0.17,0.17}
\definecolor{firebrick3}{rgb}{0.80,0.15,0.15}
\definecolor{firebrick4}{rgb}{0.55,0.10,0.10}
\definecolor{firebrick}{rgb}{0.70,0.13,0.13}
\definecolor{floralwhite}{rgb}{1.00,0.98,0.94}
\definecolor{forestgreen}{rgb}{0.13,0.55,0.13}
\definecolor{gainsboro}{rgb}{0.86,0.86,0.86}
\definecolor{ghostwhite}{rgb}{0.97,0.97,1.00}
\definecolor{gold1}{rgb}{1.00,0.84,0.00}
\definecolor{gold2}{rgb}{0.93,0.79,0.00}
\definecolor{gold3}{rgb}{0.80,0.68,0.00}
\definecolor{gold4}{rgb}{0.55,0.46,0.00}
\definecolor{goldenrod1}{rgb}{1.00,0.76,0.15}
\definecolor{goldenrod2}{rgb}{0.93,0.71,0.13}
\definecolor{goldenrod3}{rgb}{0.80,0.61,0.11}
\definecolor{goldenrod4}{rgb}{0.55,0.41,0.08}
\definecolor{goldenrod}{rgb}{0.85,0.65,0.13}
\definecolor{gold}{rgb}{1.00,0.84,0.00}
\definecolor{gray0}{rgb}{0.00,0.00,0.00}
\definecolor{gray100}{rgb}{1.00,1.00,1.00}
\definecolor{gray10}{rgb}{0.10,0.10,0.10}
\definecolor{gray11}{rgb}{0.11,0.11,0.11}
\definecolor{gray12}{rgb}{0.12,0.12,0.12}
\definecolor{gray13}{rgb}{0.13,0.13,0.13}
\definecolor{gray14}{rgb}{0.14,0.14,0.14}
\definecolor{gray15}{rgb}{0.15,0.15,0.15}
\definecolor{gray16}{rgb}{0.16,0.16,0.16}
\definecolor{gray17}{rgb}{0.17,0.17,0.17}
\definecolor{gray18}{rgb}{0.18,0.18,0.18}
\definecolor{gray19}{rgb}{0.19,0.19,0.19}
\definecolor{gray1}{rgb}{0.01,0.01,0.01}
\definecolor{gray20}{rgb}{0.20,0.20,0.20}
\definecolor{gray21}{rgb}{0.21,0.21,0.21}
\definecolor{gray22}{rgb}{0.22,0.22,0.22}
\definecolor{gray23}{rgb}{0.23,0.23,0.23}
\definecolor{gray24}{rgb}{0.24,0.24,0.24}
\definecolor{gray25}{rgb}{0.25,0.25,0.25}
\definecolor{gray26}{rgb}{0.26,0.26,0.26}
\definecolor{gray27}{rgb}{0.27,0.27,0.27}
\definecolor{gray28}{rgb}{0.28,0.28,0.28}
\definecolor{gray29}{rgb}{0.29,0.29,0.29}
\definecolor{gray2}{rgb}{0.02,0.02,0.02}
\definecolor{gray30}{rgb}{0.30,0.30,0.30}
\definecolor{gray31}{rgb}{0.31,0.31,0.31}
\definecolor{gray32}{rgb}{0.32,0.32,0.32}
\definecolor{gray33}{rgb}{0.33,0.33,0.33}
\definecolor{gray34}{rgb}{0.34,0.34,0.34}
\definecolor{gray35}{rgb}{0.35,0.35,0.35}
\definecolor{gray36}{rgb}{0.36,0.36,0.36}
\definecolor{gray37}{rgb}{0.37,0.37,0.37}
\definecolor{gray38}{rgb}{0.38,0.38,0.38}
\definecolor{gray39}{rgb}{0.39,0.39,0.39}
\definecolor{gray3}{rgb}{0.03,0.03,0.03}
\definecolor{gray40}{rgb}{0.40,0.40,0.40}
\definecolor{gray41}{rgb}{0.41,0.41,0.41}
\definecolor{gray42}{rgb}{0.42,0.42,0.42}
\definecolor{gray43}{rgb}{0.43,0.43,0.43}
\definecolor{gray44}{rgb}{0.44,0.44,0.44}
\definecolor{gray45}{rgb}{0.45,0.45,0.45}
\definecolor{gray46}{rgb}{0.46,0.46,0.46}
\definecolor{gray47}{rgb}{0.47,0.47,0.47}
\definecolor{gray48}{rgb}{0.48,0.48,0.48}
\definecolor{gray49}{rgb}{0.49,0.49,0.49}
\definecolor{gray4}{rgb}{0.04,0.04,0.04}
\definecolor{gray50}{rgb}{0.50,0.50,0.50}
\definecolor{gray51}{rgb}{0.51,0.51,0.51}
\definecolor{gray52}{rgb}{0.52,0.52,0.52}
\definecolor{gray53}{rgb}{0.53,0.53,0.53}
\definecolor{gray54}{rgb}{0.54,0.54,0.54}
\definecolor{gray55}{rgb}{0.55,0.55,0.55}
\definecolor{gray56}{rgb}{0.56,0.56,0.56}
\definecolor{gray57}{rgb}{0.57,0.57,0.57}
\definecolor{gray58}{rgb}{0.58,0.58,0.58}
\definecolor{gray59}{rgb}{0.59,0.59,0.59}
\definecolor{gray5}{rgb}{0.05,0.05,0.05}
\definecolor{gray60}{rgb}{0.60,0.60,0.60}
\definecolor{gray61}{rgb}{0.61,0.61,0.61}
\definecolor{gray62}{rgb}{0.62,0.62,0.62}
\definecolor{gray63}{rgb}{0.63,0.63,0.63}
\definecolor{gray64}{rgb}{0.64,0.64,0.64}
\definecolor{gray65}{rgb}{0.65,0.65,0.65}
\definecolor{gray66}{rgb}{0.66,0.66,0.66}
\definecolor{gray67}{rgb}{0.67,0.67,0.67}
\definecolor{gray68}{rgb}{0.68,0.68,0.68}
\definecolor{gray69}{rgb}{0.69,0.69,0.69}
\definecolor{gray6}{rgb}{0.06,0.06,0.06}
\definecolor{gray70}{rgb}{0.70,0.70,0.70}
\definecolor{gray71}{rgb}{0.71,0.71,0.71}
\definecolor{gray72}{rgb}{0.72,0.72,0.72}
\definecolor{gray73}{rgb}{0.73,0.73,0.73}
\definecolor{gray74}{rgb}{0.74,0.74,0.74}
\definecolor{gray75}{rgb}{0.75,0.75,0.75}
\definecolor{gray76}{rgb}{0.76,0.76,0.76}
\definecolor{gray77}{rgb}{0.77,0.77,0.77}
\definecolor{gray78}{rgb}{0.78,0.78,0.78}
\definecolor{gray79}{rgb}{0.79,0.79,0.79}
\definecolor{gray7}{rgb}{0.07,0.07,0.07}
\definecolor{gray80}{rgb}{0.80,0.80,0.80}
\definecolor{gray81}{rgb}{0.81,0.81,0.81}
\definecolor{gray82}{rgb}{0.82,0.82,0.82}
\definecolor{gray83}{rgb}{0.83,0.83,0.83}
\definecolor{gray84}{rgb}{0.84,0.84,0.84}
\definecolor{gray85}{rgb}{0.85,0.85,0.85}
\definecolor{gray86}{rgb}{0.86,0.86,0.86}
\definecolor{gray87}{rgb}{0.87,0.87,0.87}
\definecolor{gray88}{rgb}{0.88,0.88,0.88}
\definecolor{gray89}{rgb}{0.89,0.89,0.89}
\definecolor{gray8}{rgb}{0.08,0.08,0.08}
\definecolor{gray90}{rgb}{0.90,0.90,0.90}
\definecolor{gray91}{rgb}{0.91,0.91,0.91}
\definecolor{gray92}{rgb}{0.92,0.92,0.92}
\definecolor{gray93}{rgb}{0.93,0.93,0.93}
\definecolor{gray94}{rgb}{0.94,0.94,0.94}
\definecolor{gray95}{rgb}{0.95,0.95,0.95}
\definecolor{gray96}{rgb}{0.96,0.96,0.96}
\definecolor{gray97}{rgb}{0.97,0.97,0.97}
\definecolor{gray98}{rgb}{0.98,0.98,0.98}
\definecolor{gray99}{rgb}{0.99,0.99,0.99}
\definecolor{gray9}{rgb}{0.09,0.09,0.09}
\definecolor{gray}{rgb}{0.75,0.75,0.75}
\definecolor{green1}{rgb}{0.00,1.00,0.00}
\definecolor{green2}{rgb}{0.00,0.93,0.00}
\definecolor{green3}{rgb}{0.00,0.80,0.00}
\definecolor{green4}{rgb}{0.00,0.55,0.00}
\definecolor{greenyellow}{rgb}{0.68,1.00,0.18}
\definecolor{green}{rgb}{0.00,1.00,0.00}
\definecolor{grey0}{rgb}{0.00,0.00,0.00}
\definecolor{grey100}{rgb}{1.00,1.00,1.00}
\definecolor{grey10}{rgb}{0.10,0.10,0.10}
\definecolor{grey11}{rgb}{0.11,0.11,0.11}
\definecolor{grey12}{rgb}{0.12,0.12,0.12}
\definecolor{grey13}{rgb}{0.13,0.13,0.13}
\definecolor{grey14}{rgb}{0.14,0.14,0.14}
\definecolor{grey15}{rgb}{0.15,0.15,0.15}
\definecolor{grey16}{rgb}{0.16,0.16,0.16}
\definecolor{grey17}{rgb}{0.17,0.17,0.17}
\definecolor{grey18}{rgb}{0.18,0.18,0.18}
\definecolor{grey19}{rgb}{0.19,0.19,0.19}
\definecolor{grey1}{rgb}{0.01,0.01,0.01}
\definecolor{grey20}{rgb}{0.20,0.20,0.20}
\definecolor{grey21}{rgb}{0.21,0.21,0.21}
\definecolor{grey22}{rgb}{0.22,0.22,0.22}
\definecolor{grey23}{rgb}{0.23,0.23,0.23}
\definecolor{grey24}{rgb}{0.24,0.24,0.24}
\definecolor{grey25}{rgb}{0.25,0.25,0.25}
\definecolor{grey26}{rgb}{0.26,0.26,0.26}
\definecolor{grey27}{rgb}{0.27,0.27,0.27}
\definecolor{grey28}{rgb}{0.28,0.28,0.28}
\definecolor{grey29}{rgb}{0.29,0.29,0.29}
\definecolor{grey2}{rgb}{0.02,0.02,0.02}
\definecolor{grey30}{rgb}{0.30,0.30,0.30}
\definecolor{grey31}{rgb}{0.31,0.31,0.31}
\definecolor{grey32}{rgb}{0.32,0.32,0.32}
\definecolor{grey33}{rgb}{0.33,0.33,0.33}
\definecolor{grey34}{rgb}{0.34,0.34,0.34}
\definecolor{grey35}{rgb}{0.35,0.35,0.35}
\definecolor{grey36}{rgb}{0.36,0.36,0.36}
\definecolor{grey37}{rgb}{0.37,0.37,0.37}
\definecolor{grey38}{rgb}{0.38,0.38,0.38}
\definecolor{grey39}{rgb}{0.39,0.39,0.39}
\definecolor{grey3}{rgb}{0.03,0.03,0.03}
\definecolor{grey40}{rgb}{0.40,0.40,0.40}
\definecolor{grey41}{rgb}{0.41,0.41,0.41}
\definecolor{grey42}{rgb}{0.42,0.42,0.42}
\definecolor{grey43}{rgb}{0.43,0.43,0.43}
\definecolor{grey44}{rgb}{0.44,0.44,0.44}
\definecolor{grey45}{rgb}{0.45,0.45,0.45}
\definecolor{grey46}{rgb}{0.46,0.46,0.46}
\definecolor{grey47}{rgb}{0.47,0.47,0.47}
\definecolor{grey48}{rgb}{0.48,0.48,0.48}
\definecolor{grey49}{rgb}{0.49,0.49,0.49}
\definecolor{grey4}{rgb}{0.04,0.04,0.04}
\definecolor{grey50}{rgb}{0.50,0.50,0.50}
\definecolor{grey51}{rgb}{0.51,0.51,0.51}
\definecolor{grey52}{rgb}{0.52,0.52,0.52}
\definecolor{grey53}{rgb}{0.53,0.53,0.53}
\definecolor{grey54}{rgb}{0.54,0.54,0.54}
\definecolor{grey55}{rgb}{0.55,0.55,0.55}
\definecolor{grey56}{rgb}{0.56,0.56,0.56}
\definecolor{grey57}{rgb}{0.57,0.57,0.57}
\definecolor{grey58}{rgb}{0.58,0.58,0.58}
\definecolor{grey59}{rgb}{0.59,0.59,0.59}
\definecolor{grey5}{rgb}{0.05,0.05,0.05}
\definecolor{grey60}{rgb}{0.60,0.60,0.60}
\definecolor{grey61}{rgb}{0.61,0.61,0.61}
\definecolor{grey62}{rgb}{0.62,0.62,0.62}
\definecolor{grey63}{rgb}{0.63,0.63,0.63}
\definecolor{grey64}{rgb}{0.64,0.64,0.64}
\definecolor{grey65}{rgb}{0.65,0.65,0.65}
\definecolor{grey66}{rgb}{0.66,0.66,0.66}
\definecolor{grey67}{rgb}{0.67,0.67,0.67}
\definecolor{grey68}{rgb}{0.68,0.68,0.68}
\definecolor{grey69}{rgb}{0.69,0.69,0.69}
\definecolor{grey6}{rgb}{0.06,0.06,0.06}
\definecolor{grey70}{rgb}{0.70,0.70,0.70}
\definecolor{grey71}{rgb}{0.71,0.71,0.71}
\definecolor{grey72}{rgb}{0.72,0.72,0.72}
\definecolor{grey73}{rgb}{0.73,0.73,0.73}
\definecolor{grey74}{rgb}{0.74,0.74,0.74}
\definecolor{grey75}{rgb}{0.75,0.75,0.75}
\definecolor{grey76}{rgb}{0.76,0.76,0.76}
\definecolor{grey77}{rgb}{0.77,0.77,0.77}
\definecolor{grey78}{rgb}{0.78,0.78,0.78}
\definecolor{grey79}{rgb}{0.79,0.79,0.79}
\definecolor{grey7}{rgb}{0.07,0.07,0.07}
\definecolor{grey80}{rgb}{0.80,0.80,0.80}
\definecolor{grey81}{rgb}{0.81,0.81,0.81}
\definecolor{grey82}{rgb}{0.82,0.82,0.82}
\definecolor{grey83}{rgb}{0.83,0.83,0.83}
\definecolor{grey84}{rgb}{0.84,0.84,0.84}
\definecolor{grey85}{rgb}{0.85,0.85,0.85}
\definecolor{grey86}{rgb}{0.86,0.86,0.86}
\definecolor{grey87}{rgb}{0.87,0.87,0.87}
\definecolor{grey88}{rgb}{0.88,0.88,0.88}
\definecolor{grey89}{rgb}{0.89,0.89,0.89}
\definecolor{grey8}{rgb}{0.08,0.08,0.08}
\definecolor{grey90}{rgb}{0.90,0.90,0.90}
\definecolor{grey91}{rgb}{0.91,0.91,0.91}
\definecolor{grey92}{rgb}{0.92,0.92,0.92}
\definecolor{grey93}{rgb}{0.93,0.93,0.93}
\definecolor{grey94}{rgb}{0.94,0.94,0.94}
\definecolor{grey95}{rgb}{0.95,0.95,0.95}
\definecolor{grey96}{rgb}{0.96,0.96,0.96}
\definecolor{grey97}{rgb}{0.97,0.97,0.97}
\definecolor{grey98}{rgb}{0.98,0.98,0.98}
\definecolor{grey99}{rgb}{0.99,0.99,0.99}
\definecolor{grey9}{rgb}{0.09,0.09,0.09}
\definecolor{grey}{rgb}{0.75,0.75,0.75}
\definecolor{honeydew1}{rgb}{0.94,1.00,0.94}
\definecolor{honeydew2}{rgb}{0.88,0.93,0.88}
\definecolor{honeydew3}{rgb}{0.76,0.80,0.76}
\definecolor{honeydew4}{rgb}{0.51,0.55,0.51}
\definecolor{honeydew}{rgb}{0.94,1.00,0.94}
\definecolor{hotpink}{rgb}{1.00,0.41,0.71}
\definecolor{indianred}{rgb}{0.80,0.36,0.36}
\definecolor{ivory1}{rgb}{1.00,1.00,0.94}
\definecolor{ivory2}{rgb}{0.93,0.93,0.88}
\definecolor{ivory3}{rgb}{0.80,0.80,0.76}
\definecolor{ivory4}{rgb}{0.55,0.55,0.51}
\definecolor{ivory}{rgb}{1.00,1.00,0.94}
\definecolor{khaki1}{rgb}{1.00,0.96,0.56}
\definecolor{khaki2}{rgb}{0.93,0.90,0.52}
\definecolor{khaki3}{rgb}{0.80,0.78,0.45}
\definecolor{khaki4}{rgb}{0.55,0.53,0.31}
\definecolor{khaki}{rgb}{0.94,0.90,0.55}
\definecolor{lavenderblush}{rgb}{1.00,0.94,0.96}
\definecolor{lavender}{rgb}{0.90,0.90,0.98}
\definecolor{lawngreen}{rgb}{0.49,0.99,0.00}
\definecolor{lemonchiffon}{rgb}{1.00,0.98,0.80}
\definecolor{lightblue}{rgb}{0.68,0.85,0.90}
\definecolor{lightcoral}{rgb}{0.94,0.50,0.50}
\definecolor{lightcyan}{rgb}{0.88,1.00,1.00}
\definecolor{lightgoldenrod}{rgb}{0.93,0.87,0.51}
\definecolor{lightgoldenrod}{rgb}{0.98,0.98,0.82}
\definecolor{lightgray}{rgb}{0.83,0.83,0.83}
\definecolor{lightgreen}{rgb}{0.56,0.93,0.56}
\definecolor{lightgrey}{rgb}{0.83,0.83,0.83}
\definecolor{lightpink}{rgb}{1.00,0.71,0.76}
\definecolor{lightsalmon}{rgb}{1.00,0.63,0.48}
\definecolor{lightsea}{rgb}{0.13,0.70,0.67}
\definecolor{lightsky}{rgb}{0.53,0.81,0.98}
\definecolor{lightslate}{rgb}{0.47,0.53,0.60}
\definecolor{lightslate}{rgb}{0.47,0.53,0.60}
\definecolor{lightslate}{rgb}{0.52,0.44,1.00}
\definecolor{lightsteel}{rgb}{0.69,0.77,0.87}
\definecolor{lightyellow}{rgb}{1.00,1.00,0.88}
\definecolor{limegreen}{rgb}{0.20,0.80,0.20}
\definecolor{linen}{rgb}{0.98,0.94,0.90}
\definecolor{magenta1}{rgb}{1.00,0.00,1.00}
\definecolor{magenta2}{rgb}{0.93,0.00,0.93}
\definecolor{magenta3}{rgb}{0.80,0.00,0.80}
\definecolor{magenta4}{rgb}{0.55,0.00,0.55}
\definecolor{magenta}{rgb}{1.00,0.00,1.00}
\definecolor{maroon1}{rgb}{1.00,0.20,0.70}
\definecolor{maroon2}{rgb}{0.93,0.19,0.65}
\definecolor{maroon3}{rgb}{0.80,0.16,0.56}
\definecolor{maroon4}{rgb}{0.55,0.11,0.38}
\definecolor{maroon}{rgb}{0.69,0.19,0.38}
\definecolor{mediumaquamarine}{rgb}{0.40,0.80,0.67}
\definecolor{mediumblue}{rgb}{0.00,0.00,0.80}
\definecolor{mediumorchid}{rgb}{0.73,0.33,0.83}
\definecolor{mediumpurple}{rgb}{0.58,0.44,0.86}
\definecolor{mediumsea}{rgb}{0.24,0.70,0.44}
\definecolor{mediumslate}{rgb}{0.48,0.41,0.93}
\definecolor{mediumspring}{rgb}{0.00,0.98,0.60}
\definecolor{mediumturquoise}{rgb}{0.28,0.82,0.80}
\definecolor{mediumviolet}{rgb}{0.78,0.08,0.52}
\definecolor{midnightblue}{rgb}{0.10,0.10,0.44}
\definecolor{mintcream}{rgb}{0.96,1.00,0.98}
\definecolor{mistyrose}{rgb}{1.00,0.89,0.88}
\definecolor{moccasin}{rgb}{1.00,0.89,0.71}
\definecolor{navajowhite}{rgb}{1.00,0.87,0.68}
\definecolor{navyblue}{rgb}{0.00,0.00,0.50}
\definecolor{navy}{rgb}{0.00,0.00,0.50}
\definecolor{oldlace}{rgb}{0.99,0.96,0.90}
\definecolor{olivedrab}{rgb}{0.42,0.56,0.14}
\definecolor{orange1}{rgb}{1.00,0.65,0.00}
\definecolor{orange2}{rgb}{0.93,0.60,0.00}
\definecolor{orange3}{rgb}{0.80,0.52,0.00}
\definecolor{orange4}{rgb}{0.55,0.35,0.00}
\definecolor{orangered}{rgb}{1.00,0.27,0.00}
\definecolor{orange}{rgb}{1.00,0.65,0.00}
\definecolor{orchid1}{rgb}{1.00,0.51,0.98}
\definecolor{orchid2}{rgb}{0.93,0.48,0.91}
\definecolor{orchid3}{rgb}{0.80,0.41,0.79}
\definecolor{orchid4}{rgb}{0.55,0.28,0.54}
\definecolor{orchid}{rgb}{0.85,0.44,0.84}
\definecolor{palegoldenrod}{rgb}{0.93,0.91,0.67}
\definecolor{palegreen}{rgb}{0.60,0.98,0.60}
\definecolor{paleturquoise}{rgb}{0.69,0.93,0.93}
\definecolor{paleviolet}{rgb}{0.86,0.44,0.58}
\definecolor{papayawhip}{rgb}{1.00,0.94,0.84}
\definecolor{peachpuff}{rgb}{1.00,0.85,0.73}
\definecolor{peru}{rgb}{0.80,0.52,0.25}
\definecolor{pink1}{rgb}{1.00,0.71,0.77}
\definecolor{pink2}{rgb}{0.93,0.66,0.72}
\definecolor{pink3}{rgb}{0.80,0.57,0.62}
\definecolor{pink4}{rgb}{0.55,0.39,0.42}
\definecolor{pink}{rgb}{1.00,0.75,0.80}
\definecolor{plum1}{rgb}{1.00,0.73,1.00}
\definecolor{plum2}{rgb}{0.93,0.68,0.93}
\definecolor{plum3}{rgb}{0.80,0.59,0.80}
\definecolor{plum4}{rgb}{0.55,0.40,0.55}
\definecolor{plum}{rgb}{0.87,0.63,0.87}
\definecolor{powderblue}{rgb}{0.69,0.88,0.90}
\definecolor{purple1}{rgb}{0.61,0.19,1.00}
\definecolor{purple2}{rgb}{0.57,0.17,0.93}
\definecolor{purple3}{rgb}{0.49,0.15,0.80}
\definecolor{purple4}{rgb}{0.33,0.10,0.55}
\definecolor{purple}{rgb}{0.63,0.13,0.94}
\definecolor{red1}{rgb}{1.00,0.00,0.00}
\definecolor{red2}{rgb}{0.93,0.00,0.00}
\definecolor{red3}{rgb}{0.80,0.00,0.00}
\definecolor{red4}{rgb}{0.55,0.00,0.00}
\definecolor{red}{rgb}{1.00,0.00,0.00}
\definecolor{rosybrown}{rgb}{0.74,0.56,0.56}
\definecolor{royalblue}{rgb}{0.25,0.41,0.88}
\definecolor{saddlebrown}{rgb}{0.55,0.27,0.07}
\definecolor{salmon1}{rgb}{1.00,0.55,0.41}
\definecolor{salmon2}{rgb}{0.93,0.51,0.38}
\definecolor{salmon3}{rgb}{0.80,0.44,0.33}
\definecolor{salmon4}{rgb}{0.55,0.30,0.22}
\definecolor{salmon}{rgb}{0.98,0.50,0.45}
\definecolor{sandybrown}{rgb}{0.96,0.64,0.38}
\definecolor{seagreen}{rgb}{0.18,0.55,0.34}
\definecolor{seashell1}{rgb}{1.00,0.96,0.93}
\definecolor{seashell2}{rgb}{0.93,0.90,0.87}
\definecolor{seashell3}{rgb}{0.80,0.77,0.75}
\definecolor{seashell4}{rgb}{0.55,0.53,0.51}
\definecolor{seashell}{rgb}{1.00,0.96,0.93}
\definecolor{sienna1}{rgb}{1.00,0.51,0.28}
\definecolor{sienna2}{rgb}{0.93,0.47,0.26}
\definecolor{sienna3}{rgb}{0.80,0.41,0.22}
\definecolor{sienna4}{rgb}{0.55,0.28,0.15}
\definecolor{sienna}{rgb}{0.63,0.32,0.18}
\definecolor{skyblue}{rgb}{0.53,0.81,0.92}
\definecolor{slateblue}{rgb}{0.42,0.35,0.80}
\definecolor{slategray}{rgb}{0.44,0.50,0.56}
\definecolor{slategrey}{rgb}{0.44,0.50,0.56}
\definecolor{snow1}{rgb}{1.00,0.98,0.98}
\definecolor{snow2}{rgb}{0.93,0.91,0.91}
\definecolor{snow3}{rgb}{0.80,0.79,0.79}
\definecolor{snow4}{rgb}{0.55,0.54,0.54}
\definecolor{snow}{rgb}{1.00,0.98,0.98}
\definecolor{springgreen}{rgb}{0.00,1.00,0.50}
\definecolor{steelblue}{rgb}{0.27,0.51,0.71}
\definecolor{tan1}{rgb}{1.00,0.65,0.31}
\definecolor{tan2}{rgb}{0.93,0.60,0.29}
\definecolor{tan3}{rgb}{0.80,0.52,0.25}
\definecolor{tan4}{rgb}{0.55,0.35,0.17}
\definecolor{tan}{rgb}{0.82,0.71,0.55}
\definecolor{thistle1}{rgb}{1.00,0.88,1.00}
\definecolor{thistle2}{rgb}{0.93,0.82,0.93}
\definecolor{thistle3}{rgb}{0.80,0.71,0.80}
\definecolor{thistle4}{rgb}{0.55,0.48,0.55}
\definecolor{thistle}{rgb}{0.85,0.75,0.85}
\definecolor{tomato1}{rgb}{1.00,0.39,0.28}
\definecolor{tomato2}{rgb}{0.93,0.36,0.26}
\definecolor{tomato3}{rgb}{0.80,0.31,0.22}
\definecolor{tomato4}{rgb}{0.55,0.21,0.15}
\definecolor{tomato}{rgb}{1.00,0.39,0.28}
\definecolor{turquoise1}{rgb}{0.00,0.96,1.00}
\definecolor{turquoise2}{rgb}{0.00,0.90,0.93}
\definecolor{turquoise3}{rgb}{0.00,0.77,0.80}
\definecolor{turquoise4}{rgb}{0.00,0.53,0.55}
\definecolor{turquoise}{rgb}{0.25,0.88,0.82}
\definecolor{violetred}{rgb}{0.82,0.13,0.56}
\definecolor{violet}{rgb}{0.93,0.51,0.93}
\definecolor{wheat1}{rgb}{1.00,0.91,0.73}
\definecolor{wheat2}{rgb}{0.93,0.85,0.68}
\definecolor{wheat3}{rgb}{0.80,0.73,0.59}
\definecolor{wheat4}{rgb}{0.55,0.49,0.40}
\definecolor{wheat}{rgb}{0.96,0.87,0.70}
\definecolor{whitesmoke}{rgb}{0.96,0.96,0.96}
\definecolor{white}{rgb}{1.00,1.00,1.00}
\definecolor{yellow1}{rgb}{1.00,1.00,0.00}
\definecolor{yellow2}{rgb}{0.93,0.93,0.00}
\definecolor{yellow3}{rgb}{0.80,0.80,0.00}
\definecolor{yellow4}{rgb}{0.55,0.55,0.00}
\definecolor{yellowgreen}{rgb}{0.60,0.80,0.20}
\definecolor{yellow}{rgb}{1.00,1.00,0.00}

\newcommand{\ignore}[1]{}

 \usepackage{mathptmx}      
 \journalname{Journal of Statistical Physics}
\begin{document}

\title{Statistical Physics Analysis of Maximum a Posteriori Estimation for Multi-channel Hidden Markov Models 
}


\author{Avik Halder         \and
        Ansuman Adhikary 
}


\institute{Avik Halder \at
              Department of Physics and Astronomy, University of Southern California, Los Angeles, CA 90089, USA \\
              Tel.: +1-213-740-7914\\
              Fax: +1-213-740-6653\\
              \email{ahalder@usc.edu}           
           \and
           Ansuman Adhikary \at
              Department of Electrical Engineering, University of Southern California, Los Angeles, CA 90089, USA\\
              Tel.: +1-213-222-8519\\
              \email{adhikary@usc.edu}
}

\date{Received: date / Accepted: date}

\maketitle

\begin{abstract}

The performance of Maximum a posteriori (MAP) estimation is studied analytically for binary symmetric multi-channel Hidden Markov processes. We reduce the estimation problem to  a 1D Ising spin model and define order parameters that correspond to different characteristics of the MAP-estimated sequence. The solution to the MAP estimation problem has different operational regimes separated by first order phase transitions. The transition points for $L$-channel system with identical noise levels, are uniquely determined by $L$ being odd or even, irrespective of the actual number of channels. We demonstrate that for lower noise intensities, the number of solutions is uniquely determined for odd $L$, whereas for even $L$ there are exponentially many solutions. We also develop a semi analytical approach to calculate the estimation error without resorting to brute force simulations. Finally, we examine the tradeoff between a system with single low-noise channel and one with multiple noisy channels.

\end{abstract}

\maketitle

\section{Introduction}
\label{sec:intro}

In many dynamical systems  direct observation of the internal system state is not possible. Instead, one has  noisy observations of those states. When the internal dynamics is governed by a Markovian process, the resulting systems are known as Hidden Markov
Models, or HMMs. HMM is the simplest tool to model the systems where a correlated data passes through a noisy channel\cite{{Xu},{Bengio}}. It is used extensively to model such physical systems
and finds applications in various areas such as signal processing,
speech recognition, bioinformatics and so on \cite{{rabiner},{ephraim}}.

One of the major problems underlying HMMs is the estimation of the
hidden state sequence given the observations, which is usually done through
{\em maximum a posteriori} (MAP) estimation technique. A natural  characteristic of MAP estimation is the accuracy of estimation, i.e., the closeness of the original and estimated sequence. Another important characteristic is the number of solutions   it produces in response to a given observed sequence. This is related to the concept of {\em trackability}, which describes whether the number of hidden state sequences grows polynomially (trackable) or exponentially (non-trackable) with the length of the observed sequence. For so called {\em weak models}~\footnote{Weak Models can be described as a simplification of HMMs where one specifies possible transitions and observations from a given state without assigning  probabilities.}, it was established that there  is a sharp transition between trackable and non-trackable regimes as one varies the noise~\cite{Sheng}.  A generalization to the probabilistic case was done for a  binary symmetric HMM \cite{allahverdyan2009maximum}, where it was shown that the non-trackability is related to finite zero-temperature entropy of an appropriately defined Ising model.  In particular, \cite{allahverdyan2009maximum} demonstrated that  there is a critical noise intensity above which the number of MAP solutions is exponential in the length of the observed sequence.

In this work we study the performance of MAP estimation for the scenario where the hidden dynamics is observed via multiple observation channels,  which we refer to as {\em multi-channel} HMMs. Multi channel HMM is a more robust way of modeling
multi channel signals, such as in sensor networks where local readings from different sensors are used to make inference about the
underlying process. Note that the channels can generally have different characteristics, i.e., measurement noise. Here we assume that the readings of the channels are conditionally independent given the underlying hidden state.

One of our main results is that the presence of additional observation channels does not always improve the inference in the sense of above characteristics. In particular, in systems with even number of channels and identical noise intensities, there is always an exponential number of solutions for {\em any} noise intensity, indicated by a finite zero-temperature entropy of the corresponding statistical physics systems. Intuitively, this happens because of {\em conflicting} observations from different channels, which produces a macroscopic frustration of spins in the system. Furthermore, for two-channel systems with generally different noise characteristics, we calculate the phase boundary  between the regions of zero and non-zero entropy regimes in the parameter space. In all the systems studied with single or multiple channels, we observe discontinuous phase transitions in the thermodynamic quantities with the variation of noise. The points of phase transitions are the same for all the even number channel systems. This is also true for any of the odd number channel systems but the transition points are different from that of the even number channel systems.

For general $L$-channel systems with identical noise in the channels, we calculate the different statistical characteristics of MAP for this scenario, and find that the average error between the true and estimated sequences reduces with the addition of channels.
Furthermore, we bring in the notion of channel cost which lets us explore the tradeoff between the total cost and error that one can tolerate in the system. Finally, we also analyze the performance of MAP estimation for Gaussian noise, and find that the entropy is
always zero, so that one has exactly one solution for every observed
sequence. This suggests that the existence of exponentially degenerate solution relates to the discreteness.

Let us provide a brief outline of the structure of our paper: We start by
providing some general information about Maximum a Posteriori
estimation in Section~\ref{sec:background}. We introduce  binary
symmetric HMMs and describe the mapping to an Ising model in
Section~\ref{sec:bshmm}. In Section~\ref{sec:statphys} we describe the
statistical physics approach to MAP estimation for multi-channel
binary HMMs and define appropriate order parameters for
characterizing MAP performance.  Section~\ref{sec:recursion}
describes the solution of the model. The recurrent states are given in Section~\ref{sec:recurrent} followed by presentation of
results in Section~\ref{sec:results}. Section \ref{sec:twochannel}
focuses on a detailed analysis for two observation channel scenario and Section
\ref{sec:Lchannel} provides results for the multiple channel scenario.
In Section \ref{sec:Lchannel_cost} we discuss about the cost of designing a multiple channel system and its impact on the system performance relative to a single channel system.
Finally, we conclude by discussing our results and future work. In
the appendix, we give details about analytical calculation of MAP
accuracy and elaborate on the Gaussian observation model along with the
main differences from the binary symmetric case.

\section{MAP Estimation}
\label{sec:background}

\begin{figure}[h]
  \centering
  \includegraphics[width=20cm,bb = 50 525 620 750]{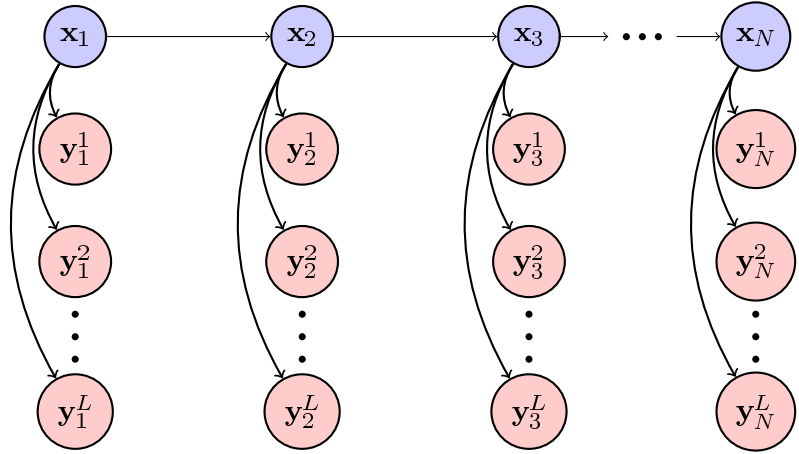}\\
  \caption{(Color online) Hidden Markov model with multiple observation sequences}
  \label{fig:fig8}
\end{figure}

The present work focuses on a specific class of stochastic
processes, namely, the binary symmetric HMMs although the techniques
of MAP estimation can be applied to general stochastic processes
too. In this section, we give a brief idea on MAP estimation for
generalized HMMs and defer to the study of binary symmetric HMMs in
Section \ref{sec:bshmm}.

Let us consider $\xv=(x_1,\ldots,x_N)$ as the signal generating sequence.  The observation in $L$ different channels at every time instant can be generically written as $\underline{\yv} =
(\yv^1,\ldots,\yv^L)$ with $\yv^i=(y_1^i,\ldots,y_N^i)\vrule_{\{i=1,\ldots,L\}}$. Here $\xv$ and $\underline{\yv}$ are the
realizations of discrete time random processes $\mathcal{X}$ and
$\mathcal{Y}$; with $i$ being the index for the observation channel
as shown in Figure \ref{fig:fig8}. $\mathbf{y}^i$'s are the noisy
observations of $\mathcal{X}$ obtained from different observation
channels. The observations from different channels are mutually
independent and are described by the conditional probability
$\textbf{p}(\yv^i|\xv)$. The observed sequences $\yv^i$'s are
considered to be given along with the probabilities
$\textbf{p}(\yv^i|\xv)$ and $\textbf{p}(\xv)$. Thus, we are required
to find the generating sequence $\xv$ from the given sets of
observation sequences.

MAP provides a method to estimate the
generating sequence ${\hat{\xv}}$ on the basis of the observations
$\yv^i$. ${\hat{\xv}}$ is found by maximizing over $\xv$ the
posterior probability,

$$\textbf{p}(\xv|\yv^1,\ldots,\yv^L)=\textbf{p}(\yv^1,\ldots,\yv^L|\xv)\textbf{p}(\xv)/\{\sum_{\yv^1,\ldots,\yv^L}
\textbf{p}(\yv^1,\ldots,\yv^L)\}$$

Since $\textbf{p}(\yv^1,\ldots,\yv^L)$ does not depend on $\xv$,
we can equivalently minimize the Hamiltonian which is given by,

\begin{equation}\label{eq:MAP1}
H(\underline{\yv},\xv) \equiv -\log[\textbf{p}(\yv^1,\ldots,\yv^L|\xv)\textbf{p}(\xv)]
\end{equation}

where by \emph{log} we imply natural logarithm. Because of the mutual independence between the observations
conditioned on $\xv$, we can rewrite $H(\underline{\yv},\xv)$ as
\begin{equation}\label{eq:MAP1a}
H(\underline{\yv},\xv) \equiv - \log[\textbf{p}(\xv)] - \sum_{i=1}^L
\log[\textbf{p}(\yv^i|\xv)]
\end{equation}
The advantage of using $H(\underline{\yv},\xv)$ is that, if
$\mathcal{Y}$ is ergodic\footnote{Ergodicity implies time average is
equal to ensemble average, i.e., an ergodic process has the same
behavior averaged over time as averaged over the space of all its
states.} (which we will assume in the rest of the paper), then for
$N>>1$, $H(\underline{\yv},\hat{\xv}(\underline{\yv}))$ will be
independent from $\underline{\yv}$, $\forall \underline{\yv} \in
\Omega_N(\mathcal{Y}) $, $\Omega_N(\mathcal{Y})$ being the typical
set of $\mathcal{Y}$\cite{cover1991elements}.
The typical set is the set of sequences
whose sample entropy is close to the true entropy, where entropy is
a measure of uncertainty of a random variable and is a function of
the probability distribution of the sequences in $\Yc$. The typical
set has total probability close to one, a consequence of the
asymptotic equipartition property (AEP) \cite{cover1991elements}.
Thus for $N
>> 1$, the sequence $(\yv^1,\ldots,\yv^L)$ will lie in the typical
set of $\Yc$ with high probability. As a result, we have
$\sum\limits_{\underline{\yv} \in \Omega_N(\mathcal{Y})}
\textbf{p}(\underline{\yv}) \rightarrow 1$ and all elements of
$\Omega_N(\mathcal{Y})$ have equal probability. We can take
advantage of this and use for
$H(\underline{\yv},\hat{\xv}(\underline{\yv}))$ the averaged
quantity
$\sum\limits_{\underline{\yv}} \textbf{p}(\underline{\yv})H(\underline{\yv},\hat{\xv}(\underline{\yv})) $.

Let us consider some extreme cases of noise in the observation channels. Since we are dealing with more than one observation channel we have to consider the overall contribution from different channels with varying noise. When we refer to the weak and strong noise in the channels we will be referring with respect to the noisiest channel of the whole set. The other channels will be relatively cleaner and modify the overall performance.  For the case $(i)$ when the noise applied to the channels is very weak, $\textbf{p}(\xv|\underline{\yv}) \cong
\prod_{i=1}^L
 \prod_{k=1}^N \delta(x_k - y_k^i)$. This results in recovering the
generating sequence almost exactly and $(ii)$ for the case of strong noise in channels the
estimation is prior dominated i.e.
$\textbf{p}(\xv|\underline{\yv}) \cong \textbf{p}(\xv)$, and hence is not informative. Without any prior imposed,
$\textbf{p}(\xv) \propto const.$. Here, the MAP estimation reduces
to the maximum likelihood (ML) estimation scheme. It should be noted that using ML estimate we can also obtain the exact sequence in the low noise regime.\\

Minimization of Hamiltonian from \eqref{eq:MAP1} for a given
$\underline{\yv}$ can be readily done using the Viterbi algorithm,
but it produces one single optimal estimate
$\hat{\xv}(\underline{\yv})$. For completeness we should also seek
for other possible sequences $\hat{\xv}^{[\gamma]}(\underline{\yv})$
for which $H(\underline{\yv},\hat{\xv}^{[\gamma]}(\underline{\yv}))$
( greater than $H(\underline{\yv},\hat{\xv}(\underline{\yv}))$) is
almost equal to the optimal estimated Hamiltonian in large $N$
limit. Under that approximation we have $\lim_{N \rightarrow \infty}
\frac{H(\underline{\yv},\hat{\xv}^{[\gamma]}(\underline{\yv}))}{N}=\lim_{N
\rightarrow \infty}
\frac{H(\underline{\yv},\hat{\xv}(\underline{\yv}))}{N}$. These
obtained sequences from the MAP estimate are equivalent for $N
\rightarrow \infty$ and can be listed as,
\begin{equation}\label{eq:MAP1a}
\hat{\xv}^{[\gamma]}(\underline{\yv}), \gamma=1,\ldots ,
\mathcal{N}(\underline{\yv})
\end{equation}
with $\mathcal{N}(\underline{\yv})$ denoting the number of such
possible sequences. If $\log \mathcal{N} (\underline{\yv}) \propto
N$, the ergodicity argument can be used to obtain the logarithm of the number of solutions
for the observed typical sequence,
\begin{equation}\label{eq:MAP2}
\Theta=\sum\limits_{\underline{\yv}}\textbf{p}(\underline{\yv})\log
\mathcal{N}(\underline{\yv})
\end{equation}
In the limit $N \rightarrow \infty$, a finite value of $\theta=\frac{\Theta}{N}$ implies that there are exponentially many
outcomes of minimizing $H(\underline{\yv},\xv)$ over
$\xv$. The term $\Theta$ is called entropy from analogy with the Ising spin model as will be explained in detail in the subsequent sections.

Various moments of
$\hat{\xv}^{[\gamma]}(\underline{\yv})$ are calculated, which are random variables
due to the dependence on $\underline{\yv}$. The knowledge of the moments along with the error analysis is employed to characterize the accuracy of the estimation. For weak noise these moments are close to the original process
$\mathcal{X}$. We also evaluate the average overlap
between estimated sequences $\hat{\xv}^{[\gamma]}(\underline{\yv})$
and the observed sequences $\underline{\yv}$. If the overlap is close to one, it would imply a observation dominated estimation of the sequence.

\section{Binary symmetric Hidden Markov Processes}
\label{sec:bshmm}

\subsection{Definition}
\label{sec:defnbinaryHMM}
\begin{figure}[h]
  \centering
   \includegraphics[width=20cm,bb = 50 650 620 720]{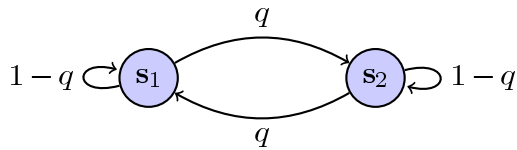}\\
  \caption{(Color online) State Diagram for a Binary symmetric Markov chain}
  \label{fig:fig6}
\end{figure}

We analyze a  binary, discrete-time Markov
stochastic process $\mathcal{X}=(x_1,x_2,\ldots,x_N)$. Each random variable
$x_k$ can have only two realizations $x_k=\pm1$. The Markov feature
implies,
\begin{equation}\label{eq:MAP3}
\textbf{p}(\xv)=\prod_{k=2}^Np(x_k | x_{k-1})p(x_1)
\end{equation}
where $p(x_k\vrule x_{k-1})$ is the time-independent transition
probability of the Markov process. The state diagram for the binary,
discrete time Markov process is shown in Figure \ref{fig:fig6}. We parameterize the
binary symmetric situation by a single number
$0<q<1$, with $p(1|1)=p(-1|-1)=1-q$ and $p(1|-1)=p(-1| 1)=q$ and the
stationary state distribution is considered to be uniform $p_{\rm
st}(1)=p_{\rm st}(-1)=\frac{1}{2}$. The noise process is assumed to
be memory-less, time independent and unbiased. Thus,
\begin{equation}\label{eq:MAP4}
\textbf{p}(\underline{\yv} |\xv)=\prod_{i=1}^L\prod_{k=1}^N\pi^i(y_k^i
| x_k),\ \ \  y_k^i=\pm 1
\end{equation}
where $\pi^i(y_k^i|x_k)$ is the probability of observing $y_k$ in
channel $i$ given the state $x_k$. For the $i^{\rm{th}}$ channel $\pi^i
(-1|1)=\pi^i (1|-1)=\epsilon_i,\pi^i (1|1)=\pi^i
(-1|-1)=1-\epsilon_i$; ${\epsilon_i}$ being the probability of error
in the $i^{\rm {th}}$ channel. Here memory-less refers to the
factorization in \eqref{eq:MAP4}, time independence refers to the
fact that $\pi^i(\ldots | \ldots)$ does not depend on $k$, and finally
unbiased means that the noise acts symmetrically on both
realizations of the Markov process, i.e., $\pi^i(1|-1)=\pi^i(-1|1)$.
Starting from this, we vary the noise between multiple observation
channels and study its effect in the sequence decoding process. In Appendix \ref{sec:gaussian} we discuss a more general case involving Gaussian distribution of noise realizations
in the observation channels. The detailed formalism is done for the case of a single observation channel.

The composite process $\mathcal{XY}$ with realizations
$(y_k^i,x_k)$ is Markov even though $\mathcal{Y}$ in general is not a Markov process. The transition probabilities for the $i^{\rm{th}}$ observation channel can be written as,
\begin{equation}\label{eq:MAP5}
p(y^i_{k+1},x_{k+1}|y^i_{k},x_{k})=\pi^i
(y^i_{k+1}|x_{k+1})p(x_{k+1}|x_k)
\end{equation}

\subsection{Mapping to the Ising model}
\label{sec:mappingising}

The problem can be efficiently mapped to the Ising spin model where we will represent the transition probabilities as,
\begin{equation}\label{eq:MAP6}
p(x_k|x_{k-1})=\frac{e^{Jx_k x_{k-1}}}{2 \cosh J}, \ \ \
J=\frac{1}{2}{\rm log} \left [ \frac{1-q}{q}\right ]
\end{equation}
The observation probabilities for binary symmetric noise are given
as,
\begin{equation}\label{eq:MAP7}
\pi^i (y^i_k|x_k)=\frac{e^{h_i y^i_k x_k}}{2 \cosh {h_i}}, \ \ \
h_i=\frac{1}{2}{\rm log} \left [
\frac{1-\epsilon_i}{\epsilon_i}\right ]
\end{equation}
and that for the Gaussian distribution of noise in the observation channels is given by,
\begin{equation}\label{eq:MAP4a}
\pi^i(y^i_k |x_k )=\frac{1}{\sqrt{2\pi}\sigma_i}e^{-\frac{(y^i_k
-x_k )^2}{2 {\sigma_i}^ 2} }
\end{equation}
where $\sigma_i^2$ is the noise variance.

Combining \eqref{eq:MAP7} with \eqref{eq:MAP1} and with the
help of \eqref{eq:MAP3} and \eqref{eq:MAP4} we represent the
log-likelihood for the case with binary symmetric noise realization
as,
\begin{equation}\label{eq:MAP8}
H_B (\underline{\yv},\xv)=-J\sum\limits_{k=2}^N x_k
x_{k-1}-\sum\limits_{i=1}^L h_i \sum\limits_{k=1}^N y^i_k x_k
\end{equation}
The same for the Gaussian distribution of noise (see \eqref{eq:MAP4a}) is given as,
\begin{equation}\label{eq:MAP8a}
H_G (\underline{\yv},\xv)=-J\sum\limits_{k=2}^N x_k
x_{k-1}-\sum\limits_{i=1}^L \frac{1}{2 {\sigma_i}^2 }
\sum\limits_{k=1}^N (y^i_k - x_k)^2
\end{equation}
which simply reduces to,
\begin{equation}\label{eq:MAP8b}
H_G (\underline{\yv},\xv)=-J\sum\limits_{k=2}^N x_k
x_{k-1}-\sum\limits_{i=1}^L \frac{1}{{\sigma_i}^2 }
\sum\limits_{k=1}^N y^i_k x_k
\end{equation}
The redundant additive factors are omitted from final expressions of the
Hamiltonian. The subscripts `$B$' and `$G$' denote the binary and
Gaussian cases respectively. $H(\underline{\yv},\xv)$ (representing the general Hamiltonian for both `$B$' and `$G$') is the
Hamiltonian of a (1d) Ising model with external random fields
governed by the probability $\textbf{p}(\underline{\yv}|\xv)$ \cite{Zuk}. The
factor $J$ is the spin-spin interaction constant, uniquely
determined by the transition probability $q$. For
$q<\frac{1}{2}$, the constant $J$ is positive. This refers to the
ferromagnetic situation where the spins tend to align in the same
direction. The random-fields (within the Ising model) obtained in the expression for $H(\underline{\yv},\xv)$ display a
non-Markovian correlation. This is different from what is known for the random fields from the literature, that are in general uncorrelated \cite{behnmarkovising},\cite{pryce}.
For all calculations, we assume $J,h_i,\sigma_i>0$. Let us also introduce another parameter $\alpha=h_2/h_1$ which represents the ratio of the noise between two observation channels (subscripts $1$ and $2$ refer to the two
observation channels) and is assumed to be a positive integer.

\subsection{Statistical Physics  of MAP estimation}
\label{sec:statphys}
We now implement the MAP estimation to minimize the Hamiltonian $H(\underline{\yv},\xv)$. In Section \ref{sec:background} we argued that this is equivalent to minimizing $\sum\limits_{\underline{\yv}} \textbf{p}(\underline{\yv})H(\underline{\yv},\hat{\xv}(\underline{\yv})) $. For this purpose let us introduce a non-zero temperature $T=\frac{1}{\beta} \ge 0$ and write the
conditional probability as,
\begin{equation}\label{eq:MAPIm1}
\rho(\xv|\underline{\yv}) \equiv \frac{e^{- \beta H(\underline{\yv},
\xv)}}{Z(\underline{\yv})};\ \ \  Z(\underline{\yv}) \equiv
\sum\limits_{\xv}e^{- \beta H(\underline{\yv},\xv)}
\end{equation}
where $Z(\underline{\yv})$ is the partition function. Using ideas from statistical physics, we find that $\rho(\xv|\underline{\yv})$
gives the probability distribution of states $\xv$ for a system
with Hamiltonian $H(\underline{\yv},\xv)$. The system is assumed to be interacting with a thermal
bath at temperature $T$, and with frozen random fields $h_i y^i_k$\cite{landau}.
For $T \rightarrow 0$, and a given $\underline{\yv}$, the individual terms of the partition function
are strongly picked at those
$\hat{\xv}(\underline{\yv})$ which minimize the Hamiltonian
$H(\underline{\yv},\xv)$ to get to the ground states. If, however,
the limit $T \rightarrow 0$ is applied after the limit $N \rightarrow
\infty$ we get,
 \begin{equation}\label{eq:MAPIm2}
\rho(\xv|\underline{\yv}) \rightarrow
\frac{1}{\mathcal{N}(\underline{\yv})}\sum\limits_{\gamma} \delta
[\xv-\hat{\xv}^{[\gamma]}(\underline{\yv})]
\end{equation}
where $\hat{\xv}^{[\gamma]}(\underline{\yv})$ and
$\mathcal{N}(\underline{\yv})$ are given by (\ref{eq:MAP1a}). We
are going to work on this low temperature regime from now on.

The average of $H(\underline{\yv},\xv)$ in the $T \rightarrow 0$ regime will equal its value minimized over $\xv$,
 \begin{equation}\label{eq:MAPIm3}
\sum\limits_{\xv
\underline{\yv}}\textbf{p}(\underline{\yv})\rho(\xv|\underline{\yv})H(\underline{\yv},\xv)=
\sum\limits_{\underline{\yv}}\textbf{p}(\underline{\yv})H(\underline{\yv},\hat{\xv}^{[1]}(\underline{\yv}))
= H(\underline{\yv},\hat{\xv}^{[1]}(\underline{\yv}))
\end{equation}
where by assumption all ground state configurations $\hat{\xv}(\underline{\yv})$ have the same energy
$H((\underline{\yv},\hat{\xv}^{[\gamma]}(\underline{\yv}))=H(\underline{\yv},\hat{\xv}^{[1]}(\underline{\yv}))$
for any $\gamma$.

The zero-temperature entropy $\Theta$ depicting the number of MAP solutions and is
given as,
 \begin{equation}\label{eq:MAPIm4}
\Theta=- \sum\limits_{\xv
\underline{\yv}}\textbf{p}(\underline{\yv})\rho(\xv|\underline{\yv}){\rm
log}\rho(\xv|\underline{\yv}) =
\sum\limits_{\underline{\yv}}\textbf{p}(\underline{\yv}){\rm
log} \mathcal{N}(\underline{\yv})
\end{equation}

Another important statistical parameter, free energy can also be defined as,
 \begin{equation}\label{eq:MAPIm5}
 F(J,h,T)=-T \sum\limits_{\underline{\yv}}\textbf{p}(\underline{\yv})\log \sum\limits_{\xv} e^{- \beta H(\underline{\yv},\xv)}
\end{equation}
where the Ising Hamiltonian $H(\underline{\yv},\xv)$ is given by
\eqref{eq:MAP8} or \eqref{eq:MAP8b}. With the help of this
definition we can now define the entropy $\Theta$ in terms of the
free energy as,
 \begin{equation}\label{eq:MAPIm6}
\Theta=- \partial _T F | _{T \rightarrow 0}
\end{equation}
We also define the order parameters which are the relevant characteristics of MAP below,
 \begin{equation}\label{eq:MAPIm7}
c=\sum\limits_{\xv
\underline{\yv}}\textbf{p}(\underline{\yv})\rho(\xv|\underline{\yv})\frac{1}{N}\sum\limits_{k=1}^{N-1}
x_k x_{k+1} = \frac{1}{N} \partial_J F
\end{equation}
 \begin{equation}\label{eq:MAPIm8}
v=\sum\limits_{\xv
\underline{\yv}}\textbf{p}(\underline{\yv})\rho(\xv|\underline{\yv})\frac{1}{N}\sum\limits_{i=1}^L\sum\limits_{k=1}^N
y^i_k x_k = \frac{1}{N} \partial_h F
\end{equation}
$c$ accounts for the correlation between the neighboring spins in the
estimated sequence and $v$ measures the overlap between the
estimated and the observed sequences for the various cases that are
studied.

The order parameters defined above are indirect measures of
estimation accuracy. Calculation of direct error involves
evaluating the overlap between the actual hidden sequence that
generates a given observation sequence, and the inferred sequence
based on the observation sequence. Since we are interested in
average quantities, it is clear that calculating the average error
amounts to finding the overlap between a {\em typical} hidden
sequence and its MAP estimate, i.e.,
\begin{equation}\label{eq:MAPo27}
\Delta = \sum\limits_{k=1}^N s_k x_k
\end{equation}
where ${\bf s} = \{s_k\}_{k=1}^N$ is a typical sequence generated by the (hidden) Markov chain. Let us define a
modified Hamiltonian,
\begin{equation}\label{eq:MAP27}
H_{g} (\underline{\yv},\xv; {\bf s})=-J\sum\limits_{k=2}^N x_k
x_{k-1}-\sum\limits_{i=1}^L h_i \sum\limits_{k=1}^N y^i_k x_k -
g\sum\limits_{k=1}^N s_k x_k
\end{equation}
Further, let  us introduce $f_{g}$ which is related to the free energy $F_g$ of the modified Hamiltonian by $f_g=\frac{F_g}{N}$. It is given by,
 \begin{eqnarray}\label{eq:MAP28f}
f_{g}=-T \frac{1}{N}\sum_{\underline{\yv},\sv} \pv(\underline{\yv},\sv)  \log \sum\limits_{\textbf{x}} e^{- \beta H_{g} (\underline{\yv},\xv; {\bf s})}
\end{eqnarray}
With this the overlap can be simply given by $\Delta=-\partial_g f_{g} |_{g
\rightarrow 0,\beta \rightarrow \infty}$, where the limits are taken in the order $g
\rightarrow 0,\beta \rightarrow \infty$.

\section{Solution of the Model}
\label{sec:recursion}
Here we will solve for the order parameters and entropy of multiple observation channels using tools that we developed using the Ising spin model. The solution of the model for $L=1$ was provided in \cite{allahverdyan2009maximum}. To make the paper self-contained, here we repeat the main steps of the derivation. Let us recall the partition function \eqref{eq:MAPIm1} which for $L$ observation sequences can be written as,
\begin{equation}\label{eq:MAP9}
Z(\underline{\yv})=\sum\limits_{x_1=\pm 1,\ldots,x_N=\pm 1} e^{\beta
J \sum\limits^{N-1}_{k=1}x_{k+1}x_k +\beta \sum\limits^L_{i=1}
h_i\sum\limits^N_{k=1}y^i_k x_k}
\end{equation}
Summing over the first spin yields  the following transformation for $Z(\underline{\yv})$,
\begin{equation}\label{eq:MAP10}
\sum\limits_{x_1,\ldots,x_N} e^{\beta J
\sum\limits^{N-1}_{k=1}x_{k+1}x_k +\beta \sum\limits^L_{i=1}
h_i\sum\limits^N_{k=1}y^i_k x_k}=\sum\limits_{x_2,\ldots,x_N}
e^{\beta J \sum\limits^{N-1}_{k=2}x_{k+1}x_k +\beta
\sum\limits^L_{i=1} h_i\sum\limits^N_{k=3}y^i_k x_k+\beta \xi_2 x_2
+ \beta B(\xi_1)}
\end{equation}
where $\xi_2=A(\xi_1) + \sum\limits^L_{i=1}h_i y^i_2,\ \
\xi_1=\sum\limits^L_{i=1}h_i y^i_1$ and,
\begin{equation}\label{eq:MAP11}
A(u)=\frac{1}{2\beta}{\rm log}\frac{\cosh[\beta J+ \beta
u]}{\cosh[\beta J- \beta u]}
\end{equation}
\begin{equation}\label{eq:MAP12}
B(u)=\frac{1}{2\beta}{\rm log}(4 \cosh[\beta J+ \beta u] \cosh[\beta
J- \beta u])
\end{equation}
Hence, once the first spin is excluded from the chain the field acting on the second spin changes
from $\sum\limits^L_{i=1}h_i y^i_2$ to $\sum\limits^L_{i=1}h_i
y^i_2+A(\xi_1)$. In the zero temperature limit ($T \longrightarrow 0$ ; $\beta\rightarrow \infty$) the functions $A(u)$ and $B(u)$ reduce to the following form,
 \begin{equation}\label{eq:MAP13}
A(u)=u\vartheta(J-|u|)+J\vartheta(u-J)-J\vartheta(-u-J)
\end{equation}
 \begin{equation}\label{eq:MAP14}
B(u)=J\vartheta(J-|u|)+u\vartheta(u-J)-u\vartheta(-u-J)
\end{equation}
where $\vartheta(x)=0$ for $(x<0)$ and $\vartheta(x)=1$ for $(x>0)$.
Thus after excluding one spin every time and repeating the above steps, the partition function for the system is obtained as,
 \begin{equation}\label{eq:MAP15}
Z(\underline{\yv})=e^{\beta \sum\limits^N_{k=1}B(\xi_k)}
\end{equation}
where $\xi_k$ is obtained from the recursion relation
 \begin{equation}\label{eq:MAP16}
\xi_k=\sum\limits^L_{i=1}h_iy^i_k +A(\xi_{k-1}),\ k=1,2,\ldots,N, \
\xi_0=0
\end{equation}
Following similar steps with $H_G$ from \eqref{eq:MAP8b} we get the
recursion relation for the Gaussian realization of noise in the observation
channel as,
 \begin{equation}\label{eq:MAP17}
\xi_k=\sum\limits^L_{i=1}\frac{1}{{\sigma_i}^2 } y^i_k
+A(\xi_{k-1}),\ k=1,2,\ldots,N,\ \xi_0=0
\end{equation}
$y^i_k$'s from the above random recursion relation are random
quantities governed by the probability
$\textbf{p}(\underline{\yv})$. As we can see from the above equations for binary realizations, $y^i_k$ can
take a finite number of values and $\xi_k$ can take an infinite
number of values. But in the asymptotic $\beta \rightarrow \infty$ limit, $A(u)$ and $B(u)$ reduce to the simple form given in \eqref{eq:MAP13} and
\eqref{eq:MAP14} respectively. Thus we obtain a finite number of values for
$\xi_k$ (the number can be large or small). However for the Gaussian distribution, we get a continuous
set of values for $\xi_k$.

The parametric form of $\xi_k$ is already explained in \cite{allahverdyan2009maximum} for a single observation channel. Here we provide the generalization to two observation channels, i.e. $L=2$, which can be easily extended to the multiple channel case. We can parameterize
$\xi_k$ as, $\xi_k(n_1,n_2,n_3)=(n_1 h_1 + n_2 h_2 + n_3 J)=([n_1 +
\alpha n_2] h_1 + n_3 J)$, where $h_2/h_1 = \alpha$, $n_1$ and $n_2$
can be positive and negative integers while $n_3$ can take only
three values $0,\pm 1$. It may be noted that the states $\xi
(n_1,n_2,0)$ are not recurrent: once $\xi_k$ takes a value with [$(
n_1 + \alpha n_2 )$ positive or negative,$n_3=\pm 1$] it never comes back to the
state $\xi (n_1,n_2,0)$. Thus in the limit $N>>1$ we can ignore the
states $\xi (n_1,n_2,0)$.

Let us now consider the problem of finding the stationary distribution of the random process given by \eqref{eq:MAP16}. Note that the process $\mathcal{Y}$ with probabilities
$\textbf{p}(\underline{\yv})$ is not Markovian, hence the process in \eqref{eq:MAP16} is not Markovian either, which slightly complicates the calculation of its stationary distribution. Towards the latter goal, let us consider an auxiliary Markov process $\mathcal{Z}$, which has  identical statistical
characteristics with the process  $\mathcal{X}$. For this auxiliary Markov process $\mathcal{Z}$ the realization
is denoted  as $\zv$, so as not to mix with the original process $\xv$. We need to include $\mathcal{Z}$ to make the
composite process Markov. Thus, to make the realizations for the composite process
$[\xi,y^1,\ldots,y^L]$ Markov, we enlarge it to
$[\xi,y^1,\ldots,y^L,z]$ (lets call it $\mathcal{C}$). The conditional probability for $\mathcal{C}$ is given by,
 \begin{equation}\label{eq:MAP18}
\omega(\xi,y^1,\ldots,y^L,z|\xi',y'^1,\ldots,y'^L,z')=p(z|z')\varphi(\xi|\xi',y^1,\ldots,y^L)\prod_{i=1}^L
\pi^i(y^i|z)
\end{equation}
For $\mathcal{C}$, $p(z|z')$ and $\pi(y^i|z)$ refer to the Markov process
$\mathcal{X}$ and the noise respectively, while $\varphi(\xi|\xi',y^1,\ldots,y^L)$
takes only two values 0 and 1, depending on whether the transition is
allowed by the recursion \eqref{eq:MAP16}. Now, we need to determine
$\varphi(\xi|\xi',y^1,\ldots,y^L)$ after finding all possible values of $\xi_k$. Let us first relate the stationary
probabilities $\omega(\xi,y^1,\ldots,y^L,z)$ of the composite Markov
process $\mathcal{C}$ to the characteristics of MAP estimation:
$\omega(\xi,y^1,\ldots,y^L,z)$, conveys the message about the stationary probabilities
$\omega(\xi)$. After this using the definition for the partition
function from \eqref{eq:MAP15} and free energy \eqref{eq:MAPIm5} for
the composite Markov process $\mathcal{C}$ which is ergodic, the
free energy is given as \cite{behnmarkovising},
 \begin{equation}\label{eq:MAP19}
-f(J,h)\equiv -F(J,h)/N=\sum\limits_{\xi} \omega(\xi) B(\xi)
\end{equation}
where the summation is taken over all possible values of $\xi$ [for
a given $(J,h)$]. We study different processes in the multi-channel scenario, generally speaking, for $L > 2$ with same noise in the channels and for $L = 2$ with varying noise in the two channels. For all these cases $h_i$'s can be represented by a single quantity $h$.
Once we find $f(J,h)$ we can apply \eqref{eq:MAPIm7},\eqref{eq:MAPIm8} to do further
calculations. To calculate the entropy we first derive a convenient
expression for free energy, which can be obtained from
\eqref{eq:MAP14}, \eqref{eq:MAP15}.
 \begin{equation}\label{eq:MAP20}
F(\underline{\yv})=-\frac{T}{2}\sum\limits_{k=1}^N \sum\limits_{s=
\pm 1} \log [2 \cosh\{\beta(\xi_k+sJ) \}]
\end{equation}
Using this we find,
 \begin{equation}\label{eq:MAP21}
\partial_T F(\underline{\yv}) | _{T \rightarrow 0}=\frac{\log 2}{2}\partial_T \left \{ T \sum\limits_{k=1}^N \delta(\xi_k \pm J )\right \} | _{T \rightarrow 0}
\end{equation}
where $\delta(.)$ is the Kronecker delta function. In the $N>>1$ limit, under the assumption that the Markov process is ergodic,
$\frac{1}{N}\sum\limits_{k=1}^N \delta(\xi_k \pm J )$ should with probability one (for the elements of the typical set
$\Omega(\mathcal{Y})$) converge to $\omega(\xi=J)+\omega(\xi=-J)$. We thus obtain \cite{behnmarkovising},
 \begin{equation}\label{eq:MAP22}
\theta \equiv \Theta/N =\frac{{\rm log} 2}{2} [\omega(J)+\omega(-J)]
\end{equation}
The above formula for an Ising spin model would imply that the zero temperature
entropy can be extensive only when the external field $\xi$ acting
on the spins would have the same energy $\xi x_k=\pm 1$ as the spin-spin
coupling constant $J$. This would imply that a macroscopic amount of
spin (from any of the models studied) is frustrated, i.e., the factors influencing those spins
compensate each other. When the entropy is not zero, there are
many sequences whose probability may still slightly differ from one
another. The MAP characteristics $(c,v)$ would refer to the
averages over all those equivalent sequences. We will discuss this effect explicitly for the various cases analyzed.

Finally, we discuss the semi-analytical error analysis that has been developed in this work. For this let us consider the overlap defined in \eqref{eq:MAPo27}. Recall that the error estimate is given by $-\partial_g f_{g}|_{g \rightarrow 0,\beta \rightarrow \infty}$ where $f_{g}$ is defined in \eqref{eq:MAP28f}. Derivation is shown explicitly for a single observation channel in Appendix \ref{sec:Analyticalerrordetails}. The results are used to analytically solve for the error in single and two observation channels and is provided in Section \ref{sec:twochannel}. The formalism can easily be generalized for multiple observation channels with some tedious algebra. In Appendix \ref{sec:Analyticalerrordetails} we show that the overlap is be given as,
\begin{equation}\label{eq:MAP29err}
\Delta_k = \sum_{\yv,\sv} \pv(\yv,\sv)
s_k[f(\xi_k)+g(\xi_k)f(\xi_{k+1}) +\ldots+ g(\xi_k) \ldots
g(\xi_{N-1})f(\xi_N)]
\end{equation}

In the limit $N\rightarrow \infty$, $\Delta_k$ is independent of $k$. The average error is related to the overlap as, $P_{MAP} \{ error\}=\frac{1-\Delta}{2}$.

\section{Characterization of the recurrent states}
\label{sec:recurrent}
For calculating the quantities of interest we
need to obtain the recurrent states for the different multi-channel
systems. The recurrent states are found out and parameterized in
a manner similar to that demonstrated for $L=1$ in \cite{allahverdyan2009maximum}. In this section
we analyze two scenarios. In Scenario I, we consider a $2$
observation channel system. The noise in either channel is varied in a certain fixed proportion.
In Scenario II, we consider the general $L>2$ channel system, each having
the same noise intensity. Below we give the parameterizations of the
recurrent states for these two cases separately. In both the
scenarios we come across multiple phase transitions which we denote
by $m$ and the noise within any of those phases varies as
$\frac{2J}{m}<h<\frac{2J}{m-1}$. The phase transitions are reflected
in the study of the order parameters and the error analysis and is discussed in detail in the next section.

\subsection{Scenario I: Two channels with different noise intensities}
\label{sec:charactwochannel}
The parameter $\alpha=h_2/h_1$ is defined as the
ratio between $h_i$'s \eqref{eq:MAP7} in the individual observation
channels. The \emph{stationary states} can be parameterized as,
$[a_i \ \ \bar{a}_i] $ which are written in terms of $h$ (depicting
the channel with higher noise),
\begin{eqnarray}\label{eq:MAP23}
a_i&=&J+(2-i)h+ \eta h \nonumber \\
\bar{a}_i&=&-J-(2-i)h+ \eta h
\end{eqnarray}
with $\eta \in \{ (\alpha +1)h, (\alpha -1)h ,-(\alpha -1)h ,-(\alpha +1)h  \}$.

The values of $\alpha=h_2/h_1$ are kept as positive integers since for an
arbitrary $\alpha$ (rational or irrational) the number of stationary
states increases very fast making the analysis complicated, but
doesn't contribute towards any additional generality of the problem. The stationary
states at all the phases can be found form the above formula by
substituting the values of $i$ from Table \ref{tab:Noise_ratio
var_states} in the above equation. The total number of states during different phases is tabulated in Table
\ref{tab:Noise ratio}.

\begin{table}[!htp]
\caption{The possible values of $i$ for different
$\alpha$ at different phases denoted by
$m$. In the table below ($\ldots$) implies for
 (i) $\alpha:even$, all integer values on the range and (ii) $\alpha:odd$, only the even values in the range.}

\centering
\begin{tabular}{lllllllllll}
  \hline \noalign{\smallskip}
  \toprule
  m & $\alpha=1$ & $\alpha=2$ & $\alpha=3$ & $\alpha=4$ & $\alpha=5$ & $\alpha=6$ & $\alpha=7$ & $\alpha=8$ & $\alpha=9$ & $\alpha=10$ \\
  \midrule
  1 & 2 & 2 & 2 & 2 & 2 & 2 & 2 & 2 & 2 & 2 \\
  2 & 2 & 2,3 & 2 & 2 & 2 & 2 & 2 & 2 & 2 & 2 \\
  3 & 2,4 & 2,\ldots,4 & 2,4 & 2 & 2 & 2 & 2 & 2 & 2 & 2 \\
  4 & 2,4 & 2,\ldots,5 & 2,4 & 2,5 & 2 & 2 & 2 & 2 & 2 & 2 \\
  5 & 2,4,6 & 2,\ldots,6 & 2,4,6 & 2,5 & 2,6 & 2 & 2 & 2 & 2 & 2  \\
  6 & 2,4,6 & 2,\ldots,7 & 2,4,6 & 2,4,5,7 & 2,6 & 2,7 & 2 & 2 & 2 & 2  \\
  7 & 2,\ldots,8 & 2,\ldots,8 & 2,\ldots,8 & 2,\ldots,8 & 2,\ldots,8 & 2,7 & 2,8 & 2 & 2 & 2  \\
  8 & 2,\ldots,8 & 2,\ldots,9 & 2,\ldots,8 & 2,\ldots,9 & 2,\ldots,8 & 2,4,7,9 & 2,8 & 2,9 & 2 & 2  \\
  9 & 2,\ldots,10 & 2,\ldots,10 & 2,\ldots,10 & 2,\ldots,10 & 2,\ldots,10 & 2,4,7,9 & 2,4,6,8 & 2,9 & 2,10 & 2  \\
  10 & 2,\ldots,10 & 2,\ldots,11 & 2,\ldots,10 & 2,\ldots,11 & 2,\ldots,10 & 2,4,6,7,9,11 & 2,4,6,8 & 2,4,9,11 & 2,10 & 2,11  \\

  \bottomrule
  \hline \noalign{\smallskip}
\end{tabular}
\label{tab:Noise_ratio var_states}
\end{table}

\begin{table}[!htp]
\caption{Number of states for different $\alpha$ at different phases denoted by $m$}
\centering
\begin{tabular}{lllllllllll}
  \hline \noalign{\smallskip}
  \toprule
  m & $\alpha=1$ & $\alpha=2$ & $\alpha=3$ & $\alpha=4$ & $\alpha=5$ & $\alpha=6$ & $\alpha=7$ & $\alpha=8$ & $\alpha=9$ & $\alpha=10$ \\
  \midrule
  1 & 8 & 8 & 8 & 8 & 8 & 8 & 8 & 8 & 8 & 8 \\
  2 & 8 & 16 & 8 & 8 & 8 & 8 & 8 & 8 & 8 & 8 \\
  3 & 16 & 24 & 16 & 8 & 8 & 8 & 8 & 8 & 8 & 8 \\
  4 & 16 & 32 & 16 & 16 & 8 & 8 & 8 & 8 & 8 & 8 \\
  5 & 24 & 40 & 24 & 16 & 16 & 8 & 8 & 8 & 8 & 8  \\
  6 & 24 & 48 & 24 & 32 & 16 & 16 & 8 & 8 & 8 & 8  \\
  7 & 32 & 56 & 32 & 56 & 32 & 16 & 16 & 8 & 8 & 8  \\
  8 & 32 & 64 & 32 & 64 & 32 & 32 & 16 & 16 & 8 & 8  \\
  9 & 40 & 72 & 40 & 72 & 40 & 32 & 32 & 16 & 16 & 8  \\
  10 & 40 & 80 & 40 & 80 & 40 & 48 & 32 & 32 & 16 & 16  \\

  \bottomrule
  \hline \noalign{\smallskip}
\end{tabular}
\label{tab:Noise ratio}
\end{table}

\subsection{Scenario II: $L-$ identical channels}
\label{sec:characLchannel}

For a multiple observation channel system we can parameterize the \emph{stationary states} for a given value of $J$ and $h$ within any phase as,
\begin{eqnarray}\label{eq:MAP24}
b_i&=&J+(2-i)h+ \sum\limits_{i=1}^L t_i h \nonumber \\
\bar{b}_i&=&-J-(2-i)h+ \sum\limits_{i=1}^L t_i h
\end{eqnarray}

with $t_i \in \{+1,-1\}$. The possible values of $i$ for different phases are given in Table
\ref{tab:Multi channel obs states}. For the various possible phases, the total number of states
are given in Table \ref{tab:Multi channel obs}. Here one can notice the significant rise in the total number of states for large $L$ as we go to a higher phase.

\begin{table}[!h]
\caption{The possible values of $i$ for multiple
observation channel systems with same $h$ at different phases denoted by $m$. In the table below ($\ldots$) implies for
 (i) $\alpha:odd$, all integer values on the range and (ii) $\alpha:even$, only the even values in the range.}
 \centering
\begin{tabular}{lllllllllll}
  \hline \noalign{\smallskip}
  \toprule
  m & $L:: {\rm even}$ & $L:: {\rm odd}$  \\
  \midrule
  1 & 2 & 2   \\
  2 & 2 & 2,3  \\
  3 & 2,4 & 2,\ldots,4 \\
  4 & 2,4 & 2,\ldots,5  \\
  5 & 2,4,6 & 2,\ldots,6 \\
  6 & 2,4,6 & 2,\ldots,7 \\
  7 & 2,\ldots,8 & 2,\ldots,8  \\
  8 & 2,\ldots,8 & 2,\ldots,9  \\
  9 & 2,\ldots,10 & 2,\ldots,10  \\
  10 & 2,\ldots,10 & 2,\ldots,11 \\

  \bottomrule
  \hline \noalign{\smallskip}
\end{tabular}

\label{tab:Multi channel obs states}
\end{table}

\begin{table}[!h]
\caption{Number of states for multi observation channels with same
$h$ at different phases denoted by $m$} \label{tab:Multi channel obs}
\centering
\begin{tabular}{lllllllllll}
  \hline \noalign{\smallskip}
  \toprule
  m & $L=1$ & $L=2$ & $L=3$ & $L=4$ & $L=5$ & $L=6$ & $L=7$ & $L=8$ \\
  \midrule
  \noalign{\smallskip}
  1 & 4 & 8 & 16 & 32 & 64 & 128 & 256 & 512  \\
  2 & 8 & 8 & 32 & 32 & 128 & 128 & 512 & 512  \\
  3 & 12 & 16 & 48 & 64 & 192 & 256 & 768 & 1024\\
  4 & 16 & 16 & 64 & 64 & 256 & 256 & 1024 & 1024 \\
  5 & 20 & 24 & 80 & 96 & 320 & 384 & 1280 & 1536   \\
  6 & 24 & 24 & 96 & 96 & 384 & 384 & 1536 & 1536  \\
  7 & 28 & 32 & 112 & 128 & 448 & 512 & 1792 & 2048   \\
  8 & 32 & 32 & 128 & 128 & 512 & 512 & 2048 & 2048 \\
  9 & 36 & 40 & 144 & 160 & 576 & 640 & 2304 & 2560  \\
  10 & 40 & 40 & 160 & 160 & 640 & 640 & 2560 & 2560  \\

  \bottomrule
  \hline \noalign{\smallskip}
  \end{tabular}

\end{table}

\section{Results}
\label{sec:results}
In this section we describe our results where we
try to analyze different multiple observation channel systems by
evaluating the order parameters and entropy. First, we consider a
$2$ observation channel system by varying the noise in the
individual channels. Next, we consider  $L$ observation channels
with identical noise intensities. Finally, we compare the
performance of a single relatively ``clean" channel with multiple
``noisy" ones by bringing in the notion of channel cost, and examine
the tradeoffs between the two setups. An attempt is made to provide
a physical intuition behind the results that we obtained using
simple thermodynamic principles
\cite{{behnmarkovising},{PhysRevB.36.285},{PhysRevE.56.6459},{PhysRevE.78.021120},{PhysRevA.2.2368}}.
The system performance is considered for both maximum likelihood (ML) as well as maximum a posteriori (MAP)
estimation. In ML, the estimation of the current state is dependent
only on the observations from the different channels at that
particular instant. In case of a tie, which arises when the number
of observation channels is even, the state is chosen randomly to be
either 0 or 1 with equal probability. For example, suppose the
number of observation channels is $L$ ($L$ being even) with $L/2$ of
the channels having an observation 1 and the remaining $L/2$ an
observation of 0. In this case, the state is chosen to be either 0
or 1 at random with probability 1/2.

\subsection{Two-Channel scenario}
\label{sec:twochannel}

In this subsection, we study the performance of a two observation
channel system with relatively different noise levels.

\begin{figure}
  \centering
  \subfigure[$c$]{
  \includegraphics[width=0.3\textwidth]{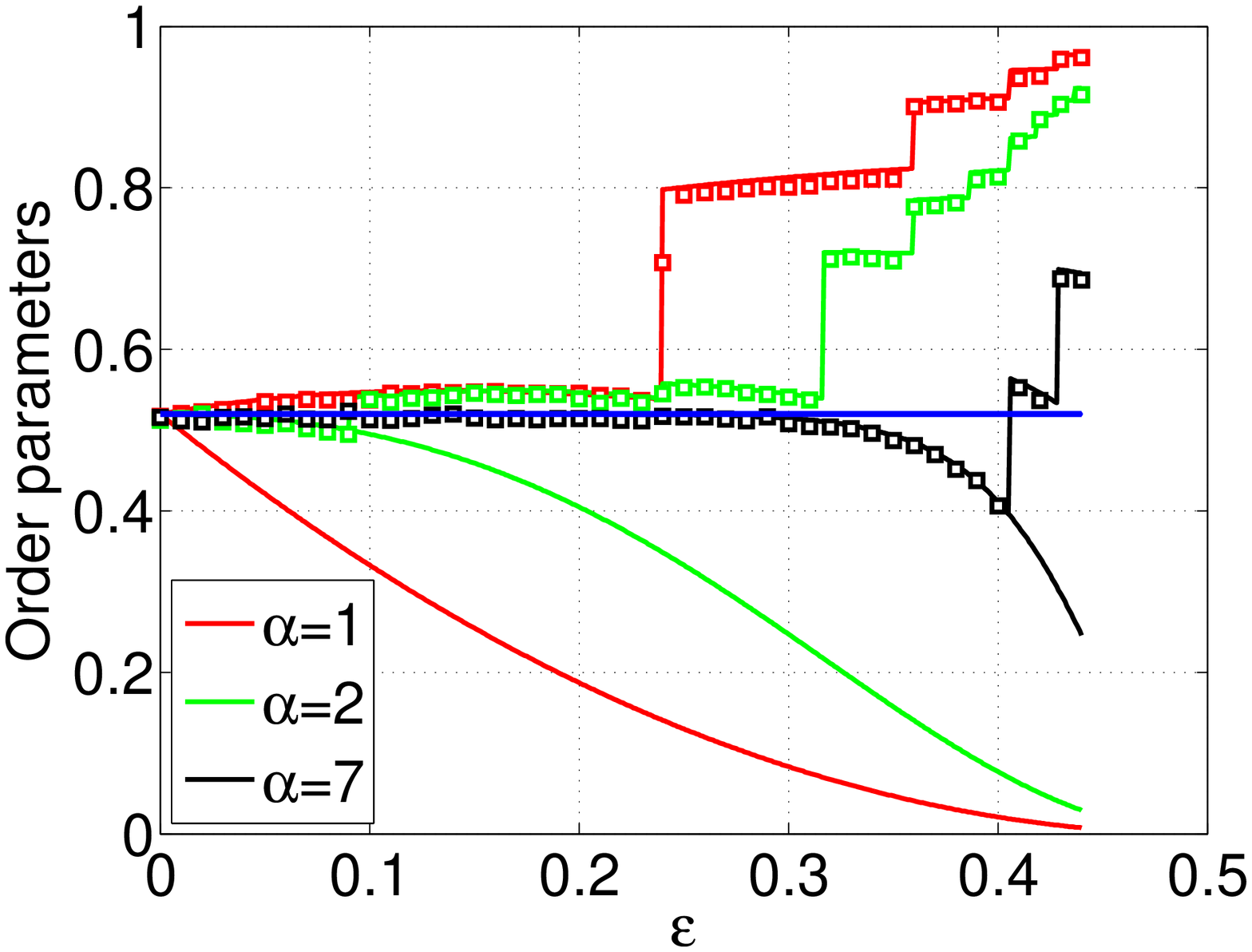} \label{fig:alpha-1}}
  \subfigure[$v$]{
  \includegraphics[width=0.3\textwidth]{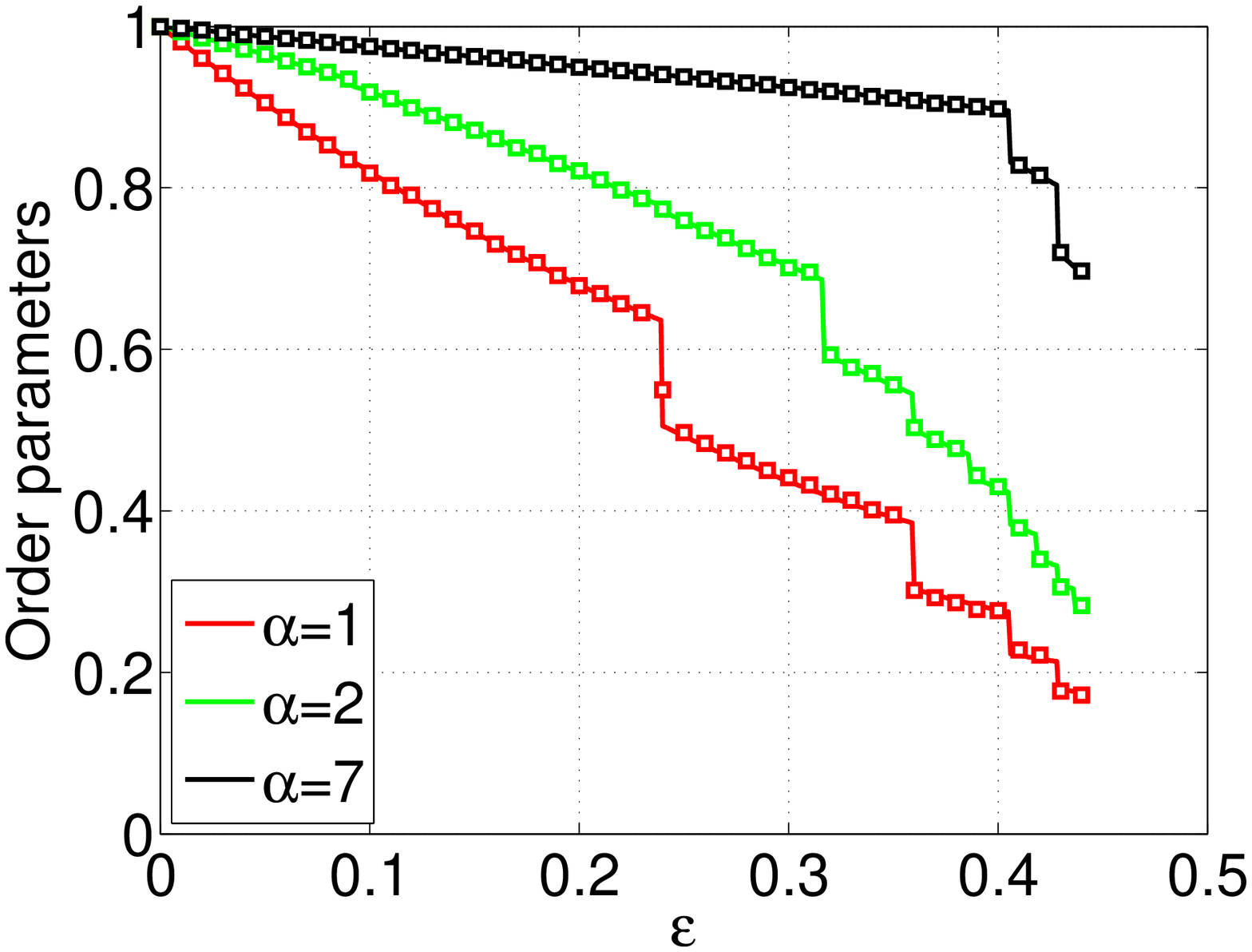} \label{fig:alpha-2}}
  \subfigure[$v_1$]{
  \includegraphics[width=0.3\textwidth]{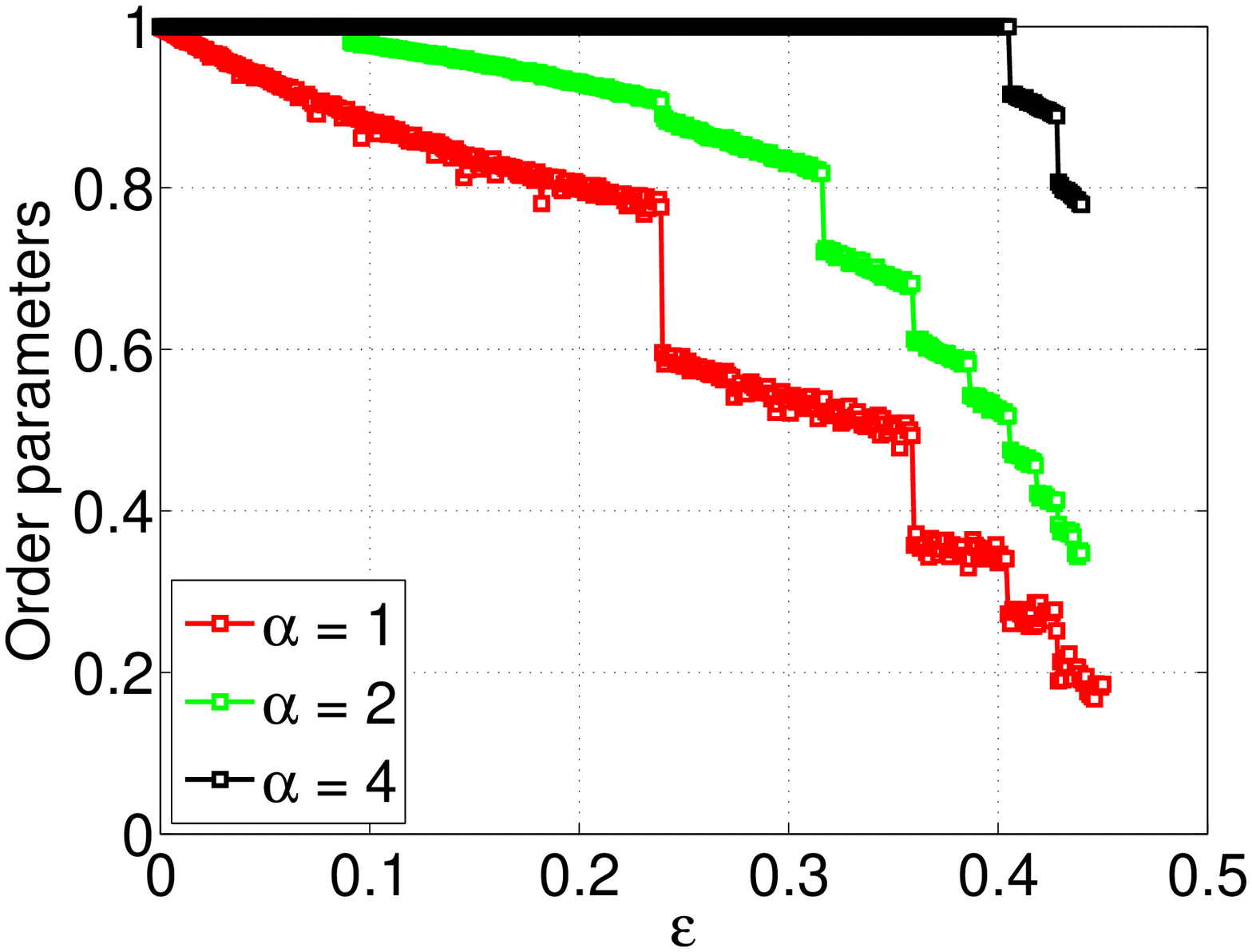} \label{fig:alpha-3}}
  \caption{(Color online) Order parameters (a) $c$ and (b) $v$ obtained analytically by Ising Hamiltonian minimization (bold line) which shows an exact superposition with the data obtained from the Viterbi algorithm (open squares) for two channels with different noise for $q=0.24$. The smooth decaying lines in the plot for $c$ are obtained via ML estimate and the horizontal blue line depicts $c_0$.(c)$v_1$ is the overlap between the MAP and ML estimated sequences. }\label{fig:comp-1c}
\end{figure}
Channel 1 has noise $\epsilon_1$ and channel 2 has noise
$\epsilon_2$. The parameter $\alpha$ defines the relation between
the noise levels in the two channels. Specifically,
\begin{displaymath}
\epsilon_2 = \frac{1}{1 +
\left(\frac{1-\epsilon_1}{\epsilon_1}\right)^\alpha}
\end{displaymath}
Since we always take $\alpha \geq 1$,\footnote{The case with $\alpha
\leq 1$ is obtained by simply interchanging the channels.} the noise
level in channel 2 is less than that in channel 1, implying channel
1 is ``noisier'' than channel 2. The order parameters $c$ and $v$,
plotted by varying the noise $\epsilon_1$\footnote{The plots are
made relative to the ``noisier'' observation channel.} in channel 1
for a fixed spin-spin correlation $J$ (a function of $q$ as given by
\eqref{eq:MAP6}), are shown in Figure \ref{fig:comp-1c}. We observe
different operational regimes, that are separated by first order
phase transitions. The point of the first phase transition
gradually moves to the right with an increase in $\alpha$. This
indicates that the overall behavior is dominated by that of the
cleaner channel, which is intuitive. In this region before the first
phase transition, the ML and MAP estimates coincide. The sequence
correlation parameter $c$ is pretty stable and noise independent
before the point of first phase transition for MAP estimation.
Afterwards the correlation shows discrete jumps and goes to the
prior dominated value of 1 as observed from MAP estimation, whereas
with the ML estimate, $c$ monotonically reduces to 0. The advantage
of MAP estimation over ML is the fact that in MAP the estimation is
supported by prior and hence, performs better at intermediate noise
ranges. For example, when $\alpha = 2$, it can be seen that the
correlation $c$ degrades faster for ML estimation and is nearly
constant for a large range of noise values for MAP estimation. The
overlap $v$ is found to gradually shift towards 1 before the first phase
transition with increase in the value of $\alpha$, implying that the
estimated sequence for all the possible noises is primarily driven
by the observations. Before first phase transition, we encounter the observation dominated regime except for $\alpha = 1$. The
value of the overlap is not 1 because of the manner the overlap is
defined in \eqref{eq:MAPIm8}. A more plausible way to see the observation dominated
regime is to consider the overlap between MAP and ML estimated
sequences, which is plotted in Figure \ref{fig:alpha-3}. This
confirms the fact that there is no observation dominated regime for
$\alpha = 1$ in 2-channel systems. After the first phase transition
the overlap becomes worse and at higher noise, $v$ decays towards 0.
Increasing $\alpha$ enlarges the observation dominated regime, or in
other words, the region of ML estimation. It is also interesting to
note that the correlation and overlap parameter are stable for a
longer range with an increase in $\alpha$, but show a rapid rise or decay respectively
after the first phase transition. The analytical results are
verified by simulations using Viterbi algorithm which are in good
agreement with each other.

\begin{figure}
  \centering
  \subfigure[$\alpha = 1$]{
  \includegraphics[width=0.3\textwidth]{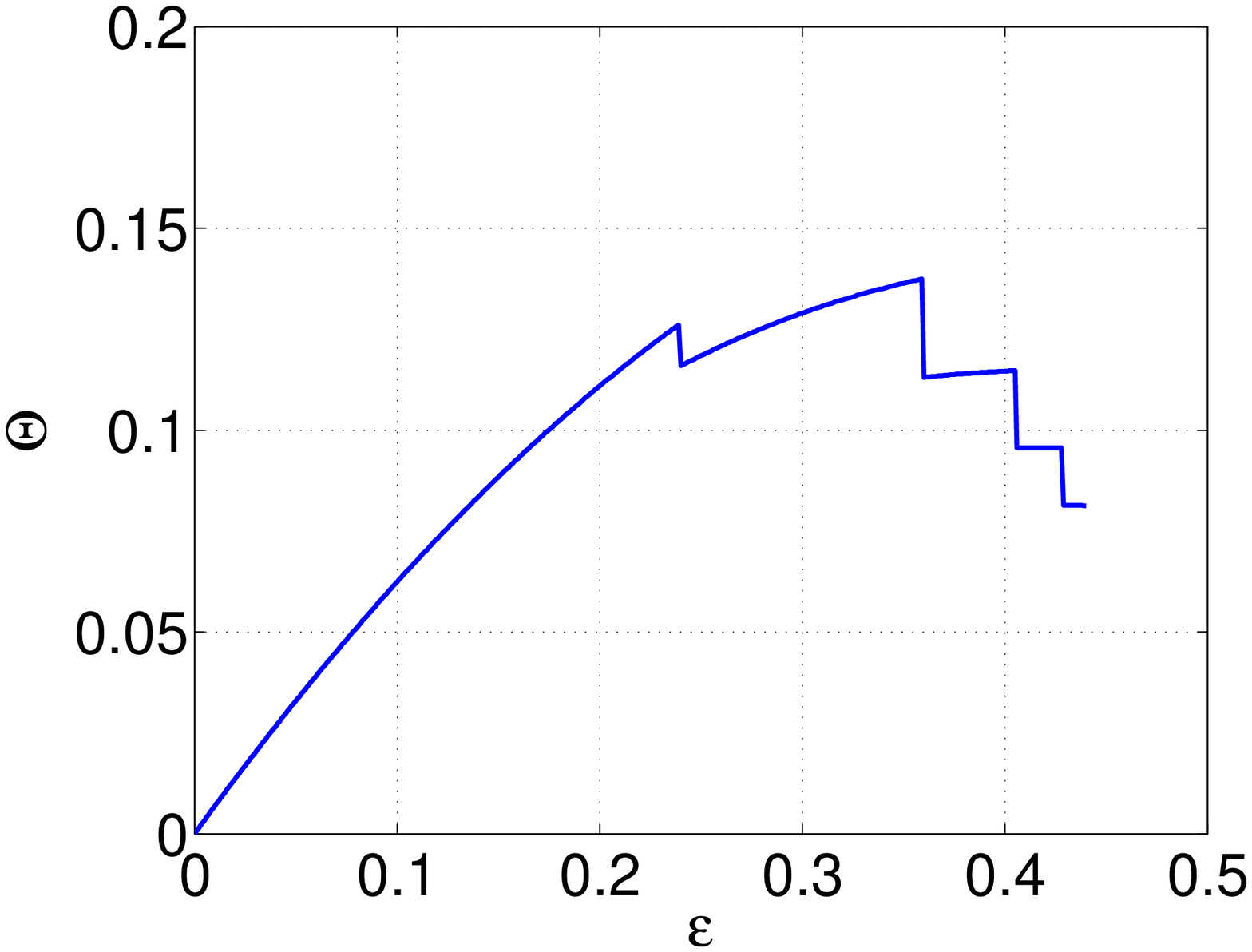} \label{fig:alpha-1}}
  \subfigure[$\alpha = 2$]{
  \includegraphics[width=0.3\textwidth]{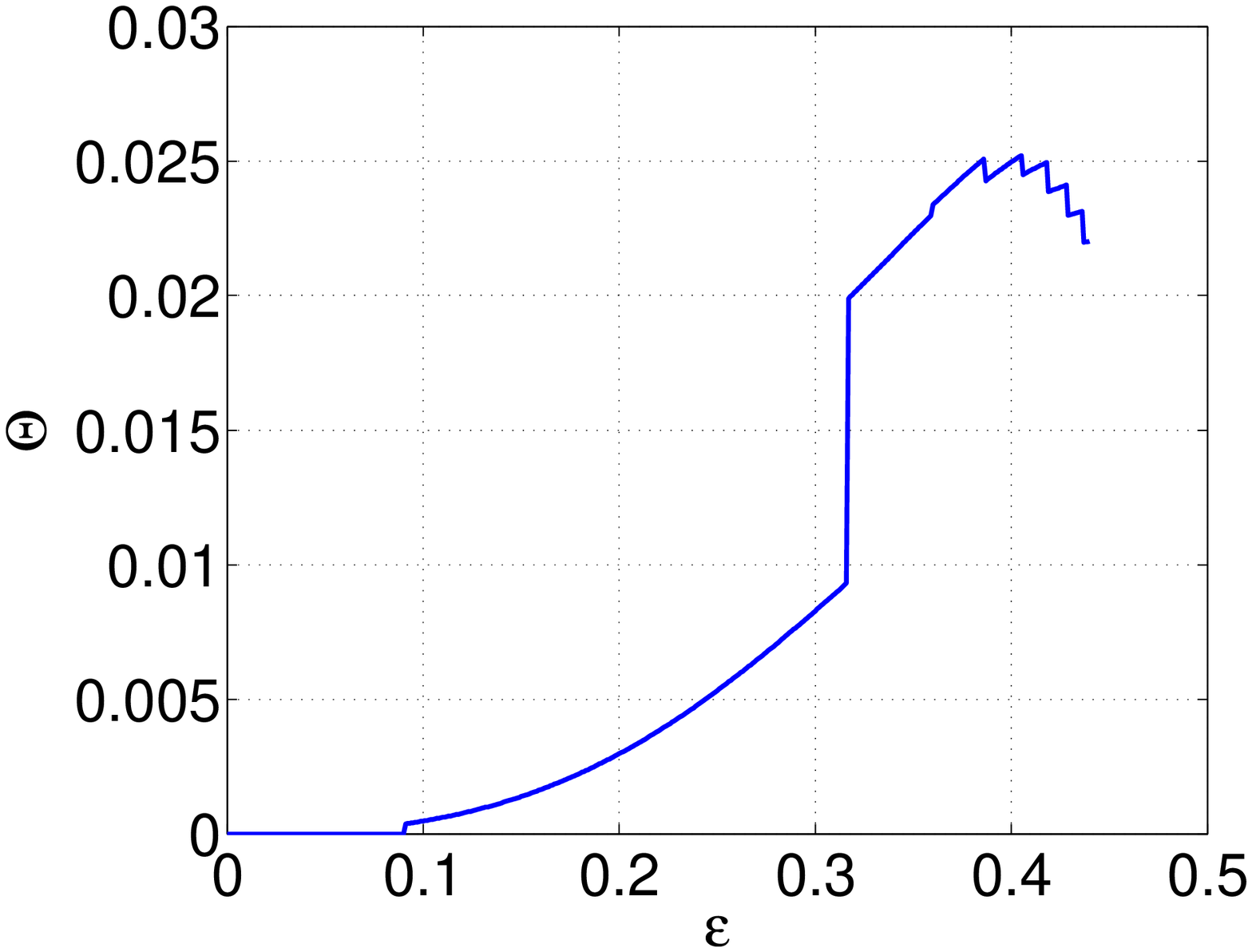} \label{fig:alpha-1}}
  \subfigure[$\alpha = 4$]{
  \includegraphics[width=0.3\textwidth]{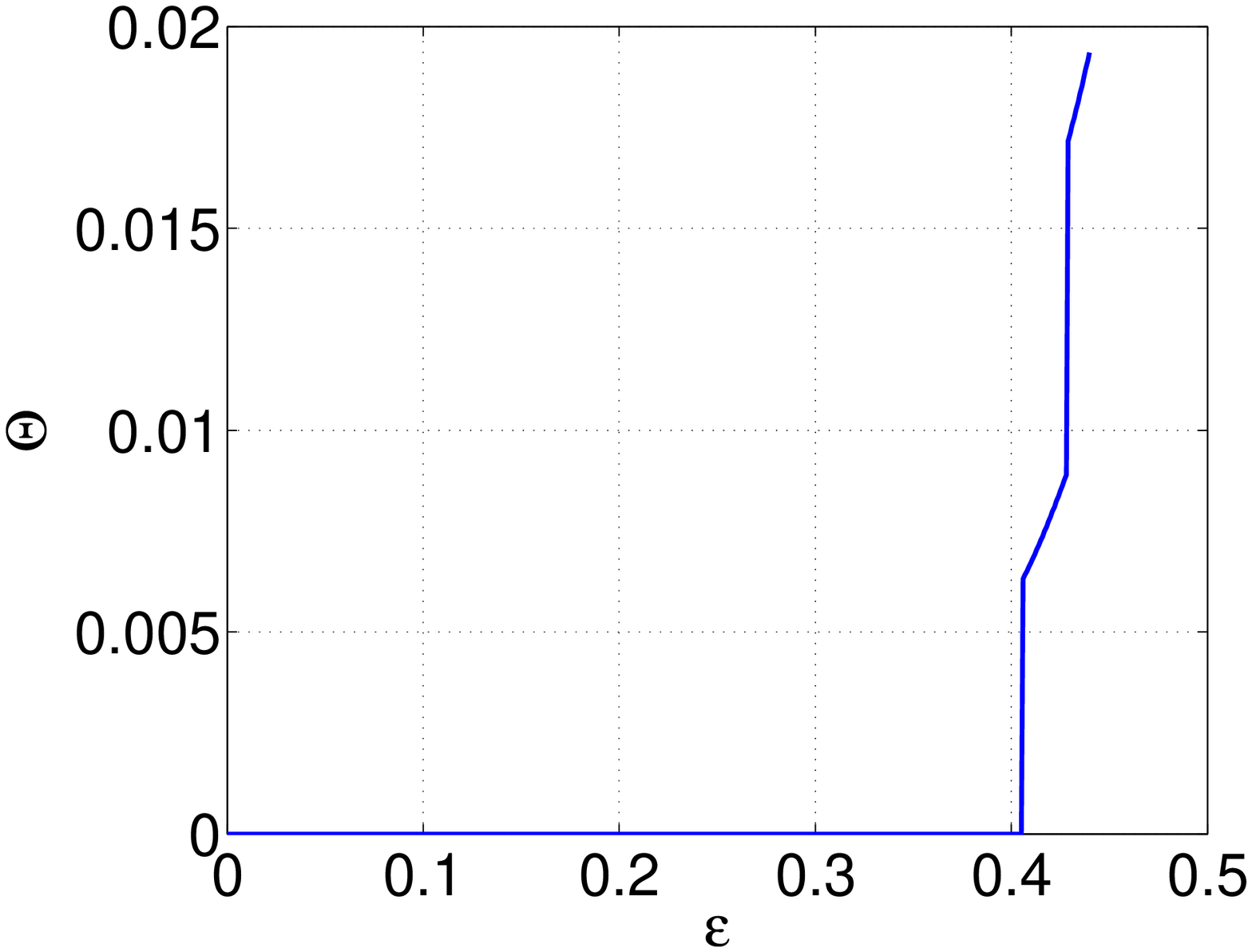} \label{fig:alpha-1}}
  \caption{(Color online) The entropy of the system obtained analytically with varying $\epsilon$ ($\epsilon$ denoting the higher noise level) for two observation channels at different noise ratios given by $\alpha$ at $q=0.24$.}\label{fig:comp-5ae}
\end{figure}

Let us now focus on entropy. We see that the addition of a second
observation channel with the same noise as the first results in
non-zero entropy for all possible values of $\epsilon$. With the
introduction of the second observation channel, the noise parameter
$h_2$ being of same strength as $h_1$ mutually cancel each other.
This gives rise to multiple degenerate ground states for the acting
external field and hence multiple solutions are obtained from MAP
estimation, resulting in non-zero entropy \cite{behnmarkovising}. However as we increase
$\alpha$, it can be seen from Figure \ref{fig:comp-5ae} that the
regions of zero entropy is obtained again and the point of first
phase transition shifts to the right. However, after the first phase
transition, the entropy rises, attains a maxima and then decays. The
reason behind this is discussed in detail below where we discuss the
regions of zero and non-zero entropies obtained for parameter $q$.

\begin{figure}
  \centering
  \subfigure[$q=0.1$]{
  \includegraphics[width=0.3\textwidth]{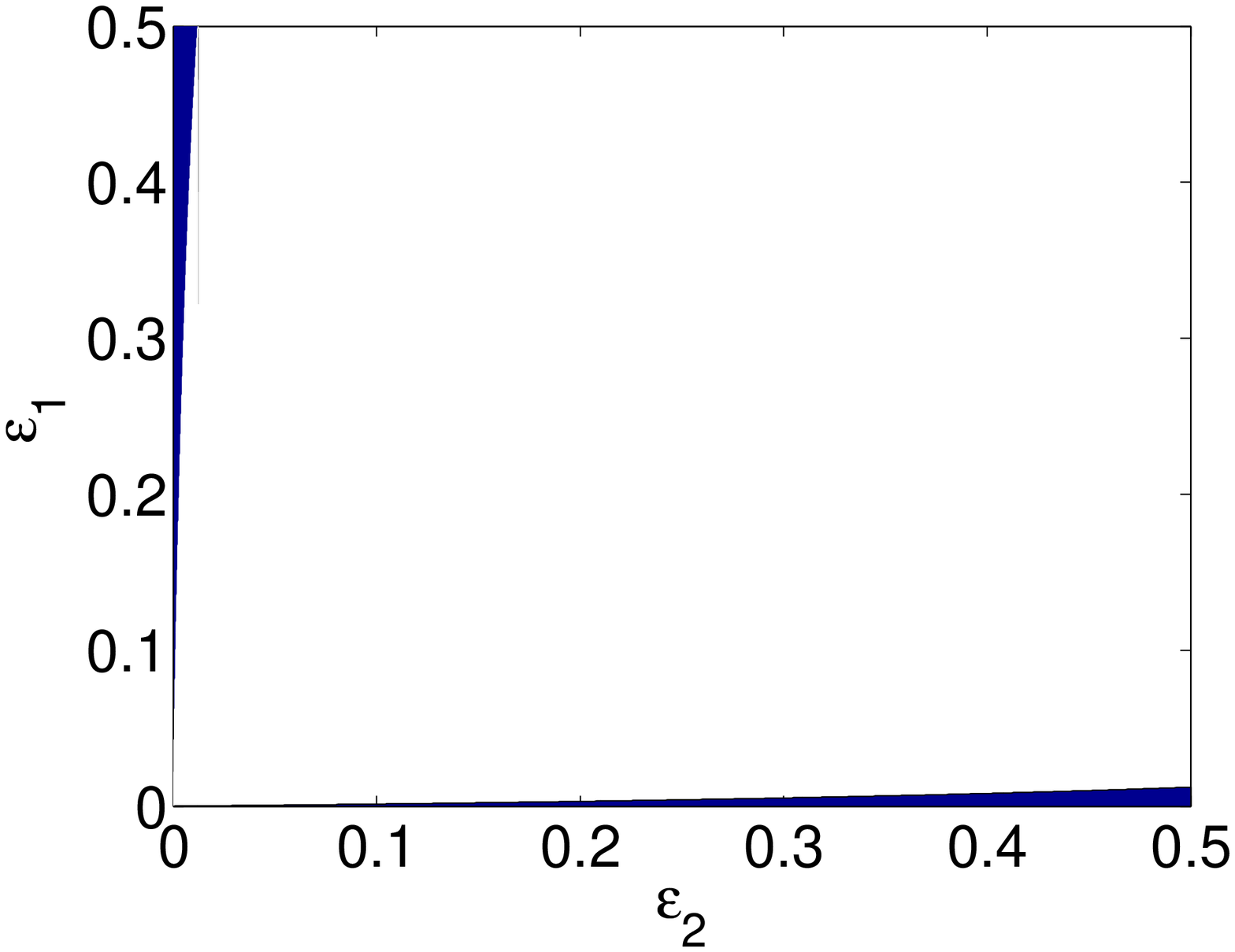} \label{fig:alpha-1}}
  \subfigure[$q=0.24$]{
  \includegraphics[width=0.3\textwidth]{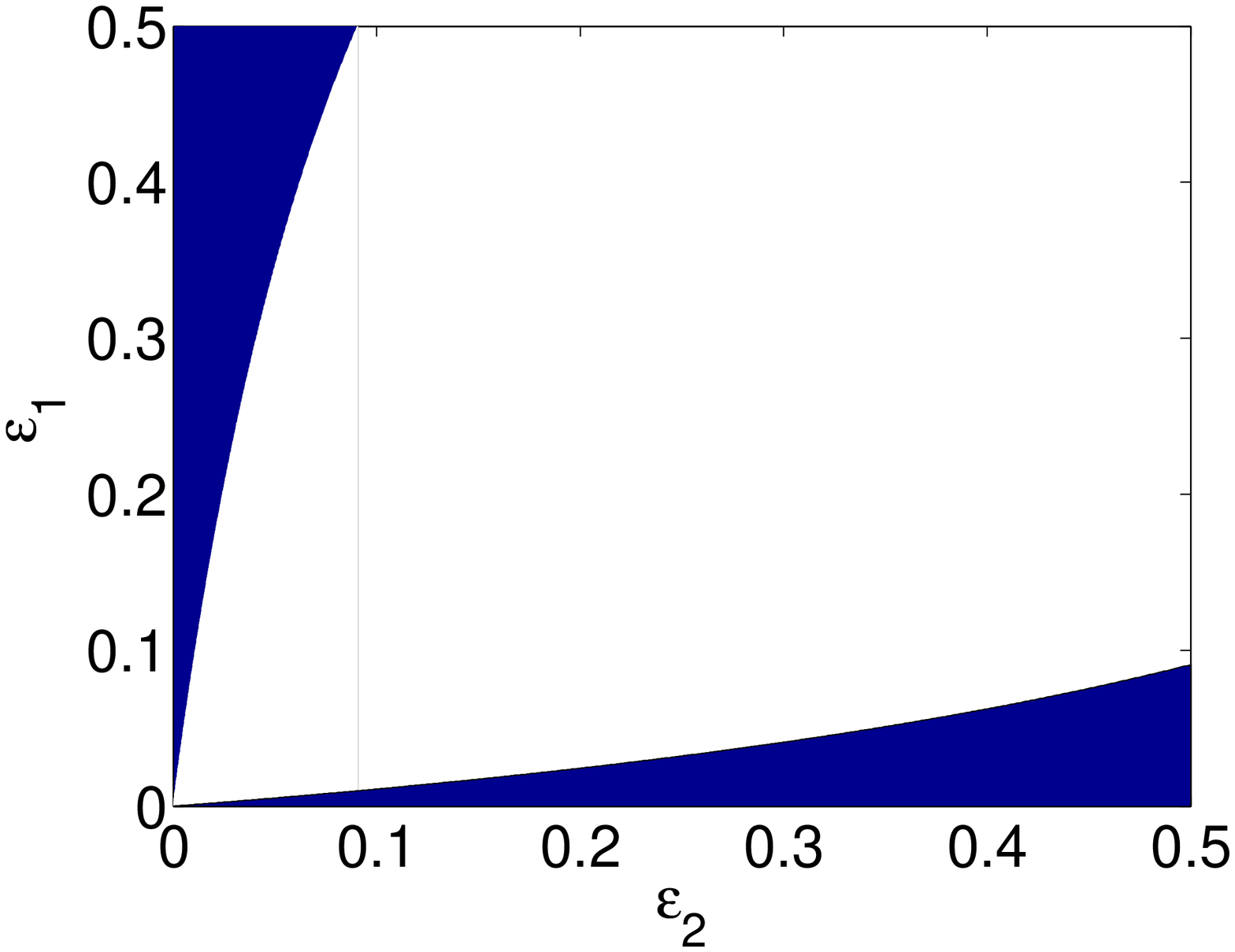} \label{fig:L-2}}
  \label{fig:comp-1}
  \subfigure[$q=0.4$]{
  \includegraphics[width=0.3\textwidth]{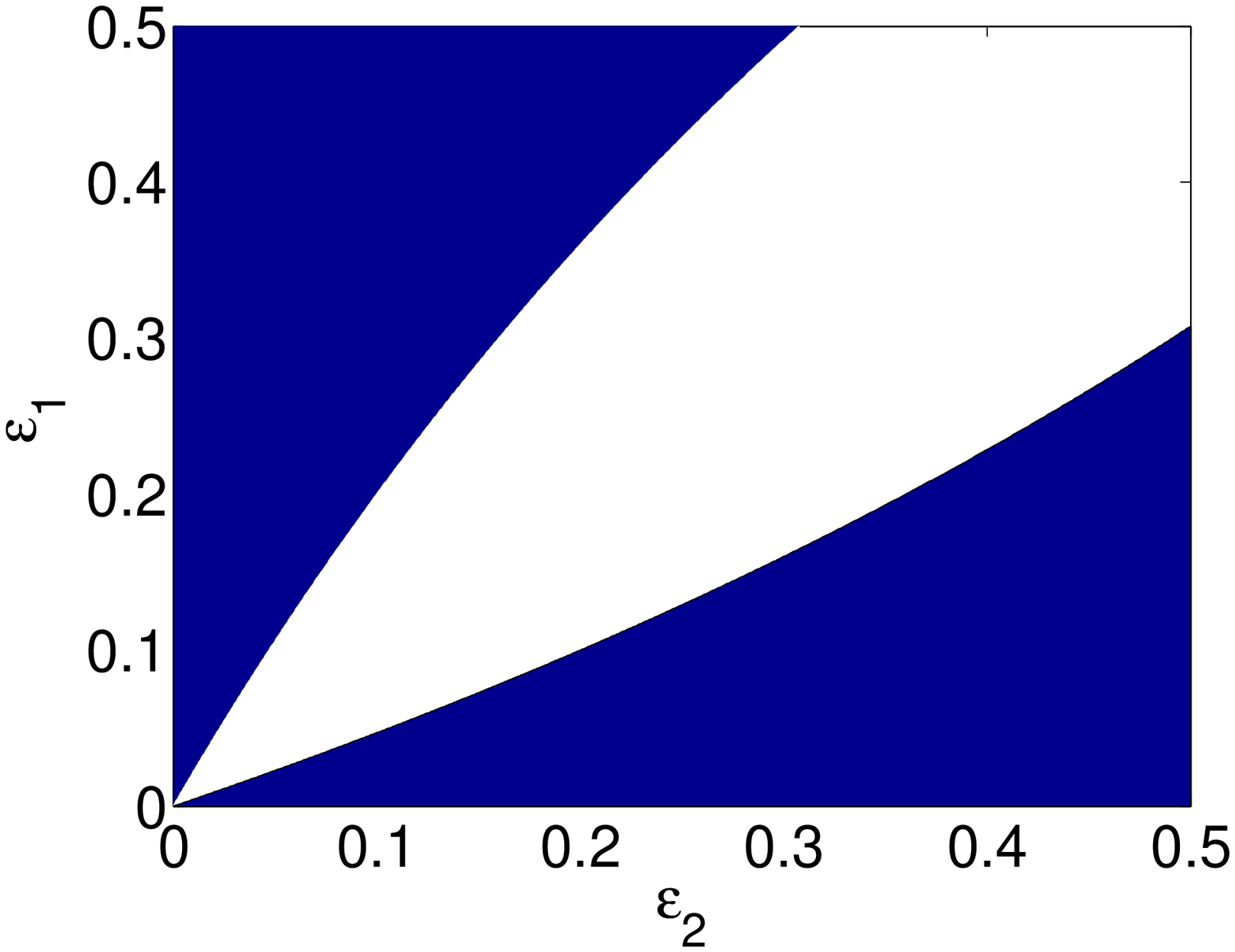} \label{fig:alpha-1}}
  \caption{(Color online) Region of zero (blue) and non-zero (white) entropy shown in the $\epsilon_1$ versus $\epsilon_2$ (noise value in channel 1 and 2 respectively) plot for different values of parameter $q$}\label{fig:comp-1p}
\end{figure}

Knowing the region of zero and non-zero entropy is of particular
interest for the two observation channel system. We derive this
region for the possible values of noise, $\epsilon$ in the two
observation channels for parameter $q$ and this is plotted in Figure
\ref{fig:comp-1p}. We can qualitatively see that for the region
corresponding to $\epsilon_1 = \epsilon_2$ line which defines
$\alpha=1$, we can never obtain a unique sequence from MAP
estimation. This is due to the prevalence of degenerate ground state
solutions obtained at zero temperature due to the mutual
nullification of the opposing field in the two observation channels.
As we perturb the external random field applied in one of the
observation channels by a certain amount (depending on the value of
$q$), we migrate to the region of zero entropy. The region above the
$\epsilon_1 = \epsilon_2$ line corresponds to $\alpha \geq 1$, and
at the zone boundary between the zero and non-zero entropy, we have
$h_1+2J=h_2$. Denoting by $\epsilon_{1b}$ and $\epsilon_{2b}$, the
values of the corresponding $\epsilon_{1}$ and $\epsilon_{2}$ at the
zone boundary\footnote{The zone boundary for the region below
$\epsilon_1 =\epsilon_2$ line can be obtained by interchanging
$\epsilon_{1b}$ and $\epsilon_{2b}$ in (\ref{eq:MAPpar})}, we have

\begin{equation}\label{eq:MAPpar}
\frac{1- \epsilon_{2b}}{1- \epsilon_{1b}} = e^{4J}
\frac{\epsilon_{2b}}{\epsilon_{1b}}
\end{equation}

 Before $h_2$ hits the point of first phase transition, the prior $2J$ and the noisier channel
 field strength $h_1$ combine to cancel out the effect of clean channel $h_2$ and the spins
 in certain clusters of the Ising chain will be frustrated which explains the reason why we
 have non-zero entropy at zero-temperature. Once $h_2$ crosses the zone boundary then the mutual
 cancelation is inefficient and we obtain the zero entropy regions. Thus, only for certain $[h_1,h_2]$ we can
 attain the zero entropy condition in the thermodynamic limit.

\begin{figure}[!h]
  \centering
  \subfigure[Error with same noise]{
  \includegraphics[width=0.4\textwidth]{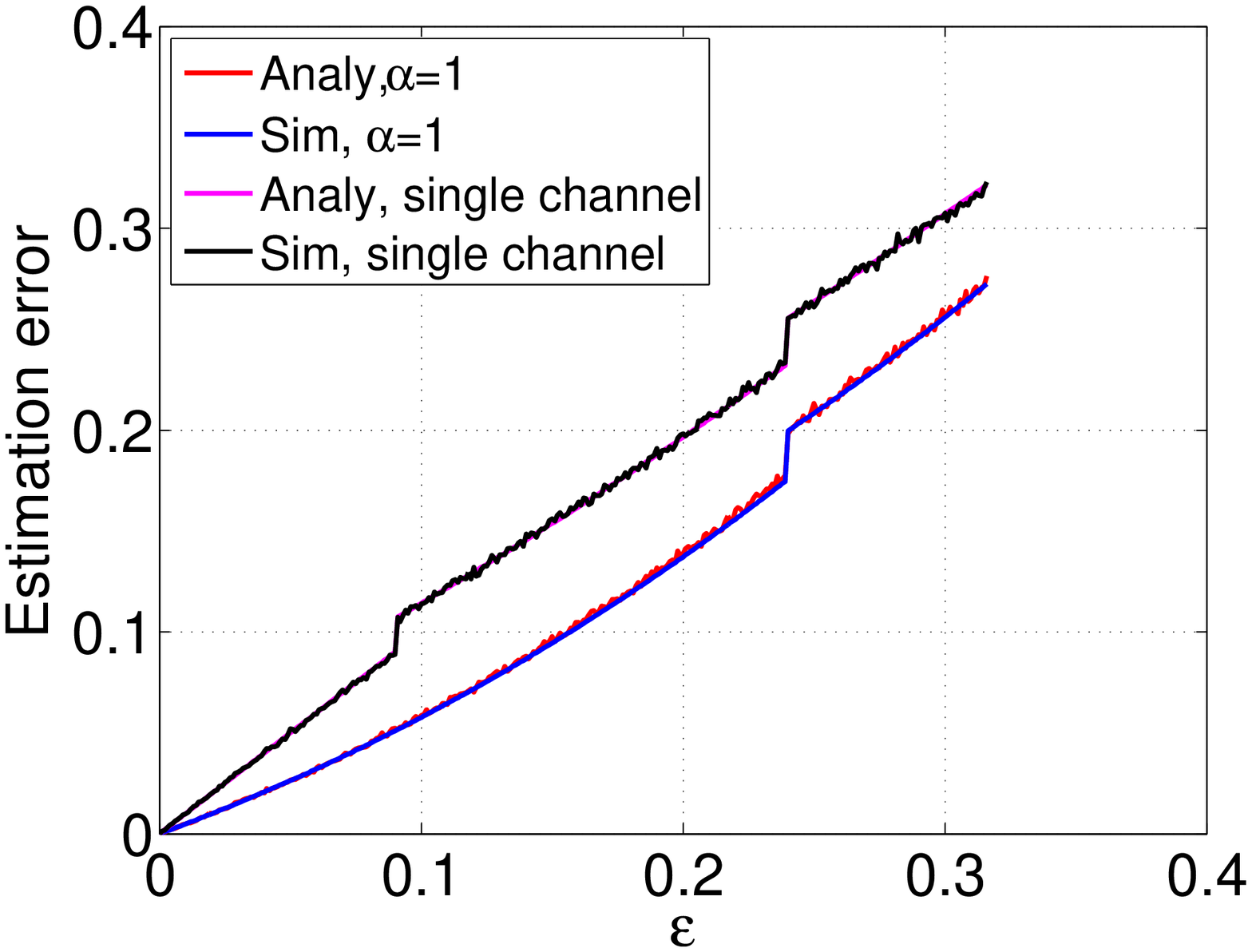} \label{fig:Analy_error_same}}
  \subfigure[Error with Varying $\alpha$]{
  \includegraphics[width=0.4\textwidth]{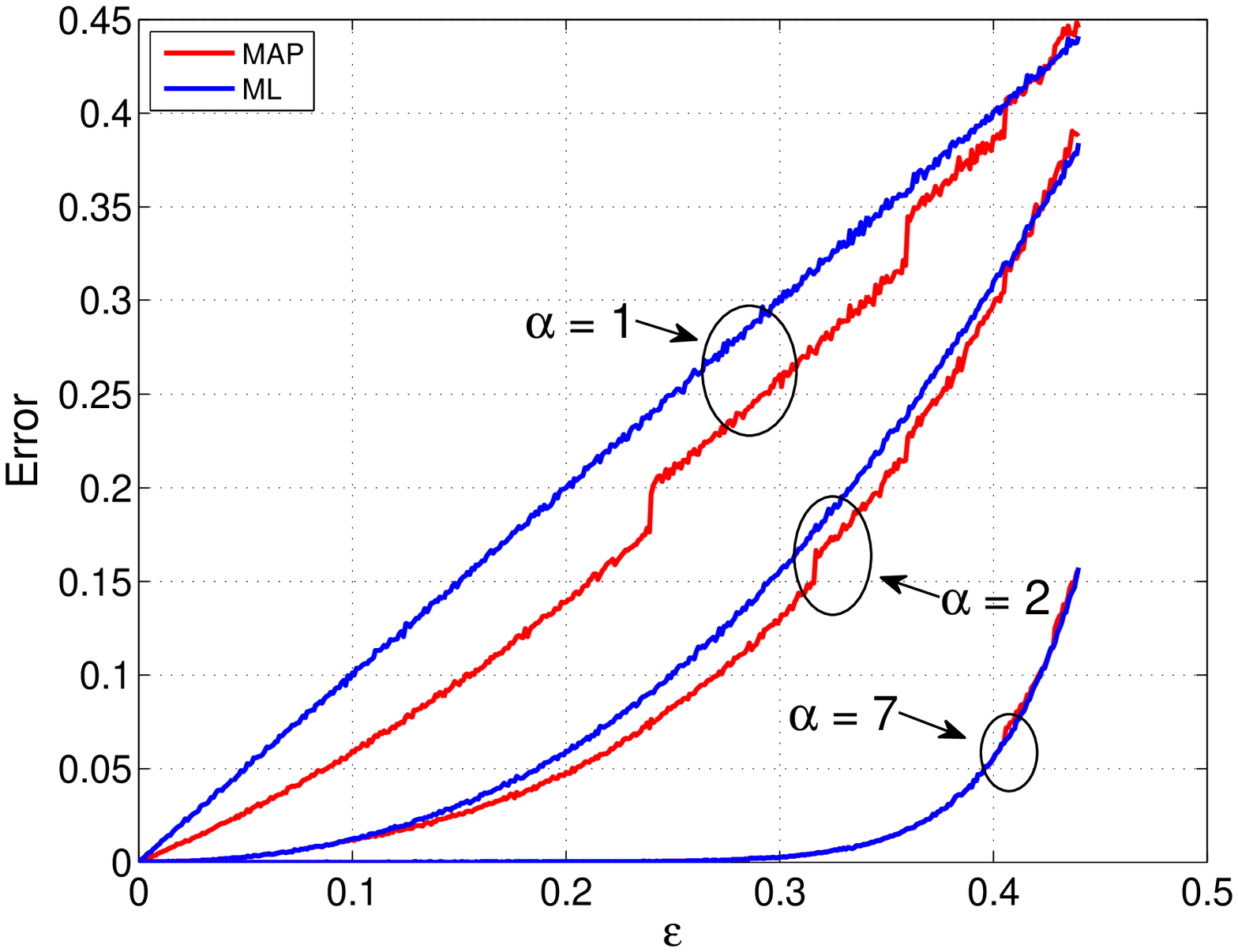} \label{fig:alpha-error}}
  \caption{(Color online) Plot of error from (a) MAP estimates obtained analytically for $1-$ and $2-$ observation channel systems with same noise in each of them (analytical and simulated) and in (b) we have the plots using ML (blue) and MAP (red) estimates obtained by varying $\alpha$ in the two channels (simulated) for $q=0.24$}\label{fig:Analytical_error}
\end{figure}

We will now find the error between the actual and estimated
sequences. In Figure \ref{fig:Analy_error_same} the error plots are made for (a) a single observation channel and (b) two observation channels with $\alpha=1$. The estimates are obtained semi-analytically using MAP technique (see \eqref{eq:MAP29err}, \eqref{eq:MAP30a}). The details of the derivation with
further analysis is provided in Appendix
\ref{sec:Analyticalerrordetails}.

In Figure \ref{fig:alpha-error} the ML and MAP error estimates are
plotted using the Viterbi algorithm. Here we have studied the
system at higher values of $\alpha$ (the semi-analytical treatment
described in Appendix \ref{sec:Analyticalerrordetails} can be
extended to study the system having high $\alpha$ with tedious
calculations). We notice from ML as well as MAP estimation (which in
this case is close to ML estimate but the performance is heavily
dependent on $q$) that addition of a cleaner channel (i.e.,
increasing $\alpha$) results in a reduction of the error. For ML,
the result is mainly driven by the cleaner channel. The error
due to ML estimation is simply the probability of error in the
cleaner channel, and is given as
  \begin{equation}\label{eq:MAP26}
P_{ML} \{ error\}=\frac{1}{1 + (\frac{1 -
\epsilon}{\epsilon})^\alpha}
\end{equation}
For higher $\alpha$, the error is small for low and intermediate
noise values but at higher noise values, the error increases rapidly
and becomes comparable to error at lower $\alpha$. We also notice that MAP estimates have lower error relative to ML at smaller $\alpha$ for the particular value of $q$ studied. However, as we increase $\alpha$, for example $\alpha=8$, the ML and MAP estimated error becomes indistinguishable for any value of $\epsilon$.

\subsection{$L$-channels with identical noise }
\label{sec:Lchannel}

\begin{figure}[!h]
  \subfigure[$c$]{
  \includegraphics[width=0.3\textwidth]{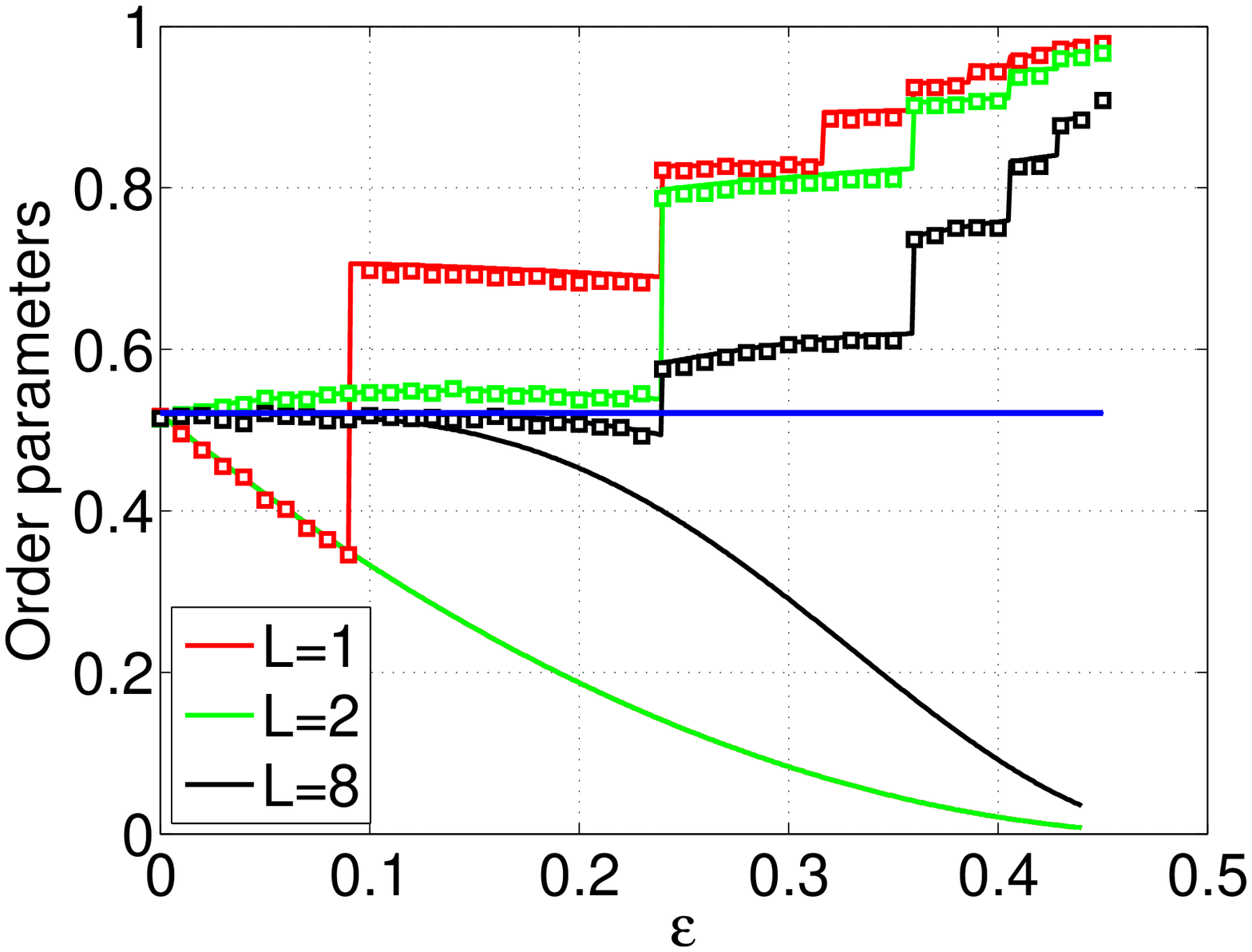} \label{fig:C_L}}
  \subfigure[$v$]{
  \includegraphics[width=0.3\textwidth]{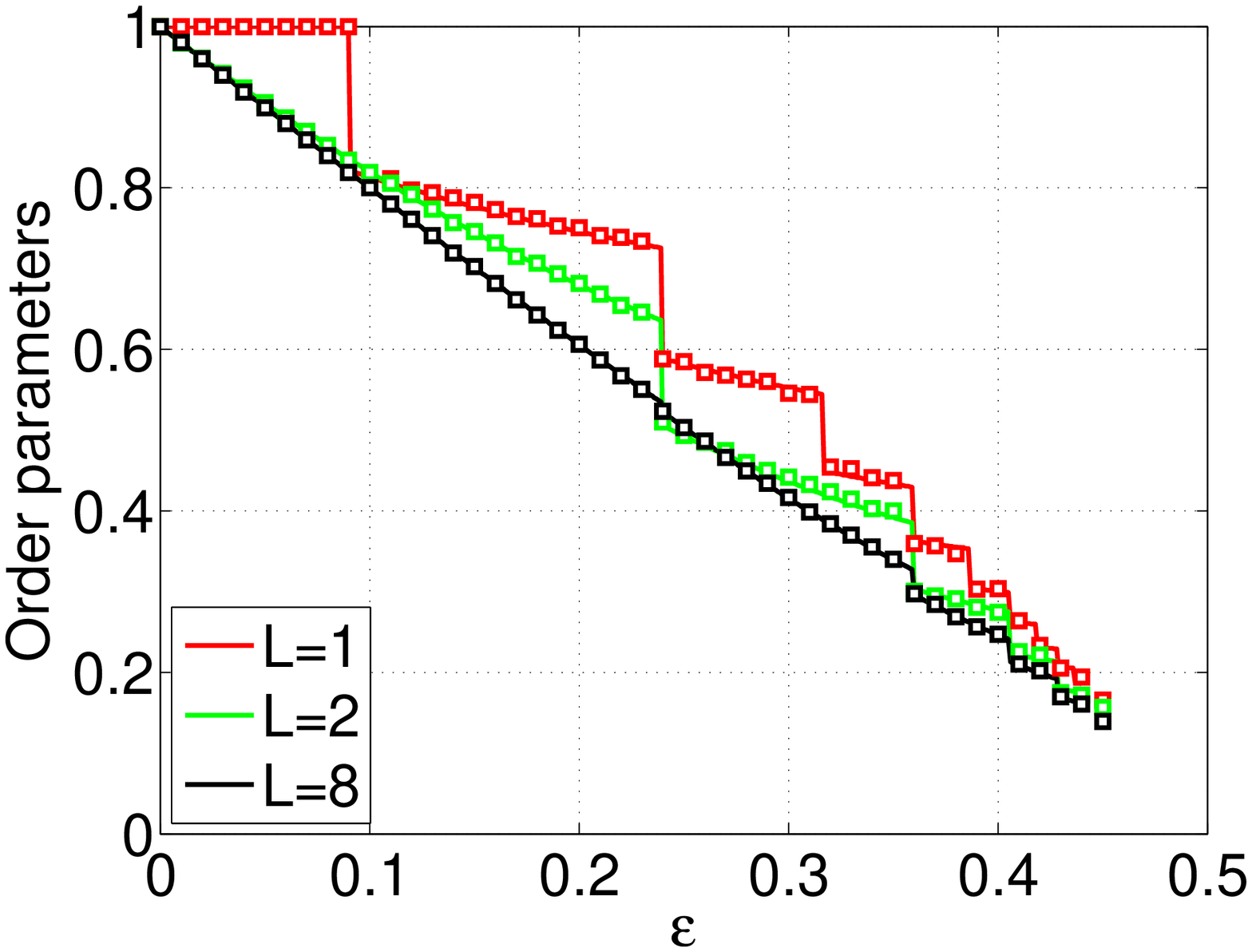} \label{fig:V_L}}
  \subfigure[$v_1$]{
  \includegraphics[width=0.3\textwidth]{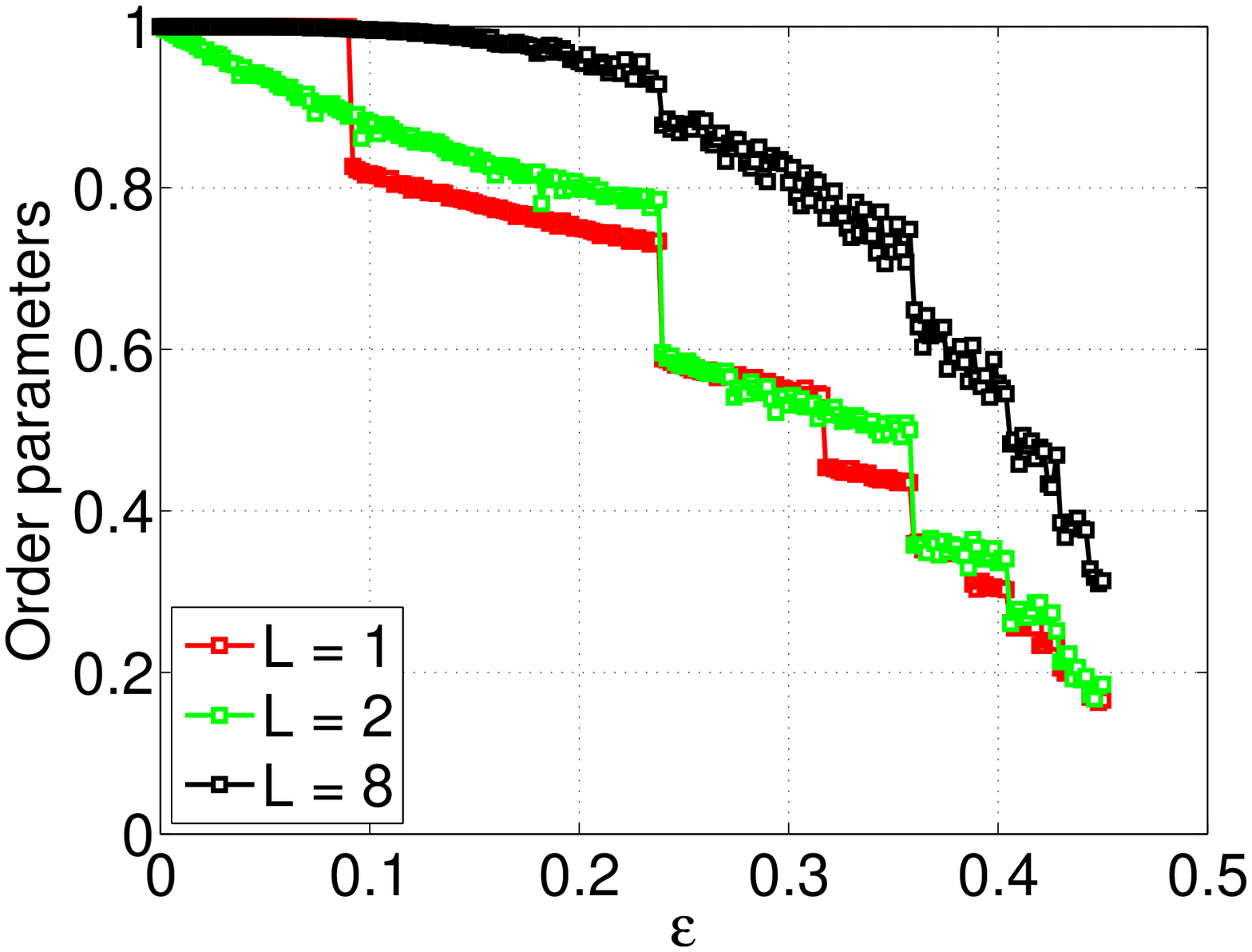} \label{fig:L-diff}}
  \caption{(Color online) Order parameters (a) $c$ and (b) $v$ obtained analytically by Ising Hamiltonian minimization (bold line) which shows an exact superposition with the data obtained from the Viterbi algorithm (open squares) for $L$ channel systems with same noise for $q=0.24$. The smooth decaying lines in the plot for $c$ are obtained via ML estimate and the horizontal blue line depicts $c_0$. (c)$v_1$ is the overlap between the MAP and ML estimated sequences. }\label{fig:all-L}
\end{figure}

In this subsection, we study a system of $L$ observation channels
with identical noise, for a fixed value of positive correlation
coefficient $J$.
The correlation $c$ found from the MAP
estimate matches exactly with the ML estimate for regions with small
noise as can be seen from Figure \ref{fig:C_L}. However for
intermediate values of noise we see that $c$ from MAP estimation is more stable for all
values of $L$ whereas its value obtained from ML estimate shows a
monotonic decrease with increase in the noise value ~\footnote{The ML estimate for $L=1$ and $L=2$ coincides as can be easily seen from \eqref{eq:MAPMLc} and \eqref{eq:MAP25ll}}.

Interestingly, for higher $L$, $c$ is stable even before the first
phase transition at lower noise regimes. This can be seen by
comparing the value of $c$ with the reference value $c_0$ of the
Markov process $\Xc$. $c_0$ and $c$ are given below,

\begin{eqnarray}\label{eq:MAPMLc}
c_0 = \sum_{x_1,x_2} x_1 x_2 p_{\rm st}(x_1) p(x_2|x_1) = 1 - 2q \nonumber \\
c_{ML}=(1-2q)(1-2P_{ML}\{ error \}|_{L})^2
\end{eqnarray}
$P_{ML} \{ error\}|_{L}$ provides the error estimate for $L$ even or odd and is given in \eqref{eq:MAP25ll}.
With increase in noise the correlation shows jumps in its value.
These jumps become smaller and appear at closer intervals with
increasing noise intensity and $c$ saturates to the prior dominated
value of 1. The overlap between the observed and estimated sequence is
plotted in Figure \ref{fig:V_L}. Similar to what was done in
the case of 2-channel systems, the overlap between the MAP and ML
estimated sequences is plotted in Figure \ref{fig:L-diff}. This
gives a more clear indication of the observation dominated regimes
for different $L$. With an increase in noise there is a
gradual drop in overlap with discrete jumps at the points of phase
transition, tending towards 0 at high noise. However, on adding
more channels to the system we find that the overlap shows a gradual
monotonic decay. All the above analytical calculations are
supported by simulations obtained by running the Viterbi algorithm
and are plotted along with the analytical data for comparison.

\begin{figure}
  \centering
  \subfigure[odd L]{
  \includegraphics[width=0.4\textwidth]{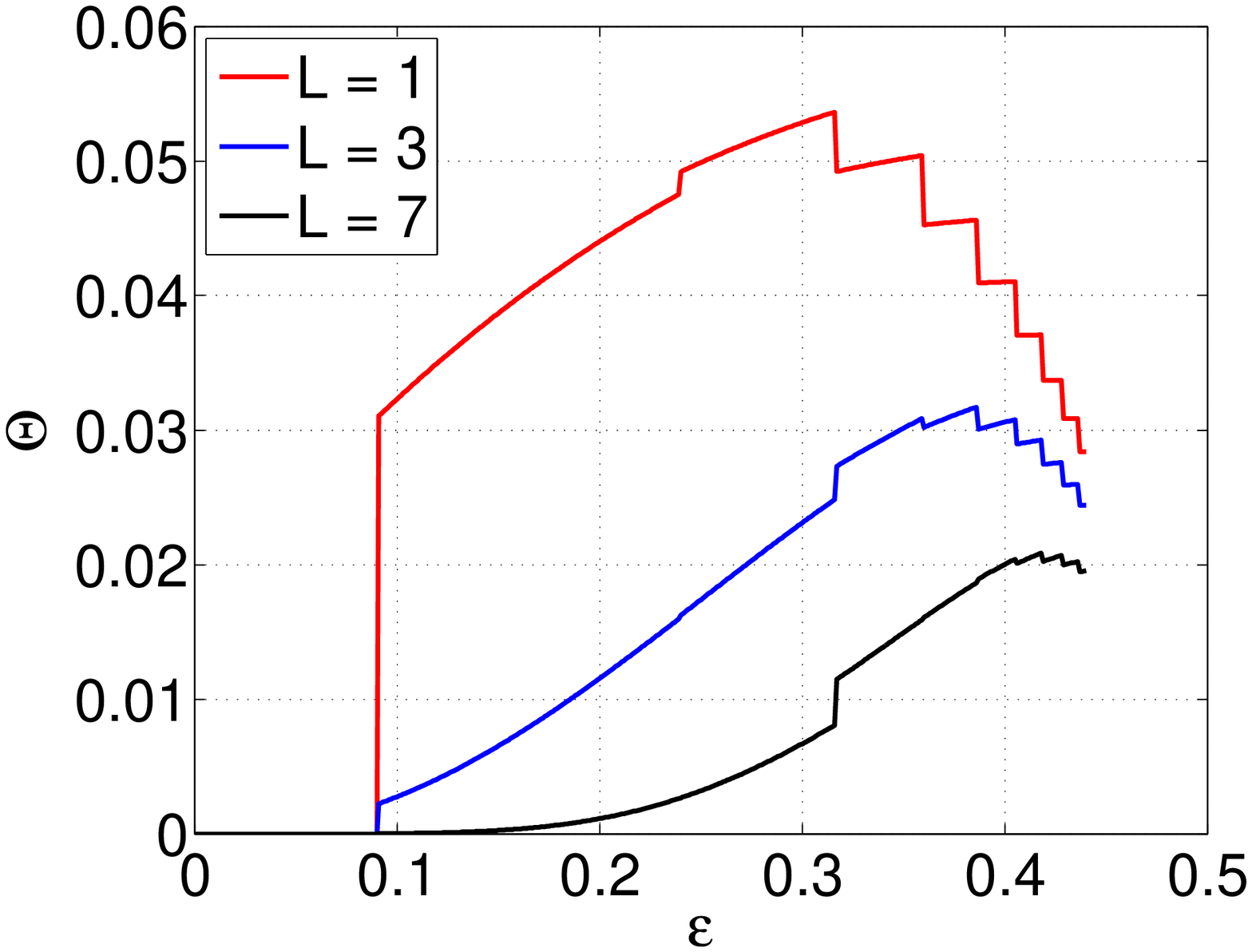} \label{fig:L-21}}
  \subfigure[even L]{
  \includegraphics[width=0.4\textwidth]{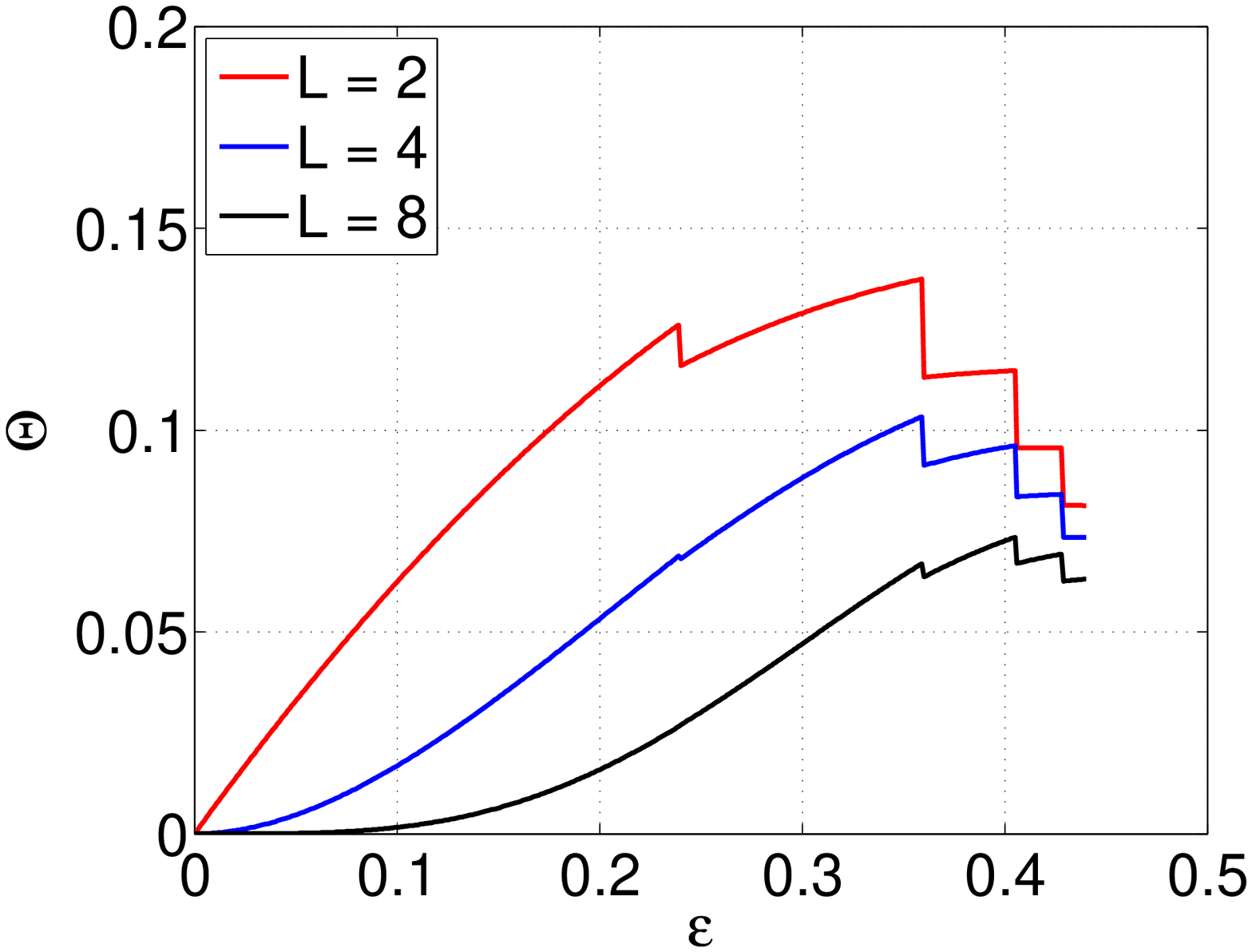} \label{fig:L-21}}
  \caption{(Color online) The entropy of (a) odd and (b) even channel systems obtained analytically with varying $\epsilon$ ($\epsilon$ is same in all the channels of the $L$ channel system) at $q=0.24$.}\label{fig:comp-5be}
\end{figure}

We now focus on the entropy of the system, defined as the natural
logarithm of the number of MAP solutions that we can possibly
obtain. The entropy is plotted in Figure \ref{fig:comp-5be}.
\begin{itemize}
\item When the number of channels in the system is odd, then  for small values of noise (in the ML dominated regime) there is a unique solution to the
MAP estimation problem.  When varying the noise, the system undergoes first-order phase transitions at the points given by  $h=\frac{2J}{m}$. In particular, the entropy becomes non-zero at the first phase transition, $h=2J$, signaling an exponentially many solutions to the MAP estimation problem.  At each phase transition point we see
that there are discrete jumps in entropy. The magnitude of those jumps at the points of phase transitions diminishes with the increase in $L$. However,  the position of phase
transitions is independent of the number of channels.
Mapping the system to an Ising spin model, we can see that there are
$L+1$ effective forces acting on a spin at a particular instant, the $L$
magnetic fields due to the observation channels and one due to the
spin-spin interaction.  At the point of first phase transition, the
prior $2J$ (quantifying the spin-spin interaction) nullifies the
effect of one of the channels, whereas the magnetic field from the remaining even
number of channels compensate each other due to mutually conflicting observations.
\item For even $L$,
there are no regions of zero entropy and thus for all possible
values of $L$, there are exponentially many MAP solutions corresponding to a typical observation sequence. This is
due to the fact that for an even number of channels with the same
noise, the magnetic fields acting on the spin
mutually compensate each other, resulting in macroscopic frustration. Due to this, for any
value of noise we have non-zero entropy \cite{behnmarkovising}. The
rise is again found to be more gradual for large even $L$ values.
The points of phase transition when $L$ is even is given by,
$h=\frac{J}{m}$. Thus for these systems the number of phase
transition points is reduced by half relative to odd $L$.
\end{itemize}

 \begin{figure}
  \centering
  \subfigure[Error with Varying $L$]{
  \includegraphics[width=0.5\textwidth]{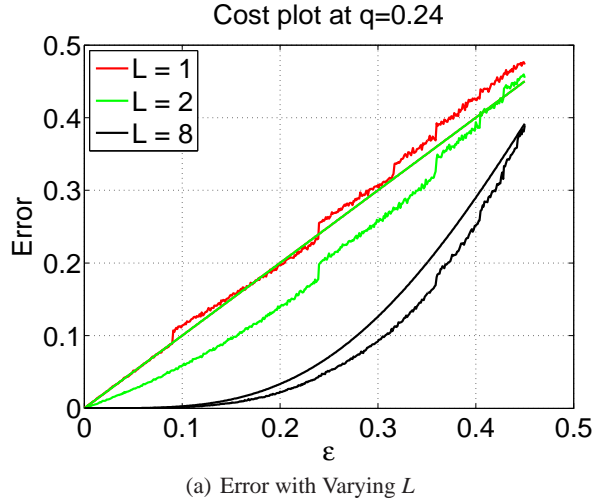} }
  \caption{(Color online) Plot of error with ML (smooth) and MAP (wiggled) estimates obtained by varying number of channels for $q=0.24$. }\label{fig:L-error}
\end{figure}

Finally, in Figure \ref{fig:L-error} we provide the accuracy of
estimation in the multi observation channel by MAP estimation using
Viterbi algorithm. The same is also calculated using ML estimate (plotted for
comparison) and is given by the following formula~\footnote{The ML estimate for $L=1$ and $L=2$ coincides as can be easily seen from the estimation error provided in \eqref{eq:MAP25ll} for $L$ odd and even},
\begin{eqnarray}\label{eq:MAP25ll}
P_{ML}\{{\rm error}\}|_{L={\rm odd}}=1-\sum\limits^{\frac{L-1}{2}}_{k=0} \binom{L}{k} \epsilon ^k (1- \epsilon)^{L-k} \nonumber \\
P_{ML}\{{\rm error}\}|_{L={\rm even}}=1-\sum\limits^{\frac{L}{2} -
1}_{k=0} \binom{L}{k} \epsilon ^k (1- \epsilon)^{L-k}-
\frac{1}{2}\binom{L}{\frac{L}{2}} \epsilon^{\frac{L}{2}} (1-
\epsilon)^{\frac{L}{2}}
\end{eqnarray}

The error in estimation is found to improve with the addition of
more observation channels.

\subsection{One ``clean" vs multiple noisy channels}
\label{sec:Lchannel_cost}

In this section, we bring in the notion of channel cost and use it
to compare the performance of a single channel system with a
multi-channel system while keeping the cost same. Channel cost can
be interpreted as a function of the channel noise $\epsilon$. In
many scenarios, it is more expensive for a system designer to build
a channel with small noise than one with higher noise. For example,
in realistic channels for data communication, thermal noise at the
receivers is a major source of channel noise and building a receiver
with low thermal noise is expensive. The binary symmetric channels
considered here can be used to model a power plant producing
equipments and the noise can be interpreted as the probability of
making a defective equipment. In order to make the equipments less
defective, a system engineer would need expensive machines and good
maintenance, which increases the cost. Because of this inverse
relation between channel cost and channel error, for simplicity, we
model the channel cost as being inversely proportional to the
channel error and linearly proportional to the number of channels in
the system. Thus, for a $L$-channel system, the channel cost is
given by $log(\frac{L}{\epsilon})$ ~\footnote{Here we take the log
of $\frac{L}{\epsilon}$ to define the cost because for small error
$\frac{L}{\epsilon}$ is exponentially large }.

\begin{figure}[!h]
  \centering
  \subfigure[$q=0.1$]{
  \includegraphics[width=0.3\textwidth]{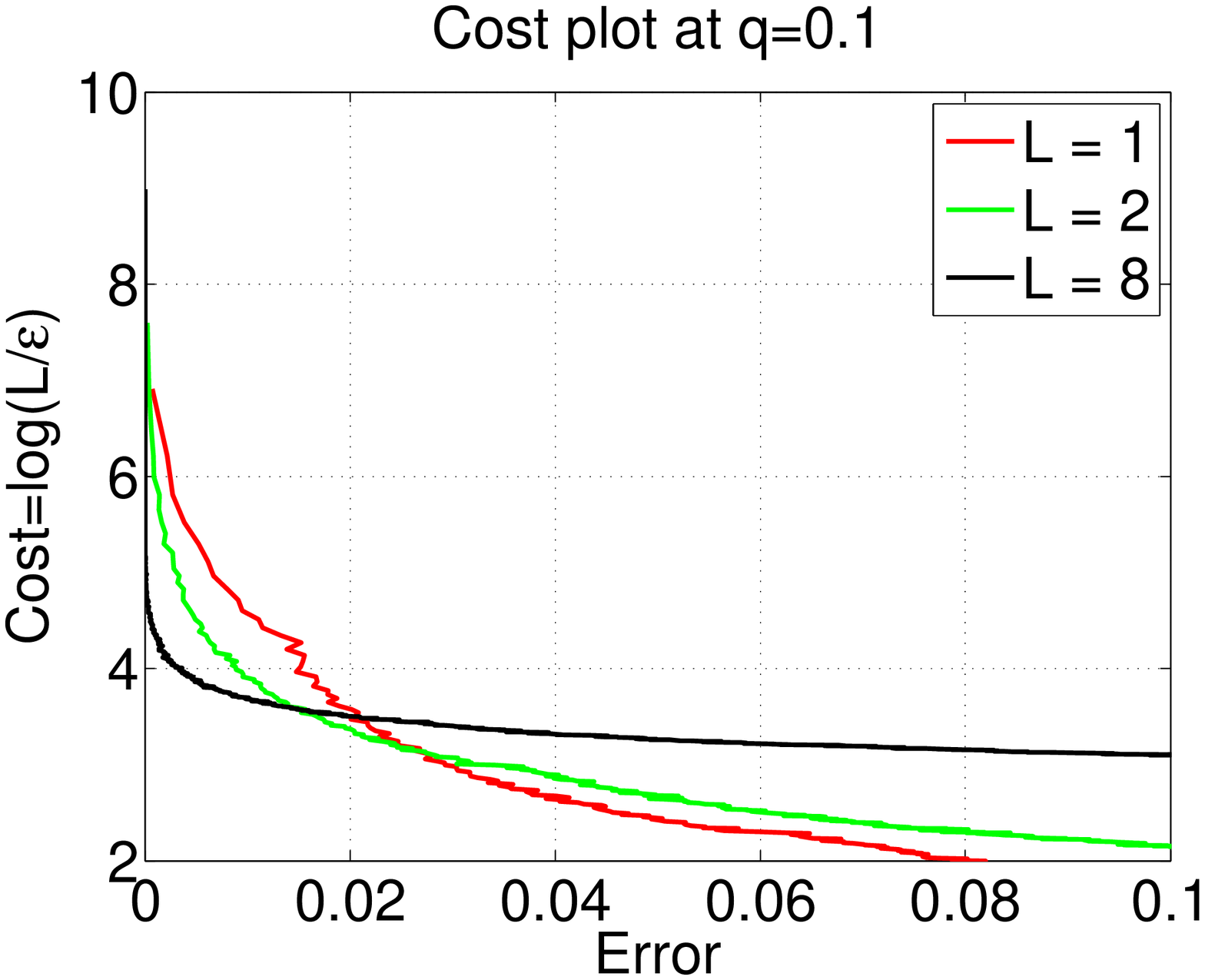}
  \label{fig:L-Errorcostq01}}
  \subfigure[$q=0.24$]{
  \includegraphics[width=0.3\textwidth]{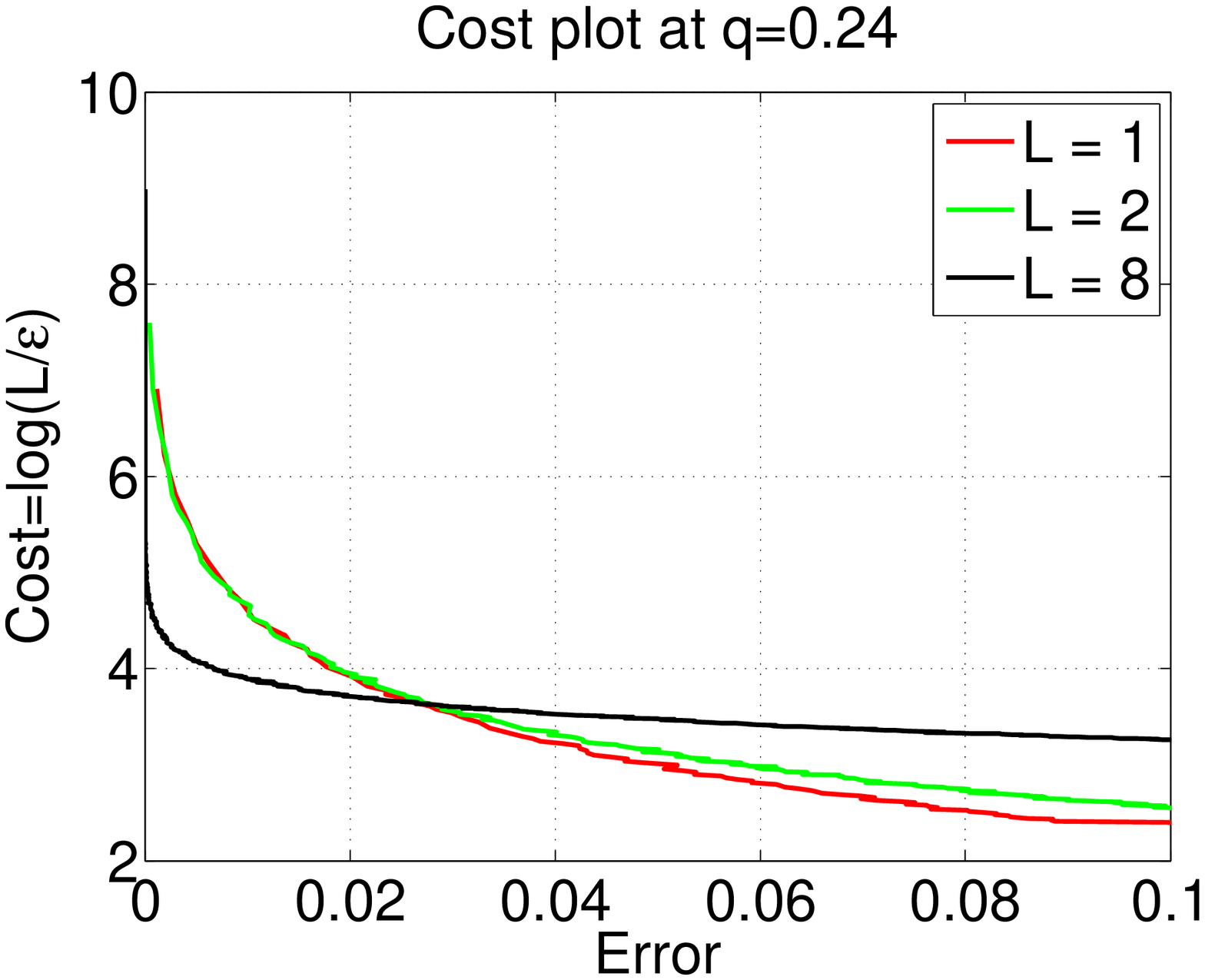} \label{fig:L-Errorcostq24}}
  \subfigure[$q=0.4$]{
  \includegraphics[width=0.3\textwidth]{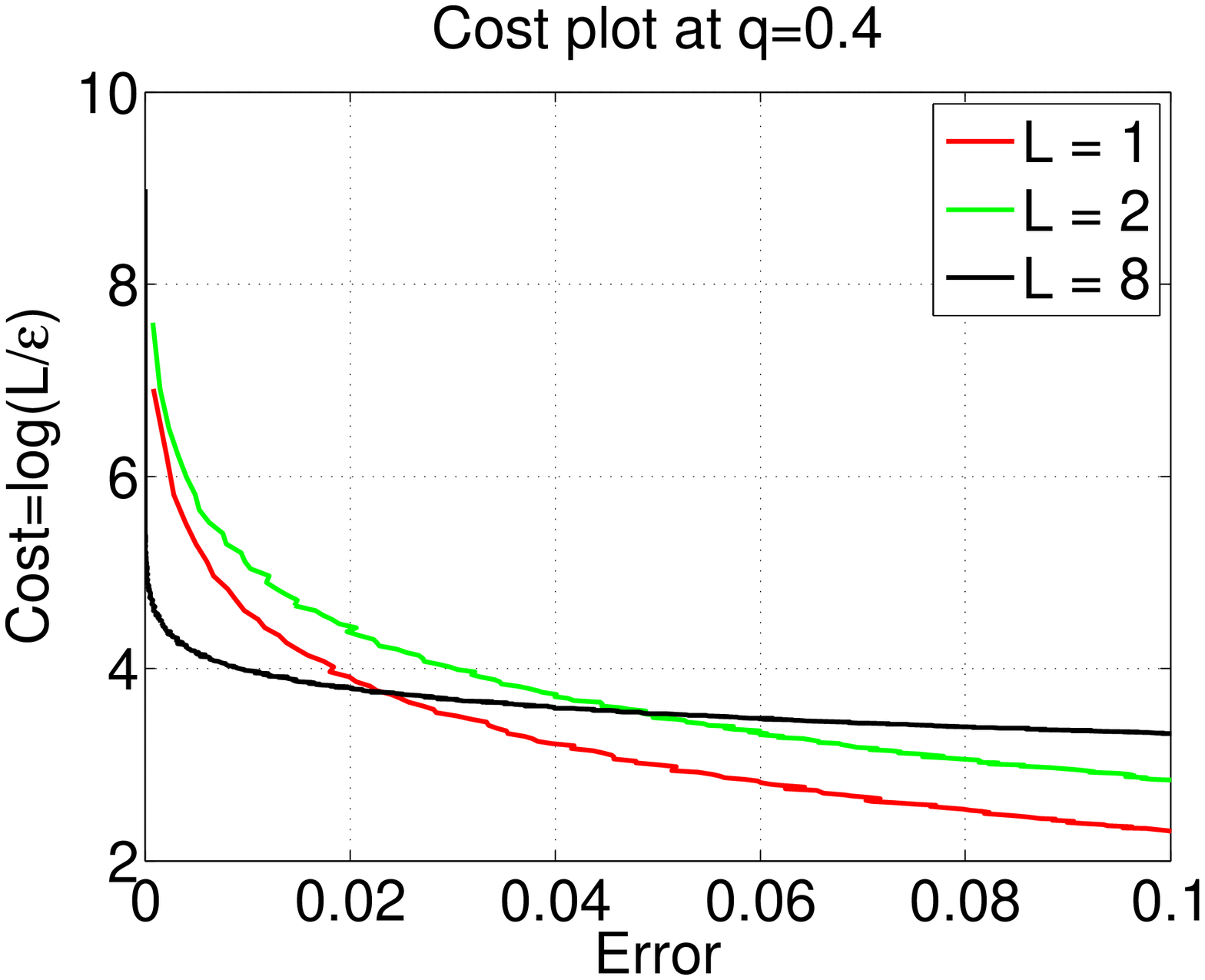} \label{fig:L-Errorcostq04}}
\label{fig:Errorcost}
  \caption{(Color online) Cost vs. Performance for multi-channel systems at (a) $q=0.10$,(b) $q=0.24$ and (c) $q=0.40$}\label{fig:orderparamter-L-cost}
\end{figure}

Figures \ref{fig:L-Errorcostq01}-\ref{fig:L-Errorcostq04} give a
comparison of cost vs. performance for various multi-channel systems
with varying $q$. The cost is plotted as
$\log\left(\frac{L}{\epsilon}\right)$ while the performance is
measured in terms of error. A point on the curve with a specific
error value, say $E$, indicates the channel cost incurred in order
to build a system that can tolerate a maximum error of $E$. It is
easy to note that if the channel cost was inversely proportional to
$\epsilon$ alone, then increasing $L$ would have always resulted in
a better system to tolerate a certain value of error. However,
because of the linear dependence on the number of channels $L$, we
see a certain value of $E$, say $E_{\rm th}$ after which the
the cost required to build an $L$ channel system becomes higher than
that required for a single channel system in order to attain the
same level of performance in terms of error requirements.

In line with our argument, we observe from the plots that in order
to build a system with small error, it is advisable to increase $L$
in order to minimize the channel cost. However, if we can tolerate a
larger error, it would be cheaper to go for a single channel system.
For example, when the error is small, the cost required to build a 8
channel system is less than that required to build a single or a
2-channel system. This is true for all values of $q$. However, when
the error is large, the single channel system is much more
preferable, since the cost required is much lesser than that needed
to build a system with $L > 1$.

\subsection{Single observation channel with gaussian noise}
\label{sec:Sunglegaussnoise}

\begin{figure}
  \centering
  \includegraphics[width=0.5\textwidth]{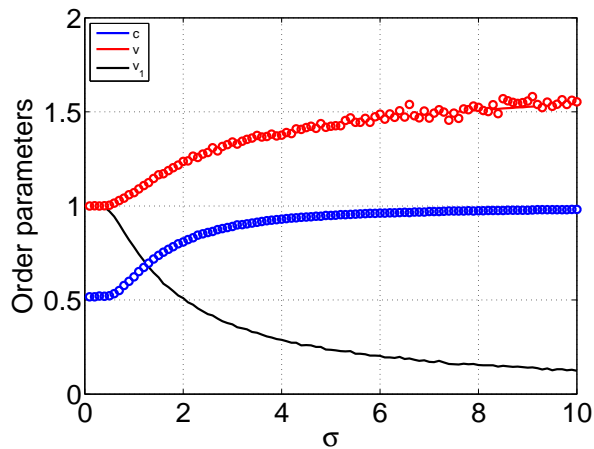}
  \caption{(Color online) Order parameters $c$ (in blue) and $v$ (in red)
  plotted from the analytically obtained data from Ising hamiltonian
  minimization (bold line) and from the Viterbi algorithm (open circles)
  for the case of single observation channel with Gaussian noise for $q=0.24$. $v_1$ is the overlap between the MAP and ML estimated sequences.}\label{fig:gaussianL}
\end{figure}

A detailed study of the single channel with a Gaussian observation noise
model is relegated to Appendix \ref{sec:gaussian}. Here, we only
provide the results for the order parameters $c$ and $v$ which are
plotted in Figure \ref{fig:gaussianL}. We also show $v_1$,
which is the overlap between the ML and MAP estimated sequences. The
plots are continuous and no first order phase transitions are
observed compared with the discrete noise model considered earlier.
We also find that the entropy is zero for this case (see Appendix
\ref{sec:gaussian} for details), meaning we do not have
exponentially many solutions for a given observation
sequence. An important point to note is that in the discrete
case, a zero entropy signified an observation dominated regime,
where the ML and MAP estimates coincided. However, in the gaussian
case, we have zero entropy at all values of $\sigma$, but this does
not mean that the ML and MAP estimates coincide everywhere. This is
justified by seeing $v_1$, which monotonically goes to zero,
indicating that the correlation between the ML and MAP estimates
reduces with an increase in the variance $\sigma^2$.

\section{Conclusion}
\label{sec:Conclusion}
In this paper we have presented an analytical study of Maximum a Posteriori (MAP) estimation for Multi-channel hidden Markov processes. We have considered a broad class of systems  with {\em odd} and {\em even} number of channels (having the same noise intensities) to understand the MAP characteristics in analogy with the thermodynamic quantities. In all the system models studied here, we observe a sequence of first-order phase transitions in the performance characteristics of MAP estimation as one varies the noise intensity. Remarkably, the position of the first phase transition depends only on whether $L$ is odd or even, but not on its value.

In the systems with {\em odd} number of channels, there is a  low noise region where the MAP estimation problem has a  a uniquely defined solution, as characterized by vanishing zero-temperature entropy of the corresponding statistical physics system. At a finite value of noise, the system experiences a first order phase transition and the number of solutions with posterior probability close to the optimal one increases exponentially. In contrast, in systems with {\em even} number of channels, the MAP estimation always yields an exponentially large number of solutions, at all noise intensities. This is explained by drawing an analogy with the thermodynamic system where the spins experience macroscopic frustration due to contradicting observations from different channels.  Our results indicate that for a system with $L=2$ observation channels, one can recover the region of zero entropy by introducing noise-asymmetry in the channels. In addition to the binary symmetric channel we have also considered the Gaussian observation channel, and demonstrated that the corresponding system has a vanishing zero-temperature entropy, indicating a unique MAP solution.

Finally, we analyze the tradeoff between system cost and estimation error for $L-$channel systems, by assuming an inversely proportional relationship between channel cost and channel noise. Our results suggest that if the objective is to achieve low estimation error, then it is more advantageous  to build a system with larger number of noisier channels, rather than having a single channel with better noise-tolerance. However if we can tolerate higher error, it is more beneficial to use a single channel system. An exception is noticed for the 2-channel system whose performance relative to a single channel system is dependent on the spin-spin interaction $(J)$. For moderate $J$ the estimation error for both the single and two channel systems are comparable while at lower $J$ the single channel system is found to perform better to achieve any degree of accuracy.

There are several directions for extending the work presented here. For instance, it will be interesting to generalize the analysis presented here beyond the binary HMMs, e.g., by reducing the problem to a generalized Potts model. Note that the critical behavior observed here  is due to two competing tendencies, $(a)$ accommodating observations and $(b)$ hidden (Markovian) dynamical model. Thus, it is natural to assume that similar behavior can be expected in non-binary systems as well.  Another interesting problem is to ``break" the macroscopic degeneracy of the MAP solution space by adding additional constraints and/or objectives. For instance, among all the MAP solutions, one might wish to select the one that has the highest overlap with the typical realization of the hidden process, which might be useful in the context of parameter learning~\cite{Allahverdyan2011NIPS}.

\begin{acknowledgements}

We would like to thank Aram Galstyan for introducing the problem to us and helping with the  preparation of the manuscript. We also thank Armen E. Allahverdyan and Greg Ver Steeg for useful discussions on various topics related to this work.

\end{acknowledgements}

\section{Appendix}
\label{sec:appendix}

\subsection{Analytical error calculation using MAP}
\label{sec:Analyticalerrordetails}

For finding the error expression for a single observation channel \footnote{The approach can be extended to multi channel system with tedious algebra}
using MAP analytically we use a modified Hamiltonian \eqref{eq:MAP27}.
To calculate the error estimate
we need to find $-\partial_g f_g|_{g \rightarrow 0,\beta
\rightarrow \infty}$ where $f_g$ represents the free energy of the
modified Hamiltonian and is given by,
 \begin{eqnarray}\label{eq:MAP28}
f_g&=&-T \frac{1}{N}\sum_{{\yv},\sv} \pv({\yv},\sv) {\rm log} \left \{ e^{\beta \sum\limits_{k=1}^N B(\xi_k)} \right \}\nonumber \\
&=& - \frac{1}{N} \sum_{{\yv},\sv} \pv({\yv},\sv)  \sum\limits_{k=1}^N B(\xi_k)
\end{eqnarray}
with $\xi_k = h y_k + g s_k + A(\xi_{k-1})$. Note that the limits
are taken in order, i.e., first we take the limit $g \rightarrow 0$
and then $\beta \rightarrow \infty$. Using the recursion relation
from \eqref{eq:MAP11} and \eqref{eq:MAP12} we get $\partial_g f_g$ as,
 \begin{equation}\label{eq:MAP29}
\partial_g f_g= - \frac{1}{N} \sum_{{\yv},\sv} \pv({\yv},\sv) \frac{1}{2} \sum\limits_{k=1}^N \{ \tanh (\beta (\xi_k + J)) + \tanh (\beta (\xi_k - J))\} \left \{ s_k + \partial_g A(\xi _ {k-1}) \right \}
\end{equation}
with the term $\partial_g A(\xi _ {k-1})$ given by,
 \begin{equation}\label{eq:MAP30}
\partial_g A(\xi _ {k-1})= - \frac{1}{2} \sum\limits_{k=1}^N \{ \tanh (\beta (\xi_{k-1} + J)) + \tanh (\beta (\xi_{k-1} - J))\} \left \{ s_{k-1} + \partial_g A(\xi _ {k-2}) \right \}
\end{equation}
Taking the limits and simplifying, we can write the terms in the
above expression as,
 \begin{eqnarray}\label{eq:MAP31}
 \tanh (\beta (\xi_{k} + J)) + \tanh (\beta (\xi_{k} - J))&=& -2 \delta (\xi_k < -J) + 2\delta (\xi_k > J) \nonumber \\
  \tanh (\beta (\xi_{k-1} + J)) - \tanh (\beta (\xi_{k-1} - J))&=&  2\delta (-J< \xi_k < J)
\end{eqnarray}
Using this we can write, {\footnotesize{
  \begin{eqnarray}\label{eq:MAP32}
-\partial_g f_g|_{g \rightarrow 0,\beta
\rightarrow \infty} &=& \frac{1}{N}\sum_{{\yv},\sv} \pv({\yv},\sv) \sum\limits_{k=1}^N [ - \delta (\xi_k>J) + \delta (\xi_k < -J) ] [ s_{k} + \delta (-J< \xi_{k-1} < J)[s_{k-1}+ \ldots ]\ldots] \nonumber \\
&=& \frac{1}{N} \sum_{{\yv},\sv} \pv({\yv},\sv) \sum\limits_{k=1}^N [f(\xi_k) ] [ s_{k} + g(\xi_{k-1})[s_{k-1}+ g(\xi_{k-2})[s_{k-2} \ldots ]\ldots] \nonumber \\
&=& \frac{1}{N} \sum_{{\yv},\sv} \pv({\yv},\sv) \sum\limits_{k=1}^N
s_k[f(\xi_k)+g(\xi_k)f(\xi_{k+1}) +\ldots+ g(\xi_k) \ldots
g(\xi_{N-1})f(\xi_N)] \nonumber \\
&=& \frac{1}{N} \sum\limits_{k=1}^N \left( \sum_{{\yv},\sv} \pv({\yv},\sv)
s_k[f(\xi_k)+g(\xi_k)f(\xi_{k+1}) +\ldots+ g(\xi_k) \ldots
g(\xi_{N-1})f(\xi_N)] \right)
\end{eqnarray}}}
where $f(\xi_k)= \delta (\xi_k>J) - \delta (\xi_k < -J)$ and
$g(\xi_k)=\delta (-J< \xi_k < J)$.
We outline the process of evaluating a particular term of the inner series, say, the term
$$ \sum_{{\yv},\sv} \pv({\yv},\sv) s_k g(\xi_k) \ldots g(\xi_{k+n-1})f(\xi_{k+n}) $$
{{
  \begin{eqnarray}\label{eq:MAP33}
&& \sum_{{\yv},\sv} \pv({\yv},\sv) s_k g(\xi_k) \ldots
g(\xi_{k+n-1})f(\xi_{k+n}) \nonumber\\ &=&
\sum_{\substack{y_1,\ldots,y_{k};s_1,\ldots,s_{k}\\y_{k+1},\ldots,y_{k+n};s_{k+1},\ldots,s_{k+n}}}
\pv({\yv},\sv) s_k g(\xi_k) \ldots
g(\xi_{k+n-1})f(\xi_{k+n})\nonumber\\
&=&
\sum_{\substack{\xi_{k},s_k;y_{k+1},\ldots,y_{k+n}\\s_{k+1},\ldots,s_{k+n}}}
\pv(\xi_{k},s_{k},y_{k+1},\ldots,y_{k+n},s_{k+1},\ldots,s_{k+n}) s_k
g(\xi_k) \ldots g(\xi_{k+n-1})f(\xi_{k+n})\nonumber\\
\label{eqn:inter} &=&
\sum_{\substack{\xi_{k},s_k;\xi_{k+1},\ldots,\xi_{k+n}\\s_{k+1},\ldots,s_{k+n}}}
\pv(s_{k+1},\ldots,s_{k+n},y_{k+1},\ldots,y_{k+n}|\xi_{k},s_{k})\pv(\xi_{k},s_{k})
s_k g(\xi_k) \ldots
g(\xi_{k+n-1})f(\xi_{k+n}) \\
&=&
\sum_{\substack{\xi_{k},s_k;\xi_{k+1},\ldots,\xi_{k+n}\\s_{k+1},\ldots,s_{k+n}}}
\prod_{i=1}^{n} p(s_{k+i}|s_{k+i-1})\pi(y_{k+i}|s_{k+i})
\varphi(\xi_{k+i}|\xi_{k+i-1},y_{k+i}) \pv(\xi_{k},s_{k})s_k
g(\xi_k) \ldots g(\xi_{n-1})f(\xi_n) \nonumber
\end{eqnarray}}}
where in (\ref{eqn:inter}) we replace $y_k$'s in the sum with $\xi_k$'s since each sequence of $y_k$'s corresponding to a sequence of $\xi_k$'s (for two channel system there will be multiple $y_k$'s corresponding to a $\xi_k$ and this is treated accordingly) and
 \begin{equation}\label{eq:MAP34a}
\pv(\xi_{k},s_{k})=\sum_{y_{k}} \omega(\xi_{k},s_{k},y_{k})
\end{equation}
It is easy to see that the term
$$ \Delta_k = \sum_{{\yv},\sv} \pv({\yv},\sv)
s_k[f(\xi_k)+g(\xi_k)f(\xi_{k+1}) +\ldots+ g(\xi_k) \ldots
g(\xi_{N-1})f(\xi_N)] $$ which is independent of $k$ for $N \rightarrow \infty$.
For our calculation we approximate the error by keeping only the first four terms
in the expression for  $\Delta_k$,
$$ \sum_{\yv,\sv} \pv(\yv,\sv)
s_k[f(\xi_k)+g(\xi_k)f(\xi_{k+1}) + g(\xi_k)
g(\xi_{k+1})g(\xi_{k+2})f(\xi_{k+3}) + g(\xi_k)
g(\xi_{k+1})g(\xi_{k+2})g(\xi_{k+3})f(\xi_{k+4})] $$

In the thermodynamic limit, we drop the subscript $k$ which gives
$$ -\partial_g f_g|_{g \rightarrow 0,\beta
\rightarrow \infty} = \Delta $$
Hence, $\Delta$ is the overlap
between $s_k$ and $y_k$, and estimation error is calculated as,

 \begin{equation}\label{eq:MAP30a}
 P_{MAP} \{ error\}=\frac{1-\Delta}{2}
\end{equation}

\begin{figure}[!h]
  \centering
  $
  \begin{array}{ccc}
  \subfigure[$q=0.1$]{
  \includegraphics[width=0.3\textwidth]{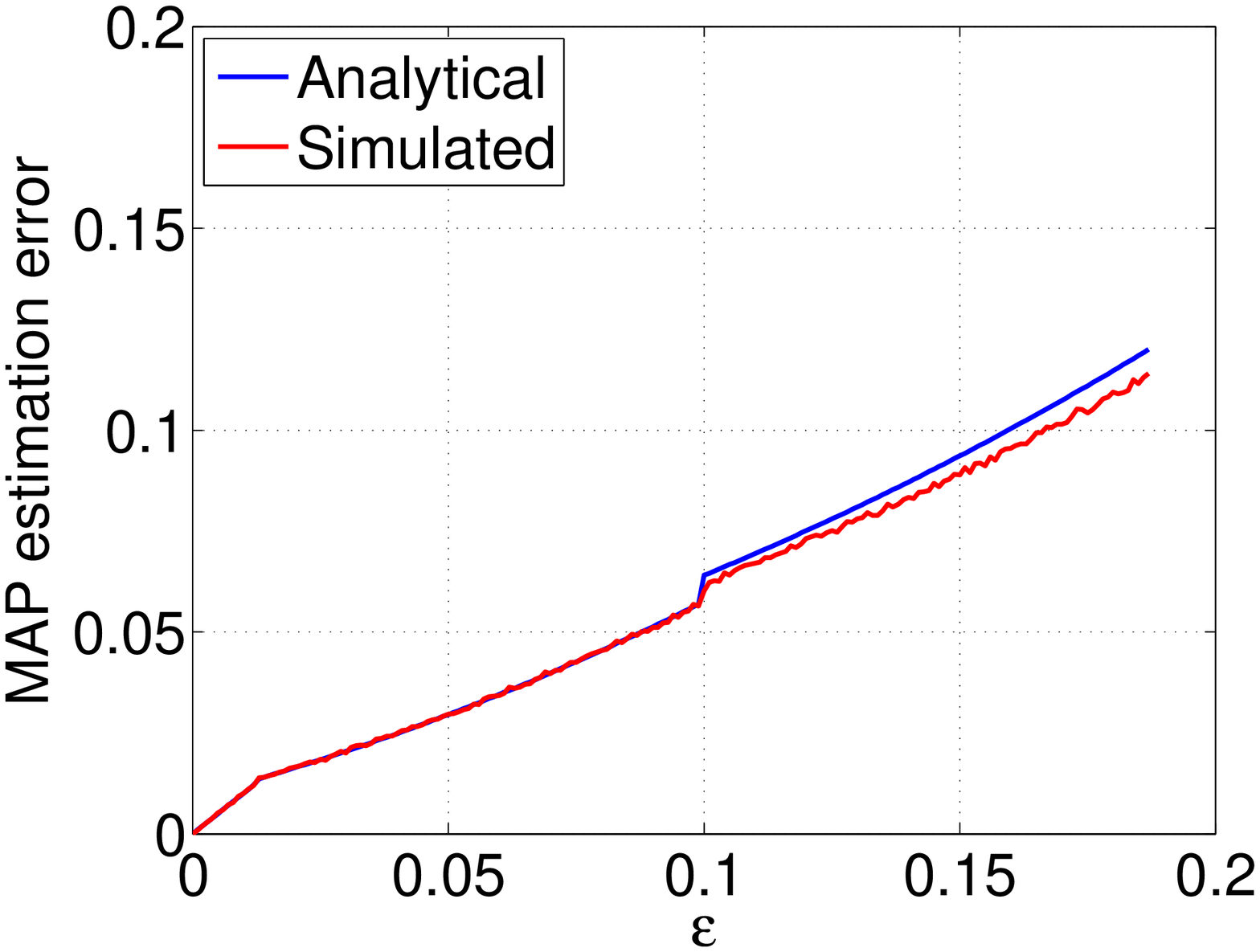} \label{fig:L-2}}
  \subfigure[$q=0.24$]{
  \includegraphics[width=0.3\textwidth]{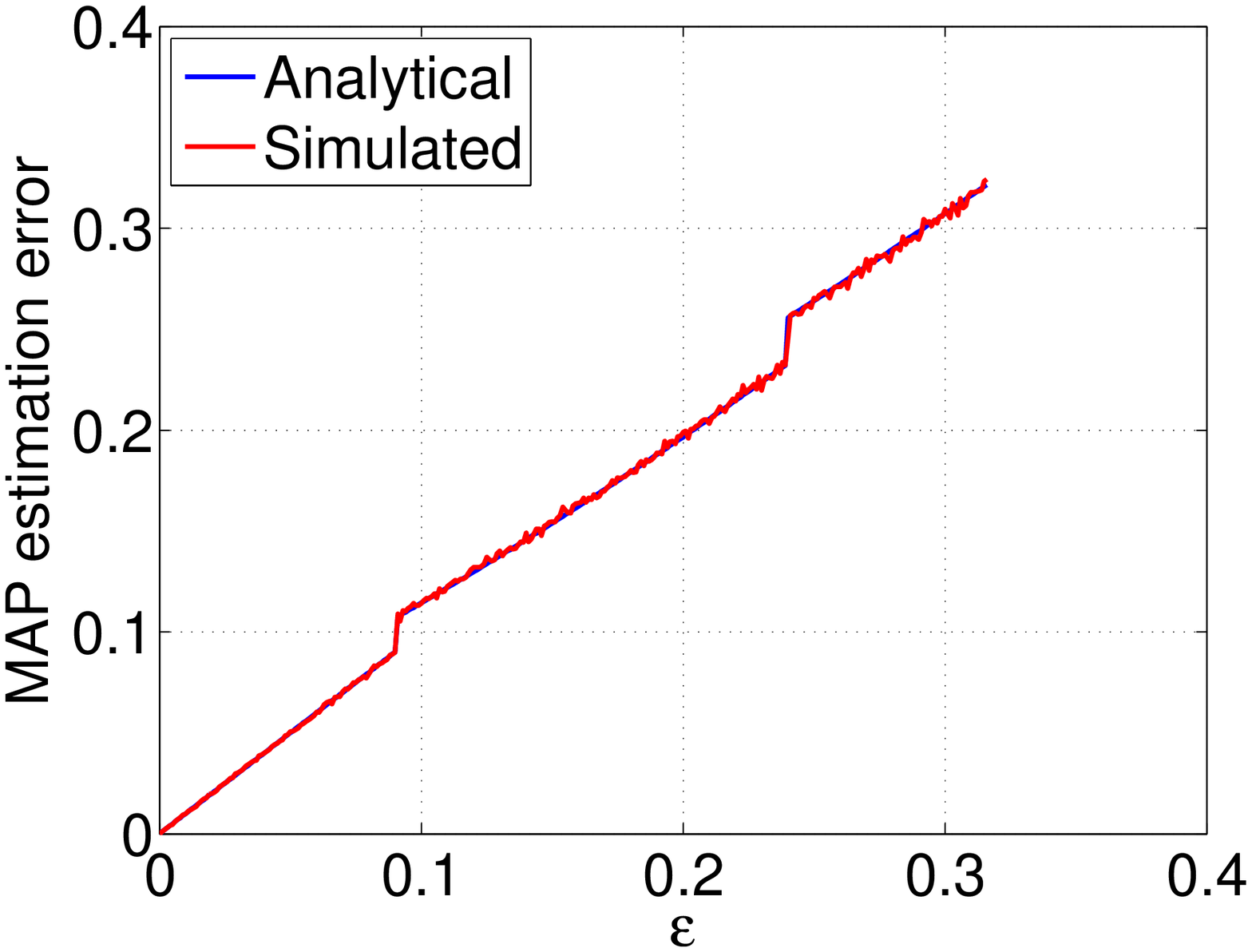} \label{fig:alpha-1}}
  \subfigure[$q=0.4$]{
  \includegraphics[width=0.3\textwidth]{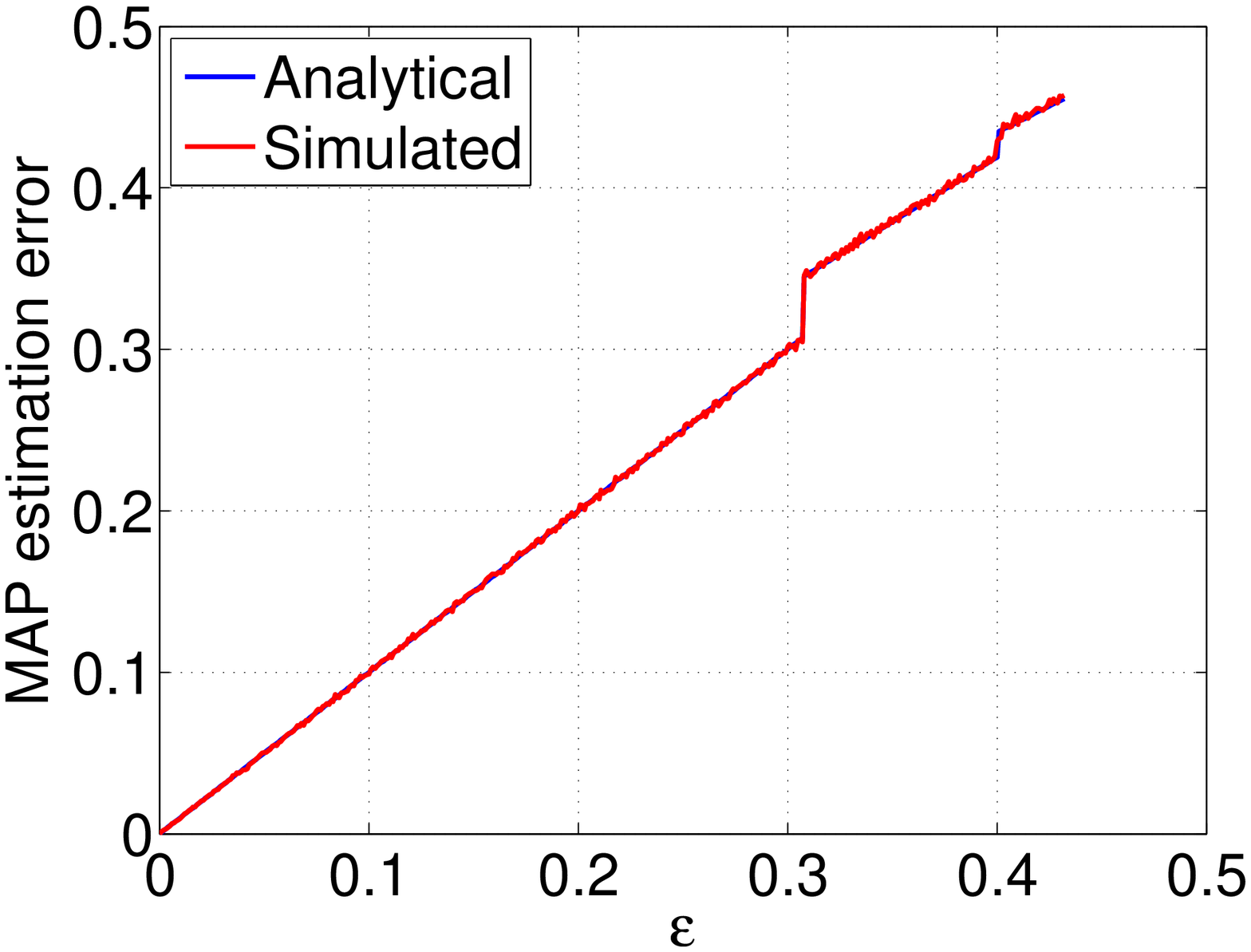} \label{fig:L-2}}
  \end{array}$
  \caption{(Color online) Estimation error plotted from MAP estimate using simulation
  (red) and analytical formula (blue) for a single observation channel}\label{fig:erroranalytical}
\end{figure}

Here we present results showing the correspondence of the simulated and the analytical data obtained
in the ranges of first three phase transitions, $h>2J$,$2J>h>J$ and $J>h>\frac{2J}{3}$. In Figure
\ref{fig:Analytical_error} we have already used this semi-analytical expression to estimate the MAP error for $1$ and $2$ channel systems at a particular value of $q$. In Figure
\ref{fig:erroranalytical} we plot the error calculated for a single observation channel at different values of $q$. On truncation of the infinite series \eqref{eq:MAP29err} at the fourth term we see an exact match between the analytical and the simulated data for larger values of $q$. However we can see that that around the third phase for $q=0.1$, the match is not exact. This implies that as we go to smaller values of $q$, for finding an exact match with the simulated data, we need to evaluate higher order terms of the infinite series.\\

\subsection{Gaussian observation model }
\label{sec:gaussian}
Now we consider a single Gaussian observation
channel. The Ising Hamiltonian for the case of Gaussian observation
channel is given by,
\begin{equation}\label{eq:MAP24aa}
H_G({\yv},\xv) = -\sum_{k=1}^{N-1} J x_k x_{k+1} - \sum_{k=1}^N h
x_k y_k
\end{equation}
where $h=\frac{1}{\sigma^2}$, $\sigma^2$ being the variance of Gaussian
noise. The noise distribution in the observation channel is given by,
\begin{equation}
\pi(y_k |x_k )=\frac{1}{\sqrt{2\pi}\sigma}e^{\frac{(y_k -x_k )^2}{2
{\sigma}^ 2} }
\end{equation}
Using the above equation with $\textbf{p}({\yv} | \xv)=\prod_{k=1}^N
\pi(y_k | x_k )$ we get after discarding the irrelevant additive
factors,
\begin{equation}
{\rm log} \textbf{p}({\yv} | \xv) = \sum_{k=1}^N \frac{1}{\sigma^2}
x_k y_k
\end{equation}

The recursion relation is given as
\begin{equation}\label{eq:MAP24ab}
\xi_k = h y_k + A(\xi_{k-1})
\end{equation}
where the generic form of the function $A(\xi)$ is defined in
(\ref{eq:MAP11}). It is easy to see that $-J \leq A(\xi) \leq J$.
Thus, $\xi$ (which we refer to as our state) takes the form $n J + h
y$ where $n = \{-1,0,1\}$. Since the states having $n = 0$ are
non-recurrent (see the argument in Section \ref{sec:recursion}), we
can quantify our state space in the form $\pm J + h y$, where $y \in
(-\infty,\infty)$. Thus, our state space is now continuous, as
opposed to being discrete in the case when the noise channel was
binary.

The conditional probabilities for the Gaussian distribution in the
observation channel can be written as,
 \begin{equation}\label{eq:MAP25}
\omega(\xi,y,z|\xi',y',z')=p(z|z')\pi(y|z)\varphi(\pm J+hy | \pm
J+hy',y)
\end{equation}
The exact expression for \eqref{eq:MAP25} is tabulated below which
is used to calculate the state transition probabilities.

\begin{table}[!htp]
\caption{Conditional probabilities for the Gaussian distribution of
the observation channels where $\omega(\xi,y,z|\xi',y',z') \equiv
\omega(aJ + hy,y,z | cJ + hy',y',z')$ and the combinations of
$(a,z),(c,z')$ are tabulated. }
\centering
\begin{tabular}{lllll}
  \hline \noalign{\smallskip}
  \toprule
  $(a,z)\downarrow , (c,z') \rightarrow $ & $(1,1)$ & $(1,-1)$ & $(-1,1)$ & $(-1,-1)$ \\
  \midrule
  $(1,1)$ & $\Sigma^1 | _{\varsigma^1 =1}$ & $\frac{q}{1-q}\Sigma^1 | _{\varsigma^1 = 1}$ & $\Sigma^2 | _{\varsigma^1 =1}$ & $\frac{q}{1-q} \Sigma^2 | _{\varsigma^1 = 1}$   \\
  $(1,-1)$ & $\frac{q}{1-q} \Sigma^1 | _{\varsigma^1 =-1}$ & $\Sigma^1 | _{\varsigma^1 = -1}$ & $\frac{q}{1-q} \Sigma^2 | _{\varsigma^1 =-1}$ & $\Sigma^2 | _{\varsigma^1 =-1}$   \\
  $(-1,1)$ & $\Sigma^4 | _{\varsigma^1 =1}$ & $\frac{q}{1-q} \Sigma^4 | _{\varsigma^1 = 1}$ & $\Sigma^3 | _{\varsigma^1 =1}$ & $\frac{q}{1-q}\Sigma^3 | _{\varsigma^1 = 1}$ \\
  $(-1,-1)$ & $\frac{q}{1-q} \Sigma^4 | _{\varsigma^1 =-1}$ & $\Sigma^4 | _{\varsigma^1 =-1}$ & $\frac{q}{1-q} \Sigma^3 | _{\varsigma^1 =-1}$ & $\Sigma^3 | _{\varsigma^1 = -1}$ \\

  \bottomrule
  \hline \noalign{\smallskip}
\end{tabular}
\label{tab:gauss-dist-table}
\end{table}

The quantities used in the table are given as,
\begin{eqnarray}\label{eq:MAP24a}
\Sigma^1 &=& \frac{1}{\sqrt{2 \pi}\sigma}(1-q)e^{-\frac{(y-\varsigma^1)^2}{2 \sigma^2}}I(y'>0)+\frac{1}{\sqrt{2 \pi}\sigma}(1-q)e^{-\frac{(y-y'-\varsigma^1)^2}{2 \sigma^2}}I \left(\frac{-2J}{h}<y'<0 \right) \nonumber \\
\Sigma^2 &=& \frac{1}{\sqrt{2 \pi}\sigma}(1-q)e^{-\frac{(y-\varsigma^1)^2}{2 \sigma^2}}I \left(y'>\frac{2J}{h} \right)\nonumber \\
\Sigma^3 &=& \frac{1}{\sqrt{2 \pi}\sigma}(1-q)e^{-\frac{(y-\varsigma^1)^2}{2 \sigma^2}}I(y'<0)+\frac{1}{\sqrt{2 \pi}\sigma}(1-q)e^{-\frac{(y- y'-\varsigma^1 )^2}{2 \sigma^2}}I \left(0<y'<\frac{2J}{h} \right)\nonumber \\
\Sigma^4 &=& \frac{1}{\sqrt{2
\pi}\sigma}(1-q)e^{-\frac{(y-\varsigma^1 )^2}{2
\sigma^2}}I \left(y'<\frac{-2J}{h} \right)
\end{eqnarray}

\begin{figure}[h]
  \centering
  \subfigure[PDF for $p_{st}$ at $\sigma=0.1$]{
  \includegraphics[width=0.4\textwidth]{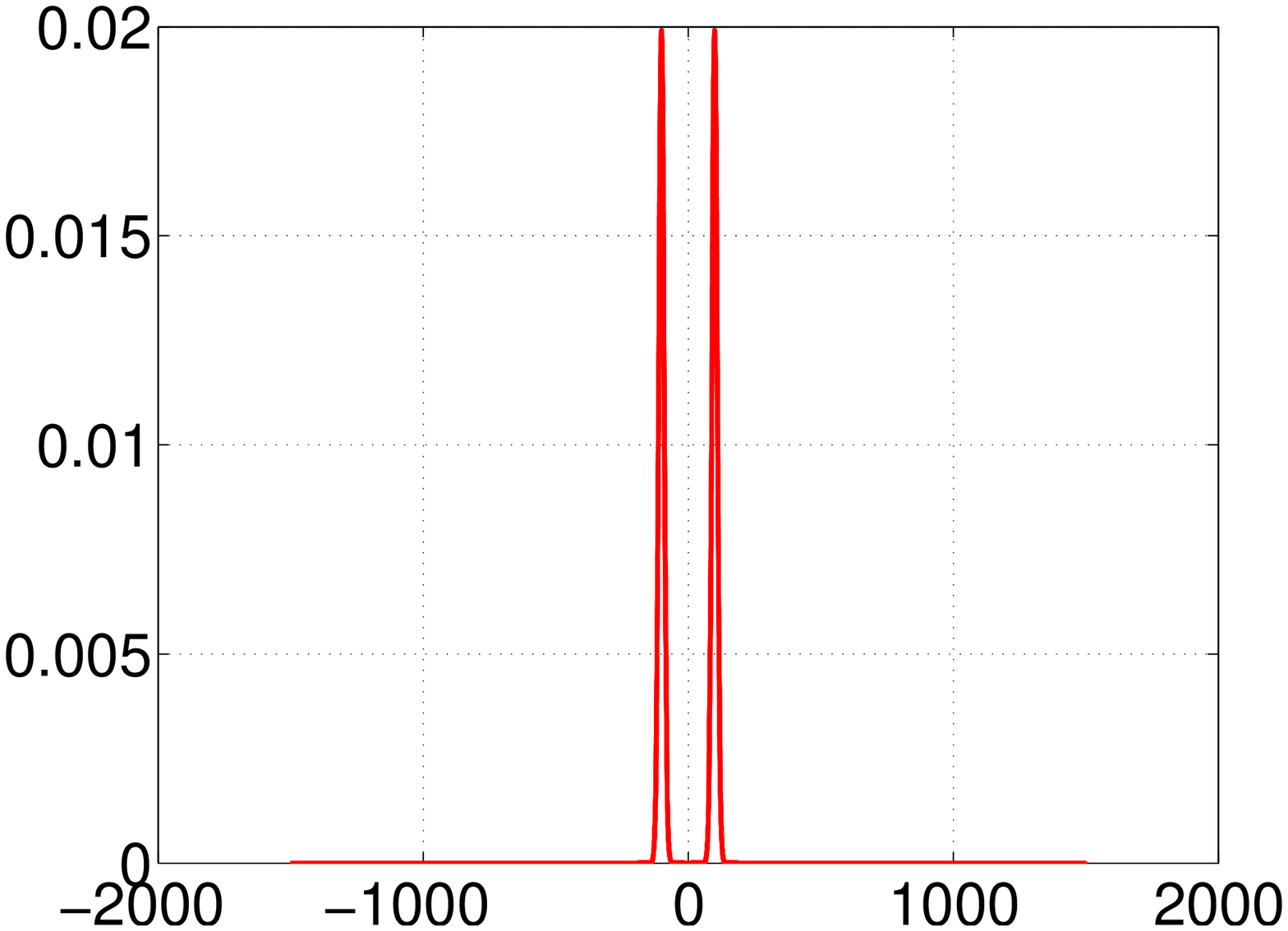} \label{fig:pdffig2}}
  \subfigure[CDF for $p_{st}$ at $\sigma=0.1$]{
  \includegraphics[width=0.4\textwidth]{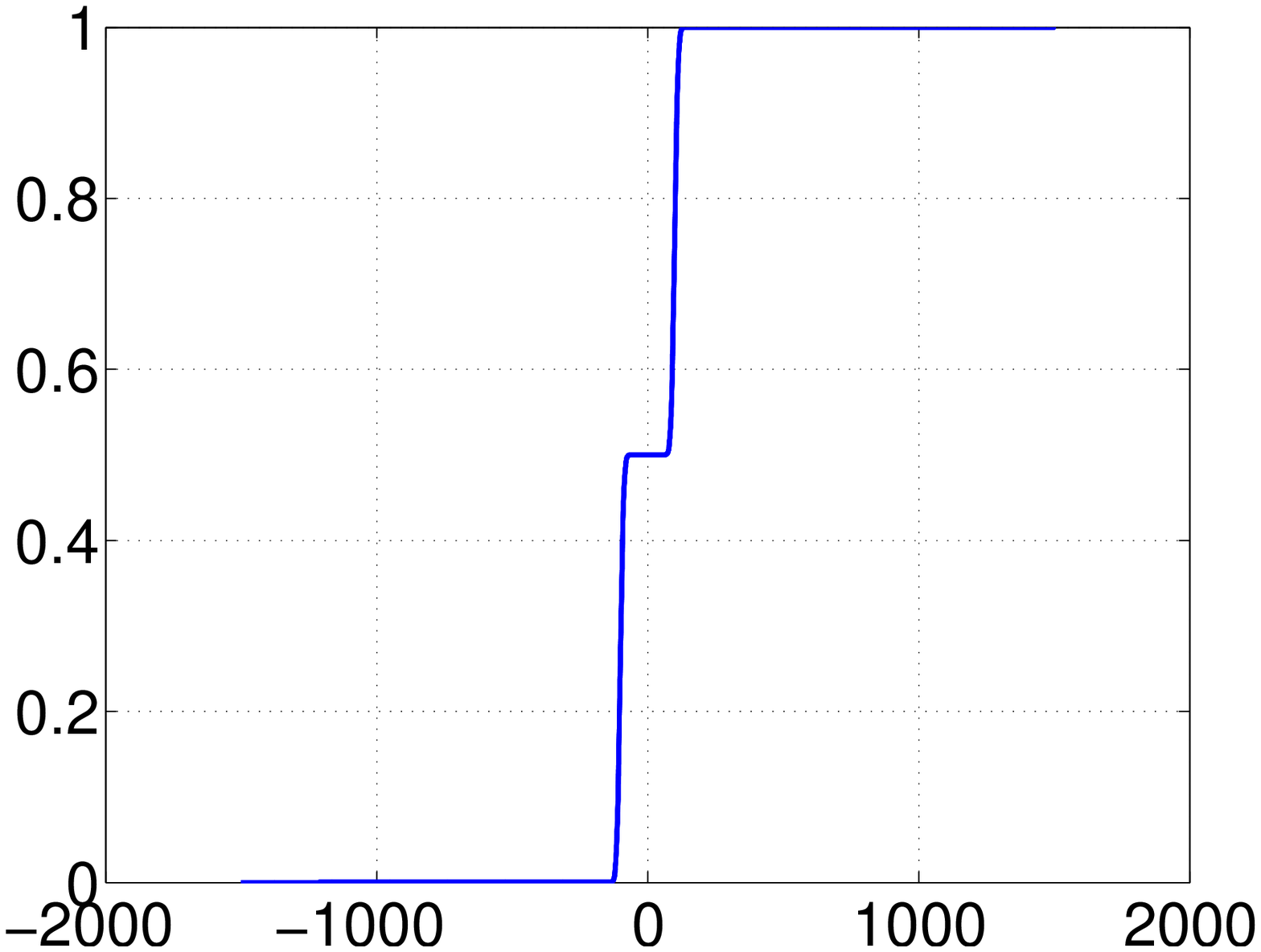} \label{fig:cdffig5}}
  \caption{(Color online) Plots for $p_{st}$ for $q=0.24$}\label{fig:error2}
\end{figure}

\begin{figure}[h]
  \centering
  \subfigure[PDF for $p_{st}$ at $\sigma=10$]{
  \includegraphics[width=0.4\textwidth]{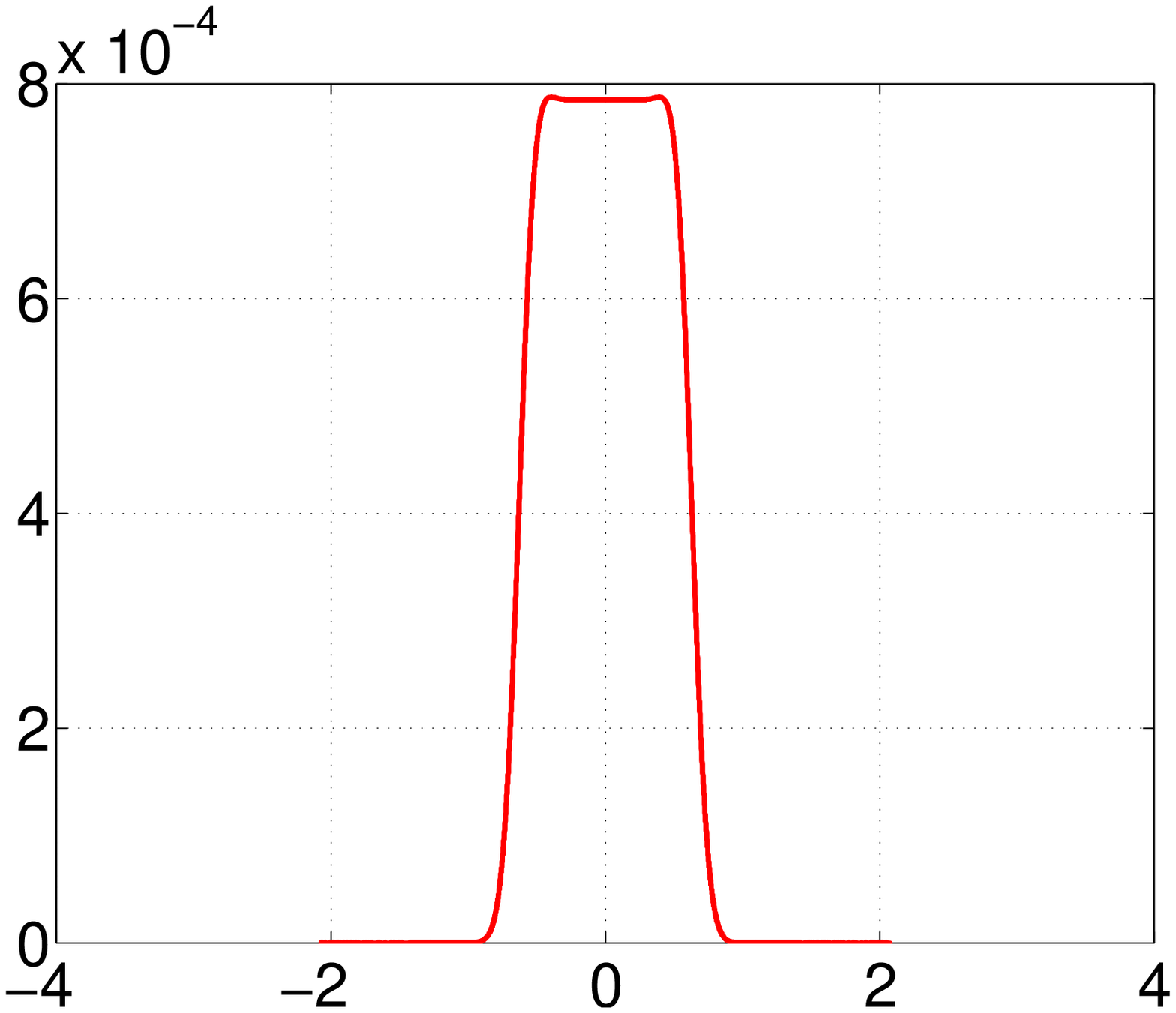} \label{fig:pdffig9}}
  \subfigure[CDF for $p_{st}$ at $\sigma=10$]{
  \includegraphics[width=0.4\textwidth]{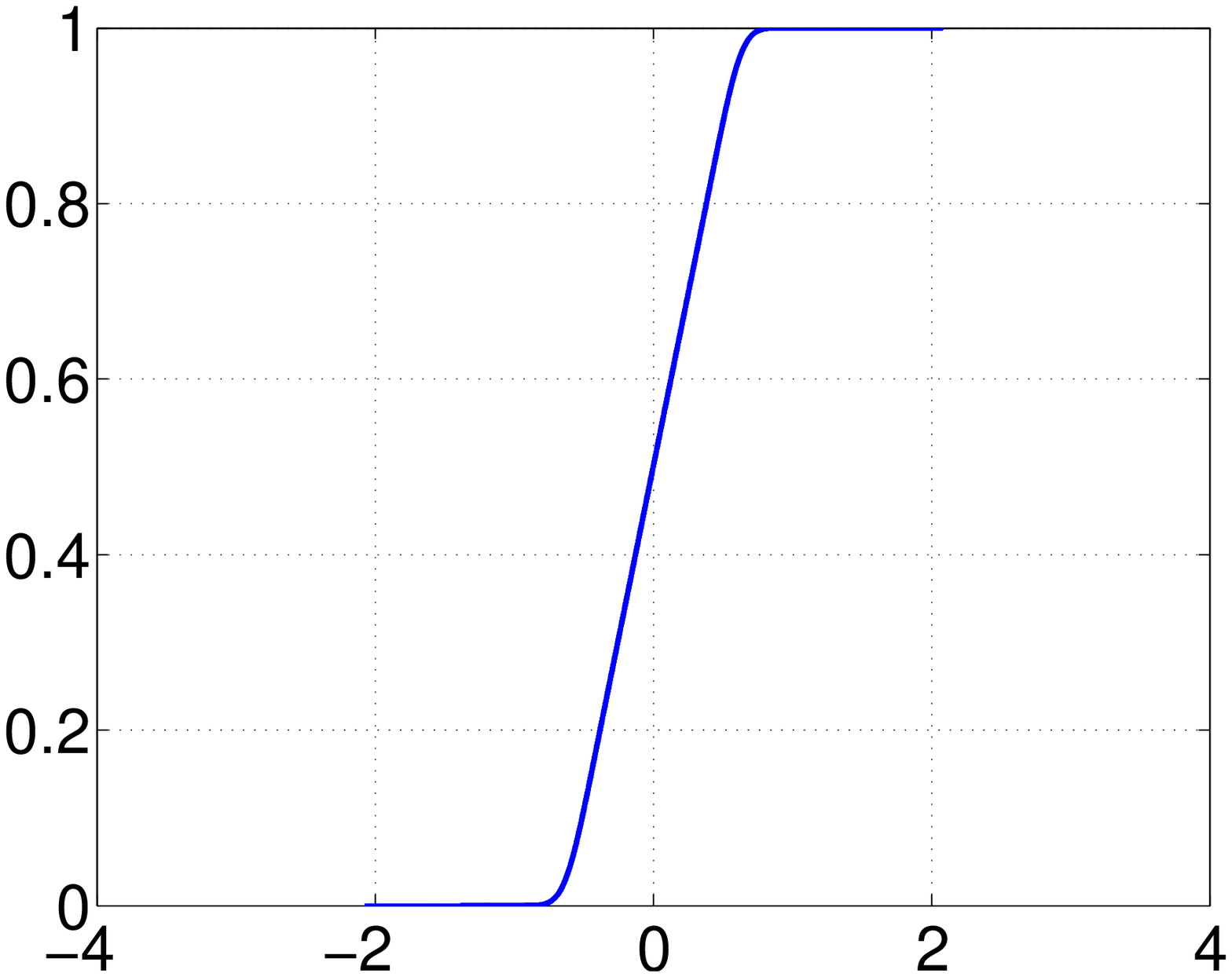} \label{fig:cdffig12}}
  \caption{(Color online) Plots for $p_{st}$ for $q=0.24$}\label{fig:error}
\end{figure}

We provide an explanation to calculate the entries corresponding to
the 1st row and the 1st and 3rd columns of Table
\ref{tab:gauss-dist-table}.
\begin{itemize}
\item $(a,z,c,z') = (1,1,1,1)$:
Qualitatively, this means that we were in a state $\xi' = J + hy'$
and $z' = 1$ and we moved to $\xi = J + hy$ and $z = 1$. From the
conditional probability expression (\ref{eq:MAP25}), we need to
compute
\begin{eqnarray*}
\omega(\xi = J+hy,y,1|\xi' = J+hy',y',1) &=& p(1|1)\pi(y|1)\varphi(J+hy | J+hy',y)\\
&=& (1-q) \pi(y|1) \varphi(J+hy | J+hy',y)
\end{eqnarray*}
Now, we can go from $J + hy'$ to $J + hy$ in two ways:
\begin{itemize}
\item When $y' > 0$ (we always assume $h > 0$), $A(J + hy') = J$ and
$\xi = A(\xi') + hy = J + hy$, so we should have $y$ as the channel
observation. Thus, we have
\begin{equation}
\label{eq:1}\pi(y|1) \varphi(J+hy | J+hy',y) = \frac{1}{\sqrt{2
\pi}\sigma}e^{-\frac{(y-1)^2}{2 \sigma^2}}
\end{equation}
\item When $\frac{-2J}{h} \leq y' \leq 0$, $A(J + hy') = J + hy'$
and $\xi = A(\xi') + hy = J + hy' + hy$. This means if we want to
have $\xi = J + hy$, our observation should be $y - y'$ since this
will give $\xi = A(\xi') + hy = J + hy' + h(y - y') = J + hy$. Thus,
we have
\begin{equation}
\label{eq:2} \pi(y|1) \varphi(J+hy | J+hy',y) = \frac{1}{\sqrt{2
\pi}\sigma}e^{-\frac{(y-y'-1)^2}{2 \sigma^2}}
\end{equation}
\end{itemize}
Combining (\ref{eq:1}) and (\ref{eq:2}), we get
\begin{eqnarray}
\omega(\xi = J+hy,y,1|\xi' = J+hy',y',1) &=& p(1|1)\pi(y|1)\varphi(J+hy | J+hy',y)\\
&=& \frac{1}{\sqrt{2 \pi}\sigma}(1-q)e^{-\frac{(y-1)^2}{2 \sigma^2}}I(y'>0) \nonumber\\
&& +\frac{1}{\sqrt{2 \pi}\sigma}(1-q)e^{-\frac{(y-y'-1)^2}{2 \sigma^2}}I\left(\frac{-2J}{h}<y'<0 \right) \nonumber \\
\end{eqnarray}

\item $(a,z,c,z') = (1,1,-1,1)$:
Qualitatively, this means that we were in a state $\xi' = -J + hy'$
and $z' = 1$ and we moved to $\xi = J + hy$ and $z = 1$. Again,
using the conditional probability expression (\ref{eq:MAP25}), we
have
\begin{eqnarray*}
\omega(\xi = J+hy,y,1|\xi' = -J+hy',y',1) &=& p(1|1)\pi(y|1)\varphi(J+hy | -J+hy',y)\\
&=& (1-q) \pi(y|1) \varphi(J+hy | -J+hy',y)
\end{eqnarray*}
Now, we can go from $-J + hy'$ to $J + hy$ in the following manner.
When $y' > \frac{2J}{h}$, $A(-J + hy') = J$ and $\xi = A(\xi') + hy
= J + hy$, so we should have $y$ as the channel observation. Thus,
we have
\begin{equation}
\label{eq:3}\pi(y|1) \varphi(J+hy | J+hy',y) = \frac{1}{\sqrt{2
\pi}\sigma}e^{-\frac{(y-1)^2}{2 \sigma^2}}
\end{equation}
We thus have from (\ref{eq:3})
\begin{eqnarray}
\omega(\xi = J+hy,y,1|\xi' = -J+hy',y',1) &=& p(1|1)\pi(y|1)\varphi(J+hy | -J+hy',y)\\
&=& (1-q) \frac{1}{\sqrt{2 \pi}\sigma}e^{-\frac{(y-1)^2}{2
\sigma^2}}I\left(y'> \frac{2J}{h} \right)
\end{eqnarray}
\end{itemize}

The other entries in Table \ref{tab:gauss-dist-table} can be
computed in a similar manner. The expressions for the stationary
states can be obtained from that of the conditional probabilities
as,
 \begin{equation}
\int_{\xi '} \int_{y '} \int_{z '}
\omega(\xi,y,z|\xi',y',z')p_{st}(\xi',y',z') d \xi' dy' dz'=
p_{st}(\xi,y,z)
\end{equation}

Now let us define for convenience,
\begin{eqnarray}
\omega(\xi = aJ+hy,y,z|\xi' = cJ+hy',y',z') &=& \Upsilon (a,y,z | c,y',z')\\
p_{st}(\xi=aJ+hy,y,z)&=& \Xi (a,y,z) \label{eq:MAP25aaa}
\end{eqnarray}
With the above notations we can conveniently write the coupled
integral expressions for finding the stationary probabilities as,
 \begin{equation}
\Xi (a,y,z)= \sum_{c=\pm 1} \sum_{z'=\pm 1} \int_{y'} \Upsilon
(a,y,z | c,y',z') \Xi (c,y',z') dy'
\end{equation}

It can be seen that there are four combinations of $a$ and $z$,
i.e., $\{(a,z) = (1,1),(1,-1),(-1,1),(-1,-1)\}$. Hence, we get four
coupled integral equations as follows {{
\begin{eqnarray*}
\Xi (1,y,1) &=& \int_{y' = 0}^{\infty} \frac{1}{\sqrt{2
\pi}\sigma}(1-q)e^{-\frac{(y-1)^2}{2 \sigma^2}} \Xi (1,y',1) dy' +
\int_{y' = \frac{-2J}{h}}^0 \frac{1}{\sqrt{2
\pi}\sigma}(1-q)e^{-\frac{(y-y'-1)^2}{2 \sigma^2}} \Xi (1,y',1) dy'
\\
&& + \int_{y' = 0}^{\infty} \frac{1}{\sqrt{2 \pi}\sigma} q
e^{-\frac{(y-1)^2}{2 \sigma^2}} \Xi (1,y',-1) dy' + \int_{y' =
\frac{-2J}{h}}^0 \frac{1}{\sqrt{2 \pi}\sigma} q
e^{-\frac{(y-y'-1)^2}{2 \sigma^2}} \Xi (1,y',-1) dy'
\\
&&   + \int_{y' = \frac{2J}{h}}^{\infty} \frac{1}{\sqrt{2
\pi}\sigma} (1 - q)
e^{-\frac{(y-1)^2}{2 \sigma^2}} \Xi (-1,y',1) dy' \\
&& + \int_{y' = \frac{2J}{h}}^{\infty} \frac{1}{\sqrt{2 \pi}\sigma}
q
e^{-\frac{(y-1)^2}{2 \sigma^2}} \Xi (-1,y',-1) dy' \\
\end{eqnarray*}
\begin{eqnarray*}
\Xi (1,y,-1) &=& \int_{y' = 0}^{\infty} \frac{1}{\sqrt{2 \pi}\sigma}
q e^{-\frac{(y+1)^2}{2 \sigma^2}} \Xi (1,y',1) dy' + \int_{y' =
\frac{-2J}{h}}^0 \frac{1}{\sqrt{2 \pi}\sigma} q
e^{-\frac{(y-y'+1)^2}{2 \sigma^2}} \Xi (1,y',1) dy'
\\
&& + \int_{y' = 0}^{\infty} \frac{1}{\sqrt{2 \pi}\sigma} (1-q)
e^{-\frac{(y+1)^2}{2 \sigma^2}} \Xi (1,y',-1) dy' + \int_{y' =
\frac{-2J}{h}}^0 \frac{1}{\sqrt{2 \pi}\sigma} (1-q)
e^{-\frac{(y-y'+1)^2}{2 \sigma^2}} \Xi (1,y',-1) dy'
\\
&& + \int_{y' = \frac{2J}{h}}^{\infty} \frac{1}{\sqrt{2 \pi}\sigma}
q
e^{-\frac{(y+1)^2}{2 \sigma^2}} \Xi (-1,y',1) dy' \\
&& + \int_{y' = \frac{2J}{h}}^{\infty} \frac{1}{\sqrt{2 \pi}\sigma}
(1-q)
e^{-\frac{(y+1)^2}{2 \sigma^2}} \Xi (-1,y',-1) dy' \\
\end{eqnarray*}
\begin{eqnarray*}
\Xi (-1,y,1) &=& \int_{y' = -\infty}^{0} \frac{1}{\sqrt{2
\pi}\sigma}(1-q)e^{-\frac{(y-1)^2}{2 \sigma^2}} \Xi (-1,y',1) dy' +
\int_{y' = 0}^{\frac{2J}{h}} \frac{1}{\sqrt{2
\pi}\sigma}(1-q)e^{-\frac{(y-y'-1)^2}{2 \sigma^2}} \Xi (-1,y',1) dy'
\\
&& + \int_{y' = -\infty}^{0} \frac{1}{\sqrt{2 \pi}\sigma} q
e^{-\frac{(y-1)^2}{2 \sigma^2}} \Xi (-1,y',-1) dy' + \int_{y' =
0}^{\frac{2J}{h}} \frac{1}{\sqrt{2 \pi}\sigma} q
e^{-\frac{(y-y'-1)^2}{2 \sigma^2}} \Xi (-1,y',-1) dy'
\\
&& + \int_{y' = -\infty}^{\frac{-2J}{h}} \frac{1}{\sqrt{2
\pi}\sigma} (1 - q)
e^{-\frac{(y-1)^2}{2 \sigma^2}} \Xi (1,y',1) dy' \\
&& + \int_{y' = -\infty}^{\frac{-2J}{h}} \frac{1}{\sqrt{2
\pi}\sigma} q
e^{-\frac{(y-1)^2}{2 \sigma^2}} \Xi (1,y',-1) dy' \\
\end{eqnarray*}
\begin{eqnarray*}
\Xi (-1,y,-1) &=& \int_{y' = -\infty}^{0} \frac{1}{\sqrt{2
\pi}\sigma} q e^{-\frac{(y+1)^2}{2 \sigma^2}} \Xi (-1,y',1) dy' +
\int_{y' = 0}^{\frac{2J}{h}} \frac{1}{\sqrt{2 \pi}\sigma} q
e^{-\frac{(y-y'+1)^2}{2 \sigma^2}} \Xi (-1,y',1) dy'
\\
&& + \int_{y' = -\infty}^{0} \frac{1}{\sqrt{2 \pi}\sigma} (1-q)
e^{-\frac{(y+1)^2}{2 \sigma^2}} \Xi (-1,y',-1) dy' + \int_{y' =
0}^{\frac{2J}{h}} \frac{1}{\sqrt{2 \pi}\sigma} (1-q)
e^{-\frac{(y-y'+1)^2}{2 \sigma^2}} \Xi (-1,y',-1) dy'
\\
&& + \int_{y' = -\infty}^{\frac{-2J}{h}} \frac{1}{\sqrt{2
\pi}\sigma} q
e^{-\frac{(y+1)^2}{2 \sigma^2}} \Xi (1,y',1) dy' \\
&& + \int_{y' = -\infty}^{\frac{-2J}{h}} \frac{1}{\sqrt{2
\pi}\sigma} (1-q)
e^{-\frac{(y+1)^2}{2 \sigma^2}} \Xi (1,y',-1) dy' \\
\end{eqnarray*}
}} These equations are solved numerically to find $p_{st}(\xi,y,z)$
using \eqref{eq:MAP25aaa}. The probability density function (PDF)
and the cumulative density function (CDF) are shown for two extreme
values of sigma in Figures \ref{fig:error2} and \ref{fig:error}.

From this, it is easy to see that the distribution is continuous and
the entropy which is given by \eqref{eq:MAP22} is found to be zero.

\end{document}